\documentclass[12pt,preprint,psfig,epsfig,graphics]{emulateapj}

\begin{document}

\title{$UV$ Continuum Slope and Dust Obscuration from
  $\lowercase{z}\sim6$ to $\lowercase{z}\sim2$: The Star Formation
  Rate Density at High Redshift$^{1}$}
\author{R.J. Bouwens$^{2,3}$, 
G.D. Illingworth$^{2}$, M. Franx$^{3}$, R-R. Chary$^{4}$, G.R. Meurer$^{5}$, 
C.J. Conselice$^{6}$, H. Ford$^{5}$, M. Giavalisco$^{7}$, P. van Dokkum$^{8}$}
\affil{1 Based on observations made with the NASA/ESA Hubble Space
  Telescope, which is operated by the Association of Universities for
  Research in Astronomy, Inc., under NASA contract NAS 5-26555. These
  observations are associated with programs \#7235, 7817, 9425, 9575,
  9797, 9803, 9978, 9979, 10189, 10339, 10340, 10403, 10504, 10530,
  10632, 10872, 10874, 11082, and 11144.}
\affil{2 Astronomy Department, University of California, Santa Cruz,
CA 95064}
\affil{3 Leiden Observatory, University of Leiden, Postbus 9513, 2300 RA Leiden, Netherlands}
\affil{4 Divison of Physics, Mathematics, and Astronomy, California Institute of Technology, Pasadena, CA 91125}
\affil{5 Department of Physics and Astronomy, Johns Hopkins University, 3400 North Charles Street, Baltimore, MD 21218}
\affil{6 University of Nottingham, School of Physics \& Astronomy, Nottinghm NG7 2RD}
\affil{7 Department of Astronomy, University of Massachusetts, Amherst, MA}
\affil{8 Department of Astronomy, Yale University, New Haven, CT}
\begin{abstract}
We provide a systematic measurement of the rest-frame UV continuum
slope $\beta$ over a wide range in redshift ($z$$\,\sim\,$2-6) and
rest-frame $UV$ luminosity (0.1$L_{z=3}^{*}$ to 2$L_{z=3}^{*}$) to
improve estimates of the SFR density at high redshift.  We utilize the
deep optical and infrared data (ACS/NICMOS) over the CDF-South and
HDF-North GOODS fields, as well as the UDF for our primary $UBVi$
``dropout'' Lyman Break Galaxy sample.  We also use strong lensing
clusters to identify a population of very low luminosity,
high-redshift dropout galaxies.  We correct the observed distributions
for both selection biases and photometric scatter.  We find that the
$UV$-continuum slope of the most luminous galaxies is substantially
redder at $z$$\,\sim\,$2-4 than it is at $z$$\,\sim\,$5-6 (from
$\sim-2.4$ at $z\sim6$ to $\sim-1.5$ at $z\sim2$).  Lower luminosity
galaxies are also found to be bluer than higher luminosity galaxies at
$z\sim2.5$ and $z\sim4$.  We do not find a large number of galaxies
with $\beta$'s as red as $-1$ in our dropout selections at $z\sim4$,
and particularly at $z\gtrsim5$, even though such sources could be
readily selected from our data (and also from Balmer Break Galaxy
searches at $z\sim4$).  This suggests that star-forming galaxies at
$z\gtrsim5$ almost universally have very blue $UV$-continuum slopes,
and that there are not likely to be a substantial number of
dust-obscured galaxies at $z\gtrsim5$ that are missed in ``dropout''
searches.  Using the same relation between $UV$-continuum slope and
dust extinction as has been found to be appropriate at both $z\sim0$
and $z\sim2$, we estimate the average dust extinction of galaxies as a
function of redshift and $UV$ luminosity in a consistent way.  As
expected, we find that the estimated dust extinction increases
substantially with cosmic time for the most $UV$ luminous galaxies,
but remains small ($\lesssim2\times$) at all times for lower
luminosity galaxies.  Because these same lower luminosity galaxies
dominate the luminosity density in the $UV$ continuum, the overall
dust extinction correction remains modest at all redshifts and the
evolution of this correction with redshift is only modest.  We include
the contribution from ULIRGs in our star formation rate density
estimates at $z$$\,\sim\,$2-6, but find that they contribute only
$\sim$20\% of the total at $z$$\,\sim\,$2.5 and $\lesssim$10\% at
$z\gtrsim4$.

\end{abstract}
\keywords{galaxies: evolution --- galaxies: high-redshift}

\section{Introduction}

\begin{deluxetable*}{ccccccc}
\tablewidth{5in}
\tabletypesize{\footnotesize}
\tablecaption{Observational data used to select $z\sim2-6$ $U$, $B$,
  $V$, and $i$ dropouts and to measure the $UV$-continuum
  slope.\label{tab:obsdata}}
\tablehead{
\colhead{} & \multicolumn{2}{c}{Optical} & \colhead{$z_{850}$-band}
& \multicolumn{2}{c}{Near-IR} & \colhead{$H_{160}$-band} \\
\colhead{Field} & \colhead{Passbands} & \colhead{Area\tablenotemark{a}} & \colhead{$5\sigma$ limit\tablenotemark{b}} 
& \colhead{Passbands} & \colhead{Area\tablenotemark{a}} & \colhead{$5\sigma$ limit\tablenotemark{b}}}
\startdata
HDF-N & $U_{300}B_{450}V_{606}I_{814}$ & 5 & 27.6 & $J_{110}H_{160}$ & 5 & 27.3 \\
HDF-S & $U_{300}B_{450}V_{606}I_{814}$ & 4 & 28.0 & --- & --- & --- \\
GOODS & $B_{435}V_{606}i_{775}z_{850}$ & 316 & 27.4 & $J_{110}H_{160}$ & 18 & 26.4 \\
GOODS & ... & ... & ... & $H_{160}$ & 36 & 26.4 \\
HUDF05 & $B_{435}V_{606}i_{775}z_{850}$ & 22 & 28.4 & $J_{110}H_{160}$ & 2.5 & 28.6 \\
HUDF & $B_{435}V_{606}i_{775}z_{850}$ & 11 & 28.8 & $J_{110}H_{160}$ & 6 & 27.2 \\
Abell 2218 & $g_{475}r_{625}i_{775}z_{850}$ & 11 & 27.2 & $J_{110}H_{160}$ & 1.5 & 26.8 \\
MS1358 & $g_{475}r_{625}i_{775}z_{850}$ & 11 & 27.2 & $J_{110}H_{160}$ & 1.5 & 26.8 \\
CL0024 & $g_{475}r_{625}i_{775}z_{850}$ & 11 & 27.2 & $J_{110}H_{160}$ & 3.0 & 26.8 \\
Abell 1689 & $g_{475}r_{625}i_{775}z_{850}$ & 11 & 27.2 & ---\tablenotemark{c} & --- & --- \\
Abell 1703 & $g_{475}r_{625}i_{775}z_{850}$ & 11 & 27.2 & ---\tablenotemark{c} & --- & --- 
\enddata
\tablenotetext{a}{Arcmin$^2$}
\tablenotetext{b}{$5\sigma$ point-source limiting magnitude in $0.6''$ diameter
  apertures.}
\tablenotetext{c}{While NICMOS $J_{110}$ observations are available over both Abell 1689 and Abell 1703, these clusters lack deep coverage at $\sim1.6\mu$m and so we do not make use of the existing NICMOS data to measure $UV$-continuum slopes $\beta$ for $z\sim5$ $V$-dropout or $z\sim6$ $i$-dropout selections.}
\end{deluxetable*}

Quantifying the star-formation rate (SFR) density at $z\gtrsim3$ is a
challenging endeavor.  While there are a wide variety of techniques to
estimate this rate at $z\lesssim2$ from light at different wavelengths
(Condon 1992; Kennicutt 1998; Madau et al.\ 1998; Ranalli et
al.\ 2003; Y{\"u}ksel et al.\ 2008; Li 2008; Reddy et al.\ 2006; Yun
et al.\ 2001; Reddy \& Steidel 2004; Reddy et al.\ 2006; Erb et
al.\ 2006b), the situation becomes considerably more uncertain at
redshift $z\gtrsim3$.  This is because most of the techniques
effective at low redshift rely upon light at wavelengths that simply
cannot be detected at high redshifts for all but the most exceptional
sources.  Determinations of the SFR density from x-ray or radio
emission appear to be just possible at $z$$\,\sim\,$2-4 by stacking a
large number of sources in very deep integrations (e.g., Seibert et
al.\ 2002; Nandra et al.\ 2002; Reddy \& Steidel 2004), but do not
work at $z\geq5$ (Lehmer et al.\ 2005; Carilli et al.\ 2008).
Similarly, while deep mid-IR data would seem to be quite effective in
estimating the SFR at $z\sim2$ (e.g., Reddy et al.\ 2006; Caputi et
al.\ 2007), at $z\geq3$ many of the key spectral features redshift to
wavelengths not very amenable to study with current instrumentation.

This leaves us with sparingly few techniques for estimating the SFR
density at high redshift, most of which have only recently been
developed.  These techniques include traditional approaches like a
consideration of the $UV$ light to estimate the SFR densities to newer
approaches that use gamma-ray burst (GRB) rate densities (e.g.,
Y{\"u}ksel et al.\ 2008; Li 2008) or possible detections of redshifted
H$\alpha$ (e.g., Chary et al.\ 2005) to make these estimates.

Perhaps, the simplest and most practical of these techniques is to
rely on the $UV$ light emanating from the high-redshift galaxies
themselves.  Young stars emit a large fraction of their energy at
these wavelengths, and it is very easy using current instrumentation
to measure this energy to very low SFR rates (i.e., $\sim$0.1
$M_{\odot}$ yr$^{-1}$ at z$\,\sim\,$4 in the Hubble Ultra Deep Field
[HUDF: Beckwith et al.\ 2006]).  One of the most significant
challenges in establishing the SFR from the UV data is estimating the
absorption by dust.  Since $UV$ light is subject to substantial
attenuation by dust, this entire issue is quite important.  One of the
most practical methods for determining the dust attenuation is through
the $UV$-continuum slopes $\beta$ ($f_{\lambda}\propto
\lambda^{\beta}$).  Since these slopes have been shown to be well
correlated with the dust extinction at $z\sim0$ (Meurer et al.\ 1995;
Meurer et al.\ 1997; Meurer et al.\ 1999; Burgarella et al. 2005; Dale
et al.\ 2007; Treyer et al.\ 2007; Salim et al.\ 2007), $z\sim1$
(Laird et al.\ 2005), and $z\sim2$ (Reddy et al.\ 2006; Daddi et
al.\ 2004), we can use these slopes to estimate the dust obscuration
at even higher redshift.

Not surprisingly, there have already been a number of attempts to
determine these slopes at high redshift ($z$$\,\sim\,$2-6) from the
available imaging data.  Adelberger \& Steidel (2000) and Meurer et
al.\ (1999) examined the UV slopes for a sample of $z\sim3$
$U$-dropouts and inferred modest (factor of $\sim$5) dust extinctions.
Other groups (Lehnert \& Bremer 2003; Ouchi et al.\ 2004; Stanway et
al.\ 2005; Yan et al.\ 2005; Bouwens et al.\ 2006) have looked at
these slopes at higher redshifts ($z\sim4-6$), again on a dropout
selection.  Broadly, it was found that higher redshift galaxies had
bluer $UV$-continuum slopes than lower redshift galaxies have,
suggesting that the typical dust extinction at high redshift is
significantly less than it is at lower redshift.  There was also some
evidence from previous analyses (e.g., Meurer et al.\ 1999 at
$z\sim2.5$) that lower luminosity galaxies had bluer $UV$-continuum
slopes $\beta$ and thus lower dust extinctions (though the emphasis in
Meurer et al.\ (1999) was the correlation between $\beta$ and the
dust-corrected luminosity, not the observed luminosity).

Regrettably, apart from these two trends (higher redshift galaxies are
bluer and lower luminosity galaxies are bluer), it has been difficult
to make quantitative statements about the distribution of
$UV$-continuum slopes or dust extinctions for galaxies at high
redshift.  In part, this can be attributed to the piecemeal way the
$UV$-continuum slopes and dust extinction have been derived at high
redshift, as different analyses have derived $UV$-continuum slopes
based upon a wide variety of disparate high-redshift galaxy samples,
using different techniques (not all of which are consistent).  In
part, this can also be attributed to the limited amounts of data
available to each particular study.  Fortunately, the situation has
begun to change, and there is now a wide variety of HST data that are
available to select high redshift galaxies and determine their slopes
over a wide range in redshift and luminosity.  Determining how these
slopes also depend upon luminosity seems particularly relevant given
the fact that these sources likely dominate the $UV$ luminosity
density and perhaps star formation rate density at $z\gtrsim2$ (e.g.,
Bouwens et al.\ 2006; Sawicki \& Thompson 2006b; Bouwens et al.\ 2007;
Reddy et al.\ 2008; Yan \& Windhorst 2004; Reddy \& Steidel 2009).

It seems clear that a comprehensive, systematic determination of the
distribution of $UV$-continuum slopes is needed, both as a function of
redshift and luminosity.  The goal of the present paper is to provide
just such an analysis, taking advantage of the considerable quantity
of deep HST data to clarify our understanding of dust obscuration and,
by consequence, the star formation rate density at high redshift.  We
select galaxies for $UV$-continuum slope measurements over a wide
range in redshift using well-tested $z\sim2.5$ $U$, $z\sim4$ $B$,
$z\sim5$ $V$, and $z\sim6$ $i$ dropout selections.  We use the
wide-area ACS Great Observatories Origins Deep Survey (GOODS) fields
(Giavalisco et al.\ 2004a) to quantify these slopes at bright
magnitudes and extend our analyses to very low luminosities by
examining very deep HST data sets like the HUDF, and also by examining
dropouts behind massive galaxy clusters where the substantial
magnification factors permit us to consider very faint sources.

The plan for this manuscript is as follows.  In \S2, we present the
observational data set.  In \S3, we describe our procedure for
generating source catalogs from these data, selecting the dropout
samples, and estimating the $UV$-continuum slopes while correcting for
the effect of object selection and photometric scatter.  Finally, we
discuss our results (\S4), examine the likely implications of these
results for the effective dust extinction and SFR density at
$z$$\,\sim\,$2-6 (\S5), and then include a summary (\S6).  Throughout
this work, we will find it convenient to quote results in terms of the
luminosity $L_{z=3}^{*}$ Steidel et al.\ (1999) derived at $z\sim3$,
i.e., $M_{1700,AB}=-21.07$, for consistency with previous work --
though we note that the Steidel et al.\ (1999) LF results are now
updated ($M_{1700,AB}=-20.97\pm0.14$: Reddy \& Steidel 2009) but still
consistent with the previous determination.  Where necessary, we
assume $\Omega_0 = 0.3$, $\Omega_{\Lambda} = 0.7$, $H_0 =
70\,\textrm{km/s/Mpc}$.  Although these parameters are slightly
different from those determined from the WMAP five-year results
(Dunkley et al.\ 2009), they allow for convenient comparison with
other recent results expressed in a similar manner.  Unless otherwise
stated, the $UV$-continuum continuum slope $\beta$ presented here all
assume a 1600\AA-2300\AA$\,\,$baseline.  We express all magnitudes in
the AB system (Oke \& Gunn 1983).

\begin{figure*}
\epsscale{1.13}
\plotone{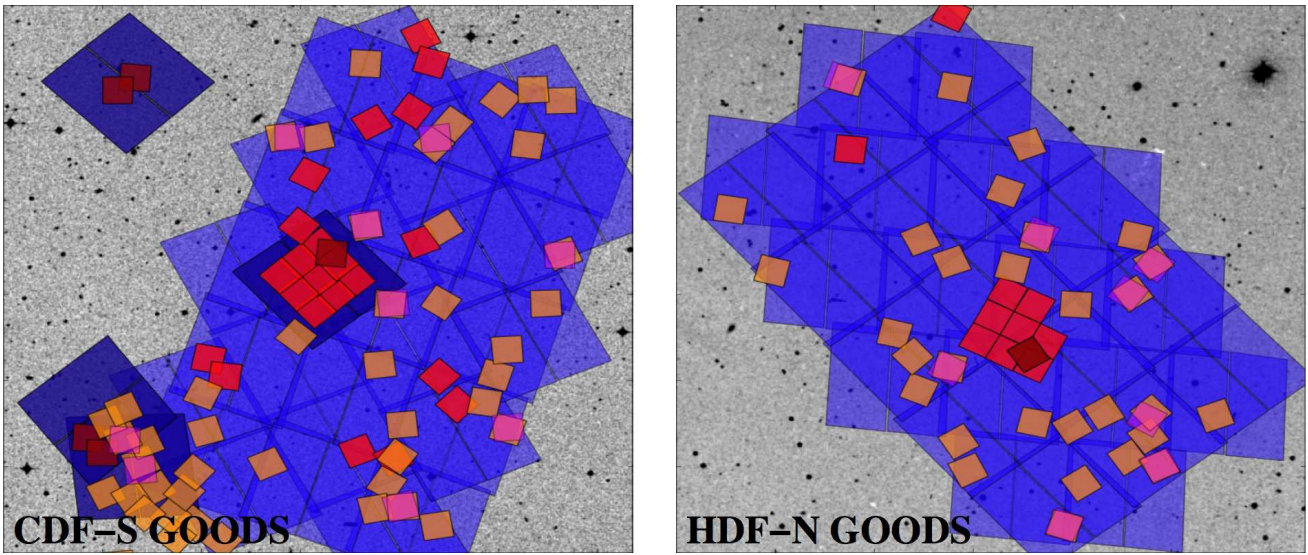}
\caption{Illustration of the deep near-IR data available over the two
  GOODS fields.  Deep near-IR data are essential for measurements of
  the UV-continuum slope of galaxies at $z\gtrsim5$ and valuable for a
  measurement of these slopes at $z\sim4$.  The red and dark red
  regions correspond to regions with deep and very deep NICMOS
  $J_{110}$ and $H_{160}$ coverage, respectively ($5\sigma$ depths
  $\gtrsim26.5$ and $\gtrsim28$ AB mag).  The light orange regions
  have deep NICMOS $H_{160}$-band coverage ($5\sigma$ depths of
  $\gtrsim26.5$: H. Teplitz et al.\ 2009, in prep; C. Conselice et
  al.\ 2009, in prep) while the magenta regions have deep NICMOS
  $J_{110}$-band coverage ($5\sigma$ depths of $\gtrsim26.5$:
  R.J. Bouwens et al.\ 2009, in prep).  The blue and dark blue regions
  correspond to regions with deep and very deep optical ACS
  $V_{606}i_{775}z_{850}$ coverage, respectively ($5\sigma$ depths of
  $\gtrsim28$ and $\gtrsim29$ AB mag).  For more details on the NICMOS
  data over the two GOODS fields, see Bouwens et al.\ (2008) and
  R.J. Bouwens et al.\ (2009, in prep).\label{fig:obsdata}}
\end{figure*}

\section{Observational Data}

The purpose of this study is to measure the $UV$-continuum slope over
as wide a range in luminosity and redshift as is practical from
current observations.  Of course, establishing this distribution even
over a limited range in redshift or luminosity can be a challenge, as
it requires very deep imaging over a wide wavelength range, both to
select galaxies through the Lyman break technique and to measure their
UV-continuum slope redward of this break.  It also requires sufficient
search area to provide a sufficiently large samples of sources to
effectively define the distribution of $UV$-continuum slopes.

At $z\sim2.5$, we use the deep $U$-band coverage available over the
WFPC2 Hubble Deep Field (HDF) North (Williams et al.\ 1996) and HDF
South (Casertano et al.\ 2000) to select $z\sim2.5$ $U$-dropouts.
Those two deep WFPC2 fields are invaluable for defining the
distribution of $UV$-continuum slopes at lower luminosities, but their
small area limits their value for establishing the distribution at the
higher luminosities.  However, as we will see, the $UV$-continuum
slope distribution for bright $z\sim2.5$ galaxies is similar to what
Adelberger \& Steidel (2000) and Reddy et al.\ (2008) derived
previously.  For our $z\sim4$ $B$, $z\sim5$ $V$, and $z\sim6$
$i$-dropout samples, we utilize the ACS $B_{435}V_{606}i_{775}z_{850}$
data available over the HUDF, HUDF05 (Oesch et al.\ 2007; Bouwens et
al.\ 2007), and GOODS fields.  The wide-area GOODS fields are used for
establishing the $UV$-continuum slope distribution for luminous
galaxies while the deeper HUDF and HUDF05 fields are used for
establishing the $UV$-continuum slope distribution for lower
luminosity galaxies.  The reduction of these latter data was performed
with \textsl{``apsis''} (Blakeslee et al.\ 2003) and is described in
Bouwens et al.\ (2006, 2007).  A summary of the properties of these
fields is given in Table~\ref{tab:obsdata}.

Importantly, the HUDF, HUDF05, and GOODS fields also have $\sim70$
arcmin$^2$ of very deep near-IR coverage with NICMOS -- reaching
$5\sigma$ limits of $\gtrsim26.5$ ($0.6''$-diameter apertures).  This
coverage is important for measurements of the $UV$-continuum slope
$\beta$ for galaxies in our $z\sim5$ $V$-dropout and $z\sim6$
$i$-dropout selections.  The NICMOS data are described in more detail
in R.J. Bouwens et al.\ (2009, in prep) and are illustrated in
Figure~\ref{fig:obsdata} (see also Bouwens et al.\ 2008).  These data
were obtained as a result of the cumulative observations of many
different HST programs (e.g., Dickinson 1999; Thompson et al. 1999;
Thompson et al. 2005; Bouwens et al.\ 2005; Oesch et al.\ 2007; Riess
et al.\ 2007; C. Conselice et al.\ 2009, in prep).  All the NICMOS
data taken since 2002 (when the NICMOS cryocooler went into operation)
were reduced with our NICMOS pipeline \textsl{``nicred.py''} (Magee et
al.\ 2007).

Finally, we take advantage of the deep HST data over three massive
galaxy clusters to identify a small number of lower luminosity
star-forming galaxies at $z\sim4-6$ behind these clusters.  Because of
lensing by the foreground cluster (magnifying faint $z$$\,\sim\,$5-6
galaxies by factors of $\gtrsim4$), it is possible to select and to
measure $UV$-continuum slopes for a small number of extremely low
luminosity galaxies.  The five galaxy clusters (MS1358, CL0024, Abell
2218, Abell 1689, Abell 1703) all have deep ACS
$B_{435}g_{475}V_{555}r_{625}i_{775}z_{850}$ data and cover 55
arcmin$^2$ in total.  These clusters also have 6 arcmin$^2$ of deep
$J_{110}H_{160}$ NICMOS data necessary to measure the $UV$-continuum
slope for $z\sim5$ $V$ dropout and $z\sim6$ $i$-dropout galaxies
behind these clusters.  The ACS data available over these clusters
were also reduced with \textsl{``apsis''} while the NICMOS data were
reduced with \textsl{``nicred.py''}.  These reductions are described
in more detail in Bouwens et al.\ (2009a) and Zheng et al.\ (2009).

\section{Analysis}

In this section, we describe our procedure for determining the
distribution of $UV$-continuum slopes $\beta$ from our observational
data as a function of redshift and $UV$ luminosity.  We begin by
discussing how we do photometry on our observational data and then
select $z$$\,\sim\,$2-6 galaxies using various dropout criteria
(\S3.1-3.2).  In \S3.3, we calibrate our optical/near-IR photometry
and then derive $UV$-continuum slopes from the measured colors
(\S3.4).  By combining measurements of the $UV$-continuum slope
determinations from both very deep and wide-area data sets, we derive
the $UV$-continuum slope distribution over a wide range in luminosity
(\S3.5).  In \S3.6, we take advantage of lensing from massive galaxy
clusters to examine the $UV$-continuum slope $\beta$ distribution for
a set of lower luminosity galaxies.  We correct the observed
distributions of $UV$-continuum slopes $\beta$ for the effect of
object selection and photometric scatter (\S3.7).  Finally, we
discuss the uncertainties that likely exist in our determinations
(\S3.8) and then give our final estimates of the $UV$-continuum slope
distributions in \S3.9.

\subsection{Photometry}

We used SExtractor (Bertin \& Arnouts 1996) to perform object
detection and photometry.  Object detection was performed using the
square root of the $\chi^2$ images (Szalay et al.\ 1999: essentially a
coaddition of the relevant images) constructed from all ACS images
with coverage redward of the dropout band (e.g.,
$V_{606}i_{775}z_{850}$ in the case of a $B_{435}$-dropout selection)
or the WFPC2 images redward of the dropout band in the case of our
$U_{300}$ dropout selection (where the $B_{450}V_{606}I_{814}$ bands
are used).  For clarity, the $\chi^2$ image is equal to
\begin{displaymath}
\Sigma _{k} (I_k (x,y) / N_k)^2
\end{displaymath}
where $I_k (x,y)$ is the intensity of image $I_k$ at pixel $(x,y)$,
where $N_k$ is the RMS noise on that image, and where the index $k$
runs over all the relevant images (i.e., those redward of the dropout
band).

Our procedure for measuring colors depends upon the PSF of the data we
are using.  For colors that include only the ACS or WFPC2 bands (where
the PSF is much sharper than for NICMOS), we measure these colors in
scalable apertures determined using a Kron (1980) factor of 1.2.  When
measuring colors that include the NICMOS near-IR bands, we first PSF
match the ACS/WFPC2 data to the NICMOS $H_{160}$-band.  Then, we use
the same apertures described above if the area in those apertures is
$\geq0.38$ arcsec$^2$ (i.e., $\geq$0.7$''$-diameter apertures).  This
only applies for the largest sources in our selection.  Otherwise (and
in most cases), the colors are measured in fixed $0.7''$-diameter
apertures.

We correct the fluxes measured in the above apertures (used for color
measurements) to total fluxes using a two step procedure.  First, the
individual flux measurements are corrected to a much larger scalable
aperture (using a Kron factor of 2.5).  These corrections are made on
a source-by-source basis using the square root of the $\chi^2$ image
(Szalay et al.\ 1999).  Second, a correction is made to account for
the light outside of these larger apertures and on the wings of the
ACS Wide Field Camera (WFC) PSF (Sirianni et al.\ 2005).  Typical
corrections are $\sim0.6$ mag for the first step and $\sim0.1-0.2$ mag
for the second step.  While there are modest uncertainties in these
corrections (typically 0.2 mag), these corrections only affect the
total magnitude measurements and do not affect the colors.

\begin{figure*}
\epsscale{1.15}
\plotone{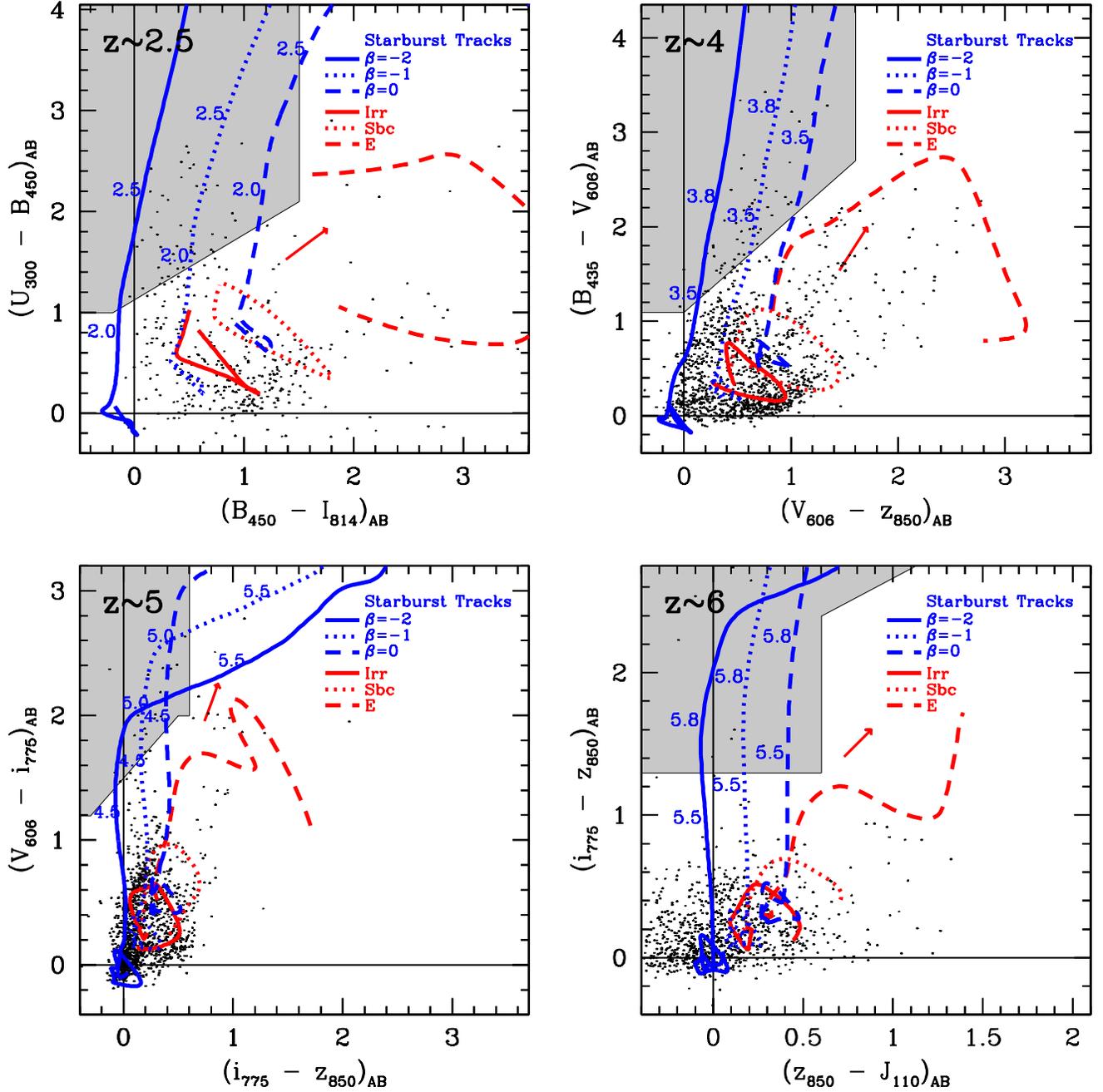}
\caption{Two-color diagrams showing the selection criteria (see \S3.2)
  we use to find $z\sim2.5$ $U$, $z\sim4$ $B$, $z\sim5$ $V$, and
  $z\sim6$ $i$ dropouts in our search fields (upper left, upper right,
  lower left, and lower right panel, respectively).  Sources that fall
  within the shaded regions would be included in our selections.  The
  blue tracks show the expected colors of starbursts with different
  $UV$-continuum slopes as a function of redshift, while the red lines
  show the colors of low-redshift interlopers (Coleman et al. 1980).
  The black points illustrate the range of colors that real sources in
  our search fields possess.  To ensure that these plotted
  distributions of colors are as realistic as possible, we only
  include sources from our deepest fields (the HDFs for our
  $U$-dropout selection and HUDF for our $B$, $V$, and $i$-dropout
  selection) and at relatively bright magnitudes ($I<26$ for our
  $U$-dropout selection, $z<26.5$ for our $B$-dropout selection, and
  $z<27$ for our $V$,$i$-dropout selections).  The red arrows show the
  Calzetti et al.\ (2000) reddening vectors.  These diagrams show that
  our selection criteria are effective in identifying high-redshift
  galaxies with $UV$-continuum slopes $\beta$ bluer than 0.5, but less
  effective in identifying galaxies redward of this limit
  (Figures~\ref{fig:seleffect} and \ref{fig:complete} show the
  effective selection volume calculated to be available to galaxies
  with various $UV$-continuum slopes).\label{fig:sel}}
\end{figure*}

\subsection{Sample Selection}

We select star-forming galaxies over the redshift range $z\sim6$ to
$z\sim2$ using the well-established dropout technique (e.g., Steidel
et al.\ 1996; Bunker et al.\ 2003; Vanzella et al.\ 2006; Dow-Hygelund
et al.\ 2007; Stanway et al.\ 2007; Vanzella et al.\ 2009).  Our
$U$-dropout criterion is similar to those used by Bouwens et
al.\ (2003) while our $B$, $V$, and $i$ dropout criteria are almost
identical to those adopted by Bouwens et al.\ (2007) except that our
$V$-dropout criterion uses a sharper $(i_{775}-z_{850})<0.6$ cut to
exclude dusty $z\sim1$ interlopers.  These criteria are
\begin{eqnarray*}
(U_{300}-B_{450} > 1.1)\wedge(B_{450}-I_{814} < 1.5) \\ 
\wedge (U_{300}-B_{450} > 0.66 (B_{450} - I_{814}) + 1.1)
\end{eqnarray*}
for our $U$-dropout sample
\begin{eqnarray*}
(B_{435}-V_{606} > 1.1) \wedge (B_{435}-V_{606} > (V_{606}-z_{850})+1.1) \\
\wedge (V_{606}-z_{850}<1.6)
\end{eqnarray*}
for our $B$-dropout sample,
\begin{eqnarray*}
[(V_{606}-i_{775} > 0.9(i_{775}-z_{850})) \vee (V_{606}-i_{775} > 2))]
\wedge \\ (V_{606}-i_{775}>1.2) \wedge (i_{775}-z_{850}<0.6)
\end{eqnarray*}
for our $V$-dropout sample, and
\begin{displaymath}
(i_{775}-z_{850}>1.3) \wedge ((V_{606}-i_{775} > 2.8) \vee (S/N(V_{606})<2))
\end{displaymath}
for our $i$-dropout sample, where $\wedge$ and $\vee$ represent the
logical \textbf{AND} and \textbf{OR} symbols, respectively, and
\textbf{S/N} represents the signal-to-noise.  These criteria are
illustrated in two color diagrams in Figure~\ref{fig:sel} and are very
similar to those considered by Giavalisco et al.\ (2004b) and Beckwith
et al.\ (2006).

To ensure that our selections did not contain any spurious sources, we
required that sources in our $U$, $B$, $V$, and $i$ dropout selections
be $\geq$5.5$\sigma$ detections in the $I_{814}$, $z_{850}$,
$z_{850}$, and $z_{850}$ bands, respectively.  We inspected all
sources in our dropout samples by eye to purge them of any obvious
contaminants (diffraction spikes, irregularities in the background).
Pointlike sources (SExtractor stellarity S/G $>$0.8) were also removed
($\lesssim2$\% of the relevant sources).

\begin{deluxetable}{cccc}
\tablecaption{Wavebands used to derive the $UV$ continuum slope for
  individual galaxies in high-redshift $U$, $B$, $V$, and $i$ dropout
  samples.\tablenotemark{a,b}\label{tab:pivotbands}} \tablehead{
  \colhead{Dropout} & \colhead{Mean} & \multicolumn{2}{c}{Filters used
    for $\beta$ determination} \\ 
\colhead{Sample} &
  \colhead{Redshift} & \colhead{Lower} & \colhead{Upper}} 
\startdata
$U$-dropout & 2.5 & $V_{606}$ (1730\AA) & $I_{814}$ (2330\AA) \\ 
$B$-dropout & 3.8 & $i_{775}$ (1610\AA) & $z_{850}$ (1940\AA) \\ 
$V$-dropout & 5.0 & $z_{850}$ (1550\AA) & $(J_{110}+H_{160})/2$ (2250\AA) \\ 
$V$-dropout & 5.0 & $z_{850}$ (1550\AA) & $H_{160}$ (2660\AA) \\ 
$i$-dropout & 5.9 & $J_{110}$ (1570\AA) & $H_{160}$ (2290\AA) \\ 
\enddata 
\tablenotetext{a}{See \S3.4 for discussion on
  the filter choices.}
\tablenotetext{b}{Throughout this work, our
  $\beta$ measurements are based upon the relationship $f _{\lambda}
  \propto \lambda^{\beta}$.  Since real spectra will not have perfect
  power law SEDs (and therefore depend upon the rest-frame wavelengths
  used to estimate $\beta$), small corrections are made to the
  estimated $\beta$'s based upon a fiducial starburst SED
  ($e^{-t/\tau}$ star formation history with $t=70$ Myr, $\tau=10$
  Myr, $[Z/Z_{\odot}]=-0.7$: from Papovich et al.\ 2001).  See \S3.4
  and Appendix A for details.  The filters that define the wavelength
  range of the $\beta$ measurements for our different dropout samples
  are presented here.}
\end{deluxetable}

\subsection{Calibration of the NICMOS Magnitudes}

To accurately estimate the $UV$-continuum slopes $\beta$ for galaxies
in our sample, it is absolutely essential that the colors that we
measure are accurate.  For example, for a source with $UV$ colors
measured across the baseline 1600\AA$\,\,$to 2300\AA, a 0.1 mag error
in this color translates into an error of 0.3 in the estimated slope
$\beta$.

Measuring colors to this level of accuracy is challenging for a number
of reasons, particularly in the near-IR.  First and foremost, the
NICMOS detector (essential for measuring the fluxes of faint sources
in the near-IR) has been shown to suffer from significant
non-linearity in its count rate.  While there are now procedures
(e.g., de Jong et al.\ 2006) to correct for these non-linearities, as
utilized here, it is unlikely that the corrections are perfect in the
magnitude range we are considering.  This is because there are
comparatively few deep fields available where the near-IR fluxes can
be measured and calibrated against other data.  Secondly, the colors
we are measuring are derived from both optical (ACS) and near-IR
(NICMOS) data where the PSF is very different.  While such differences
can be effectively addressed by carefully matching the PSFs of the two
data sets before measuring fluxes (as we do here), PSF matching can be
challenging to perform perfectly and so it is easy to introduce small
systematic errors at this stage.

As a result of these issues, we took great care in deriving correction
factors that we could apply to the measured NICMOS near-IR fluxes.
For simplicity, we assumed that these correction factors did not
depend on the luminosity of the sources we were considering -- but
rather were simple offsets we could add to the measured magnitudes.
We established these correction factors by examining the photometry of
a sample ($\sim$50) of relatively faint ($22<z_{850,AB}<26$)
point-like stars (SExtractor stellarity $>$0.8) over the HUDF/GOODS
fields, and then compared their observed colors with that expected
from the Pickles et al.\ (1998) atlas of stellar spectra.  Given the
very limited range in SED shape found in real stars, these comparisons
should allow us to make reasonable corrections to the $J$ and $H$-band
photometry.  We determined the average differences between the
measured and best-fit near-IR fluxes (both in the $J_{110}$ and
$H_{160}$ bands) and found corrections of $-$0.13 mag and $-$0.07 mag
for the $J_{110}$ and $H_{160}$ band, respectively.

To ensure that these corrections were as accurate as possible, we also
identified a sample of $\sim60$ highest S/N $B$ dropouts from our
search fields (GOODS and HUDF) and then fit their observed SEDs with
Bruzual \& Charlot (2003) $\tau$ models.  Since the $1.1\mu$ $J$ and
$1.6\mu$ $H$ band fluxes of these galaxies probe the light blueward of
the age sensitive break at 3600\AA$\,\,$(probing 2300\AA$\,\,$and
3300\AA$\,\,$rest-frame, respectively), it should be reasonable to use
the available $i$ (1600\AA\, rest-frame) and $z$-band (1900\AA$\,
\,$rest-frame) fluxes to establish a $UV$-continuum slope $\beta$ and
extrapolate to these redder wavelengths.  For our fiducial model fit
parameters, we assumed a $e^{-t/\tau}$ star formation history (with
$t=70$ Myr, $\tau=10$ Myr), a Salpeter IMF, and $[Z/Z_{\odot}]=-0.7$,
and let the dust extinction and redshift be free parameters (the
fiducial model considered by Papovich et al.\ 2001 in their stellar
population modelling of $z\sim3$ $U$-dropouts).  Comparing the
predicted near-IR fluxes (from the models) with the measured fluxes,
we computed corrections to the $J$ and $H$-band fluxes.  The
corrections were $-$0.08$\pm$0.02 mag and 0.04$\pm$0.2 mag,
respectively (the uncertainties were derived using other SED models
[Appendix A] to determine these corrections).  Because of the size of
the uncertainties in the $H$-band corrections (from the $z\sim4$
$B$-dropouts), we gave those results very low weight overall for the
$H$-band correction.

Combining the constraints we derived from our sample of stars and
$z\sim4$ LBGs -- which were consistent within $0.05$ mag -- we derived
a $-$0.10 mag correction to the measured $J_{110}$-band magnitudes.
For the NICMOS $H_{160}$-band, we derived a $-$0.05 mag correction
from the above fits -- using the SED fit results to stars as our
primary constraint.\footnote{Note that the corrections we derive here
  are very similar ($\lesssim0.04$ mag) to what we derive matching
  sources in the HUDF with both NICMOS and WFC3/IR data (e.g., Bouwens
  et al.\ 2009b).}  The measured $J_{110}$ and $H_{160}$-band fluxes
were therefore somewhat too faint (before correction).  A brief
discussion of the effect that uncertainties in the photometry/zero
points could have on our derived $UV$-continuum slopes $\beta$ is
given in \S3.8.

\subsection{Measuring the $UV$-continuum slope $\beta$}

Our principal interest is to determine the rest-frame $UV$-continuum
slope $\beta$ for the galaxies in our dropout samples.  As noted in
the introduction, the $UV$-continuum slope $\beta$ specifies how the
flux density of a galaxy varies with wavelength (i.e.,
$f_{\lambda}\propto\lambda^{\beta}$) in the $UV$-continuum (i.e., from
$\sim$1300\AA$\,\,$to $\sim$3500\AA).

We estimate the $UV$-continuum slopes $\beta$ from the broadband
colors available for each galaxy in our sample.  To make our estimates
of these slopes as uniform as possible, we selected filters at each
redshift where the rest-frame $UV$ wavelengths were as very close to
1650\AA$\,\,$and 2300\AA$\,\,$as possible.  Of course, the actual
effective wavelengths of these filters will vary somewhat from sample
to sample -- depending on the passbands in which broadband imaging is
available for our different samples or the precise redshift of
specific galaxies in our samples.  Fortunately, the wavelength range
1650\AA$\,\,$to 2300\AA$\,\,$is a good match to the wavelength range used
for studies of the dust extinction in galaxies at $z\sim0$ (e.g.,
Meurer et al.\ 1995; Meurer et al.\ 1999; Burgarella et al.\ 2005).
Table~\ref{tab:pivotbands} summarizes the passbands we use to probe
these wavelengths for each of our dropout samples.

Some additional clarification may be helpful regarding specific filter
choices.  For our $B$ dropout sample, for example, we use the
$i_{775}-z_{850}$ colors for these estimates to take advantage of the
much higher quality (deeper, wider area) ACS data available.  For our
$V$-dropout sample, we use $z_{850}-(J_{110}+H_{160})/2$ colors
predominantly for these estimates, but we also make use of
$z_{850}-H_{160}$ colors for these estimates to take advantage of the
large areas within the GOODS fields which have only NICMOS
$H_{160}$-band coverage (primarily from the Conselice et al.\ 2009, in
prep, NICMOS program: see the \textit{light orange squares} in
Figure~\ref{fig:obsdata}).

In principle, deriving the $UV$-continuum slopes $\beta$ from the
measured $UV$ colors is straightforward.  One simply takes power-law
spectra $f_{\lambda}\propto \lambda^{\beta}$, calculates their colors
at the mean redshift of each of our dropout samples, and then
determines for which $\beta$ the model spectrum reproduces the
observed colors.  In practice, however, since the spectrum of
star-forming galaxies in the $UV$-continuum is not expected to be a
perfect power law (as noted above), small differences in the
wavelength range over which the $UV$-continuum slope $\beta$ is
measured will have an effect of the derived results.

To correct for these passband and wavelength-dependent effects, we
base our estimates of $\beta$ on a small number of more realistic SEDs
calculated from stellar population models (Bruzual \& Charlot 2003).
These model SEDs are calibrated to have $UV$-continuum slopes $\beta$
of $-$2.2, $-$1.5, and $-$0.8 over the wavelength range and assume a
$e^{-t/\tau}$ star formation history ($t=70$ Myr, $\tau=10$ Myr,
$[Z/Z_{\odot}]=-0.7$) with varying amounts of dust extinction
calculated according to the Calzetti et al.\ (2000) law.  The formulae
we use to convert between the measured colors and the $UV$-continuum
slope $\beta$ are presented in Appendix A.  The formulae we use to do
the conversion depend somewhat on the star formation history assumed
and likely result in an additional uncertainty in the derived
$UV$-continuum slopes $\beta$ of $\sim0.1-0.2$.

\begin{deluxetable}{cccc}
\tablecaption{Dropout samples used to measure the distribution of $UV$-continuum slopes $\beta$ as a function of redshift and $UV$ luminosity.\label{tab:lumrange}}
\tablehead{
\colhead{Dropout} &  \colhead{} & \colhead{Luminosity} & \colhead{\# of} \\
\colhead{Sample} & \colhead{Field} & \colhead{Range\tablenotemark{a}} & \colhead{Sources}}
\startdata
$U$-dropout & HDF-North & $-22<M_{UV,AB}<-18$ & 97\\
 & HDF-South & $-22<M_{UV,AB}<-18$ & 71 \\
$B$-dropout & GOODS-North & $-23<M_{UV,AB}\lesssim-20.5$ & 218 \\
& GOODS-South & $-23<M_{UV,AB}<-20.5$ & 251 \\
& HUDF & $-20.5<M_{UV,AB}<-17$ & 470 \\
& Cluster & $-23<M_{UV,AB}\lesssim-20.5$ & 195 \\
$V$-dropout & GOODS-North & $-23<M_{UV,AB}\lesssim-20.0$ & 22 \\
            & GOODS-South & $-23<M_{UV,AB}\lesssim-20.0$ & 19 \\
            & HUDF & $-23<M_{UV,AB}\lesssim-19.5$ & 10 \\
            & HUDF05 & $-23<M_{UV,AB}\lesssim-19.5$ & 16 \\
            & Cluster & $-23<M_{UV,AB}\lesssim-20.0$ & 12 \\
$i$-dropout & GOODS-North & $-23<M_{UV,AB}\lesssim-20.5$ & 2 \\
            & GOODS-South & $-23<M_{UV,AB}\lesssim-20.5$ & 2 \\
            & HUDF & $-23<M_{UV,AB}\lesssim-20.0$ & 5 \\
            & HUDF05 & $-23<M_{UV,AB}\lesssim-20.0$ & 3 \\
            & Cluster & $-23<M_{UV,AB}\lesssim-20.5$ & 2
\enddata

\tablenotetext{a}{Luminosity limits for $z\sim4$ $B$-dropout samples
  are chosen to minimize the effects of photometry scatter and
  selection on the $UV$-continuum slope distribution.  Luminosity
  limits for $z\sim5$ $V$-dropout and $z\sim6$ $i$-dropout samples are
  set according to the depth of the near-IR NICMOS data.  See \S3.5.}
\end{deluxetable}

\subsection{Constructing the $UV$-continuum Slope Distribution from
  Multiple (Deep or Wide-Area) Dropout Selections} 

In order to accurately quantify the distribution of $UV$-continuum
slopes $\beta$ over a wide range in luminosity, we alternatively used
our deepest selections and our wide-area selections.  We used the
deepest selections to define the $UV$-continuum slope distribution at
the lowest luminosities to maximize the S/N on the derived colors
(while minimizing the importance of selection effects).  Meanwhile, we
used our widest-area selections to define this distribution at higher
luminosities.  This allows us to find a sufficient number of sources
to map out the distribution of $UV$-continuum slopes (which is helpful
for determining the mean and $1\sigma$ scatter).  In
Table~\ref{tab:lumrange}, we provide a list of the fields we use to
quantify the distribution of $UV$-continuum slopes at a given
luminosity and redshift.

In Figure~\ref{fig:colmag}, we plot the $UV$-continuum slopes $\beta$
observed for galaxies in our four dropout samples versus their $UV$
luminosities.  The mean $UV$-continuum slope $\beta$ and $1\sigma$ for
galaxies is also included on this figure (blue squares) as a function
of luminosity for each of the dropout samples.  We adopt finer 0.5 mag
bins for our $B$ dropout selections than the 1.0 mag bins we adopt for
our $U$ and $V$ dropout selections or 2.0 mag bins we adopt for our
$i$ dropout selection.  The width of the bins depends upon the number
of sources present in each of our dropout selections.  The $UV$
luminosity as shown on this diagram is the geometric mean of the two
$UV$ luminosities used to establish the $UV$ slope (see
Table~\ref{tab:pivotbands}).  Use of the geometric mean of the two
$UV$ luminosities is preferred here for examining the correlation of
$\beta$ with luminosity since it allows us to evaluate the correlation
without introducing any artificial correlations.  Had we, for example,
elected to use the bluer bands for examining this correlation, the
$UV$-continuum slope $\beta$ would be bluer as a function of the
bluer-band luminosity, simply as a result of our examining the
relationship as a function of this quantity.  A similar (but opposite)
bias would be introduced, in examining the $UV$-continuum slope
$\beta$ as a function of the redder band.

A clear trend is seen towards bluer $UV$-continuum slopes $\beta$ at
lower luminosities.  This trend is similar to that found by Meurer et
al.\ (1999) in their analyses of $z\sim2.5$ $U$-dropout galaxies in
the HDF North and also noted by Overzier et al.\ (2008) in their
analysis of $z\sim4$ $g$-dropouts in the TN1338 field (Figure 4 from
that work). 

\subsection{$UV$-continuum Slopes Derived for $z\sim4-6$ Galaxies
  Gravitationally Lensed By Galaxy Clusters}

In \S3.5, we considered galaxies from both wide-area and deep surveys
to establish the distribution of $UV$-continuum slopes $\beta$ over a
wide range in luminosity.  We can increase the number of sources in
our lower luminosity $z\sim4-6$ samples, by considering
gravitationally lensed sources behind high-redshift galaxy clusters.
The clusters under consideration include Abell 2218, MS1358, CL0024,
Abell 1689, and Abell 1703, and all allow us to select very faint
star-forming galaxies at $z\sim4-6$ from the available HST ACS
data.

These clusters substantially amplify the flux from distant galaxies,
making it possible to measure the $UV$-continuum slope of galaxies at
lower luminosities than would otherwise be possible.  Of course, in
the case of $B$-dropout selections from very deep optical data like
HUDF, we are able to reach to the same intrinsically luminosities as
we reach in the $B$-dropout samples we compile from searches behind
lensing clusters.  We correct the luminosities of the dropout galaxies
identified behind these clusters by applying gravitational lensing
models from the literature.  For Abell 2218, MS1358, CL0024 (sometimes
known as CL0024+1652), Abell 1689, and Abell 1703, we adopt the models
constructed by El{\'{\i}}asd{\'o}ttir et al.\ (2007), Franx et
al.\ (1997), Jee et al.\ (2007), Limousin et al.\ (2007), and Limousin
et al.\ (2008), respectively.  Typical magnification factors for the
$z\sim4-6$ galaxies we find behind galaxy clusters are $\sim$5-10.  We
do not include the $UV$-continuum slope results for sources where the
model magnification factors are substantially greater than 10 -- due
to sizeable model-dependent uncertainties in the actual magnification
and hence their luminosities (the uncertainties in the magnification
factor for sources with such high predicted magnifications can be very
large, i.e., greater than factors of $\sim$3-4: see Appendix A of
R.J. Bouwens et al.\ 2009, in prep).  Even for sources where the
predicted magnification factors are smaller, the model magnification
factors (and hence the luminosities corrected for lensing) are likely
uncertain at the factor of $\sim$2 level (R.J. Bouwens et al.\ 2009,
in prep).

\begin{figure*}
\epsscale{1.13}
\plotone{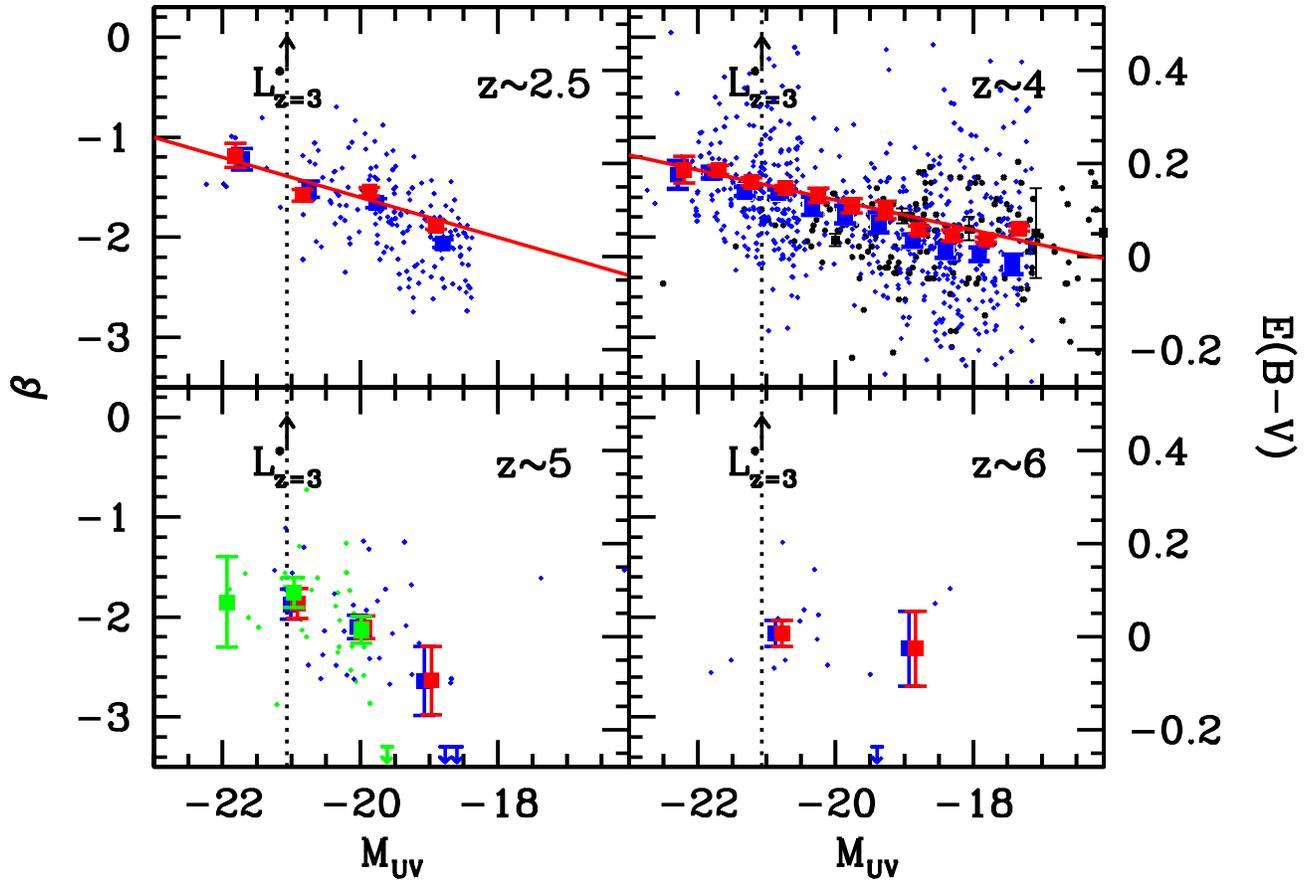}
\caption{Determinations of the $UV$-continuum slope $\beta$
  (1600\AA$\,\,$to 2300\AA) versus rest-frame $UV$ luminosity for the
  present selection of $z\sim2.5$ $U$, $z\sim4$ $B$, $z\sim5$ $V$, and
  $z\sim6$ $i$ dropouts.  The $UV$ luminosity here is the geometric
  mean of the $UV$ luminosities measured in the bands used to
  establish the $UV$ slope (see Table~\ref{tab:pivotbands}).  On the
  right axes are the equivalent $E(B-V)$ extinctions (Calzetti et
  al.\ 2000 law) to a given $\beta$ if the base spectrum is a young
  star-forming galaxy ($\sim$100 Myr of constant star formation) --
  which is only $\Delta E(B-V) = 0.02$ bluer than our fiducial $\tau$
  model (i.e., the one referred to in footnote b of
  Table~\ref{tab:pivotbands}).  The small points represent the
  $UV$-continuum slopes measured for individual galaxies in our
  samples while the open blue squares and vertical bars show the mean
  $UV$-continuum slope $\beta$ and $1\sigma$ scatter (shown in a
  darker blue in the $z\sim4$ panel for clarity).  Blue upper limits
  are shown for a few $z\sim5$ and $z\sim6$ sources with $\beta$'s
  bluer than $-$3.5.  The green points on the $z\sim5$ panel show the
  $UV$-continuum slope determinations derived from our $V$-dropout
  samples where only NICMOS $H_{160}$-band data are available (i.e.,
  with no deep NICMOS $J_{110}$-band coverage).  The black points on
  the $z\sim4$ panel show the $UV$-continuum slope determinations
  derived from our $B$-dropout samples behind lensing clusters (where
  we have corrected their luminosities for model lensing magnification
  factors: see \S3.6).  It is encouraging that the $UV$-continuum
  slopes we derive from both of the latter $V$ and $B$ dropout
  selections agree with the determinations derived from our primary
  selections.  The open red squares and vertical bars show the mean
  $UV$-continuum slope and $1\sigma$ error in this slope, after
  correcting for the effect of object selection and photometric
  scatter (\S3.7).  See Table~\ref{tab:uvslope} for a tabulation of
  these slopes.  The red lines show the best fit relationship between
  $UV$ continuum slope and the observed magnitude of a galaxy in the
  $UV$ (see \S3.9 and \S4.5).  It is also clear that the
  $UV$-continuum slope $\beta$ of star-forming galaxies is much redder
  at higher luminosities than it is at lower luminosities
  (particularly at $z\sim2.5$ and $z\sim4$).\label{fig:colmag}}
\end{figure*}

We include the $UV$-continuum slope determinations of these galaxies
with those estimated from our field samples on Figure~\ref{fig:colmag}
(\textit{black points} on $z\sim4$ panel).  As a result of a
substantial magnification from gravitational lensing (typical factors
of $\sim$5-10), the intrinsic luminosities of galaxies found behind
clusters are much lower on average than those found in our field
samples.  In general, we observe good agreement between the
distribution of $UV$-continuum slopes $\beta$ derived from our field
samples and those inferred from our selections around clusters.

\begin{figure*}
\epsscale{1.13} 
\plottwo{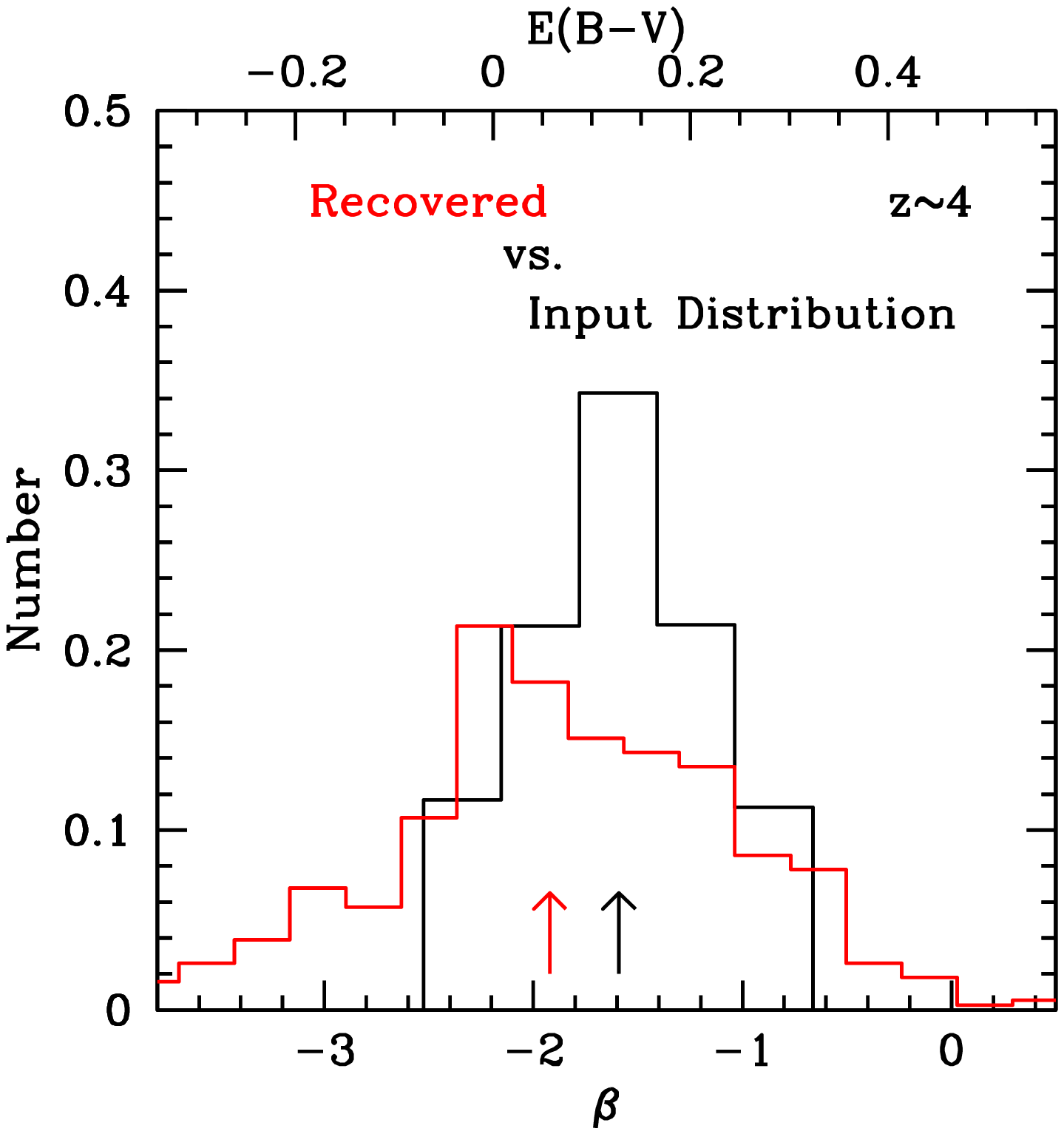}{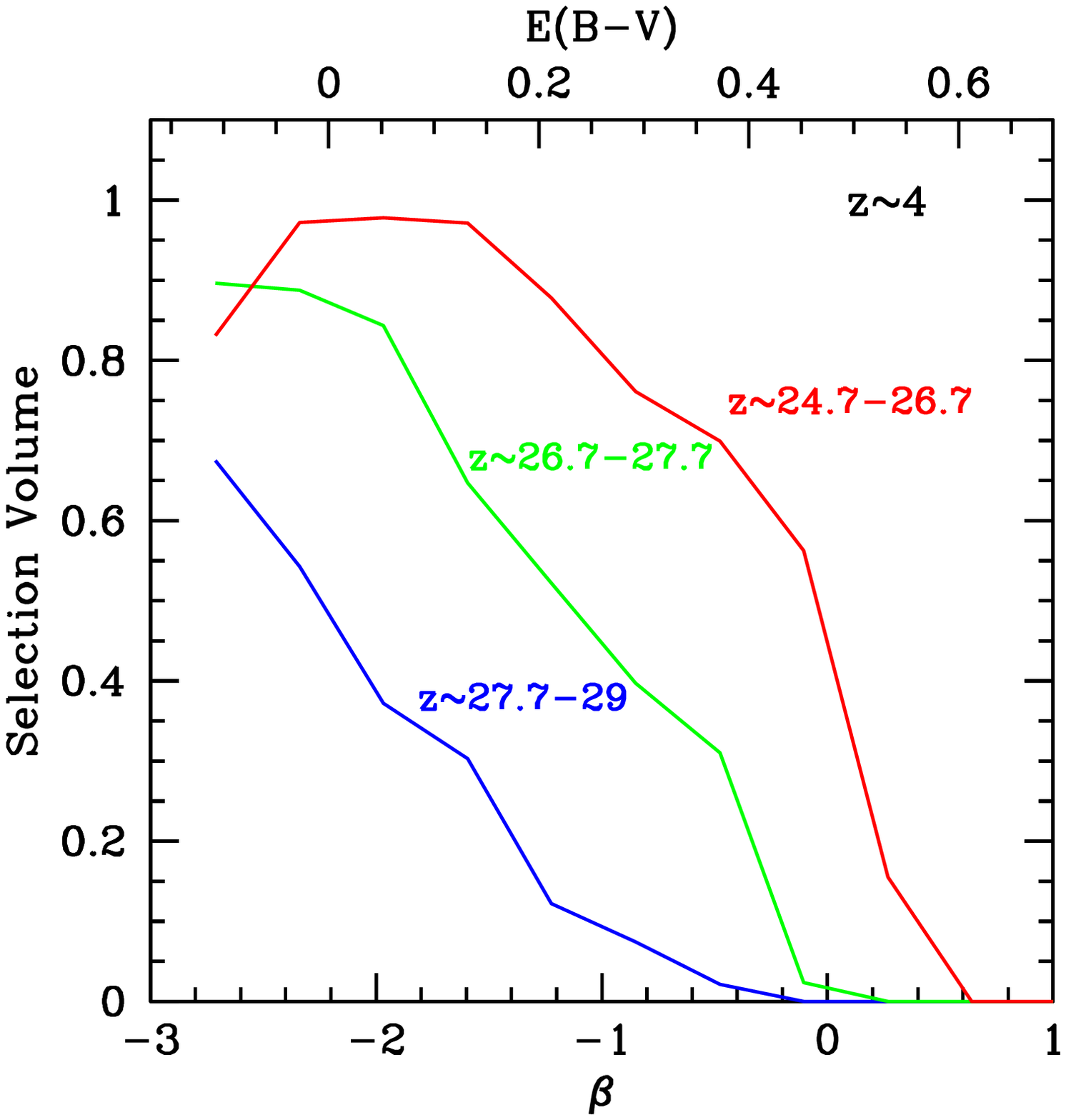}
\caption{(\textit{left}) Comparison between the $UV$-continuum slope
  $\beta$ distribution input into the simulations (\textit{black
    histogram}) and that directly recovered from the simulations
  (\textit{red histogram}: which included the effect of object
  selection and measurement error).  This panel is for our faintest,
  lowest S/N $z\sim4$ $B$-dropout selection over the HUDF
  ($z_{850}$-band magnitudes of 28-29).  Both the input (and
  recovered) distributions are normalized to have the same area.  On
  the top axes are the $E(B-V)$ extinctions (Calzetti et al.\ 2000
  law) that are equivalent to a given $\beta$ if the base spectrum is
  a young star-forming galaxy ($\sim$100 Myr of constant star
  formation).  The median $UV$-continuum slope $\beta$ input into the
  simulations and that recovered are shown with the black and red
  arrows, respectively.  The recovered distribution of $UV$-continuum
  slopes is broader and peaked to somewhat bluer values than the input
  distribution.  For our higher S/N selections, the differences
  between the input and recovered distributions are much smaller (the
  offsets between the red and blue squares in Figure~\ref{fig:colmag}
  show the size of the corrections).  For our faintest selections (as
  shown here) the differences can be larger.  It is nonetheless
  reassuring that when these corrections are applied there is
  agreement between the mean $UV$-continuum slopes estimated in our
  shallower selections (where the corrections are larger) and in our
  deeper selections (where the corrections are smaller).  For example,
  the mean $UV$-continuum slope derived at $z_{850,AB}$$\sim$26.5-27
  for our GOODS $B$-dropout selections (i.e., $-1.81\pm0.04$ at
  $-19.2$ AB mag) agrees very well with that derived from our much
  higher S/N HUDF selections for the GOODS fields (i.e.,
  $-1.73\pm0.07$ for the HUDF).  (\textit{right}) The selection
  volumes (arbitrary units) available to galaxies for our $B$-dropout
  selection over the HUDF versus $UV$-continuum slope $\beta$.  Shown
  are the relative selection volumes for galaxies with
  $z_{850}\sim$24.7-26.7 \textit{(red line)}, $z_{850}\sim$26.7-27.7
  \textit{(green line)}, and $z_{850}\sim$27.7-29.0 \textit{(blue
    line)}.  We emphasize that the same normalization for the
  selection volume is used for all the three magnitude selections,
  demonstrating that it is much easier to select galaxies with red
  $UV$-continuum slopes at bright magnitudes than it is at faint
  magnitudes.  The reason bright $z\sim4$ galaxies are easier to
  select is that one needs to confirm that a source has a large Lyman
  break at $>$1$\sigma$, and this is easier to do at bright
  magnitudes.  Selection biases and measurement errors are less for
  brighter samples (where the S/N is higher) and greater for fainter
  samples.  The two panels here demonstrate how important it is for us
  to correct for selection biases and measurement uncertainties as we
  do in \S3.7.\label{fig:seleffect}}
\end{figure*}

\subsection{Selection Biases and Measurement Uncertainties}

The distribution of $UV$-continuum slopes that we derive is quite
clearly affected by the manner in which sources are selected.  Dropout
criteria include galaxies with bluer $UV$-continuum slopes more
efficiently (i.e., over a larger range in redshift) than they do for
galaxies with other colors.  It is much easier to identify a sharp
break in the SED of a blue galaxy than it is for a galaxy that is
somewhat redder.  For galaxies with red enough colors (i.e.,
$UV$-continuum slopes $\beta$ larger than 0.5), it is essentially
impossible to robustly select high-redshift galaxy using the dropout
criteria, and in fact the only sources that would satisfy the
selection criteria would do so because of photometric scatter.  This
is illustrated in Figures~\ref{fig:sel} and \ref{fig:seleffect}.

To control for this effect, we construct models of the $UV$-continuum
slope $\beta$ distribution, use these models to add artificial
galaxies to real data, select sources from these data, and measure
their $UV$-continuum slopes in the same way as from the real data.
The goal is to construct a model that when ``observed'' reproduces the
distribution of $UV$-continuum slopes $\beta$ measured from the data.
For the sizes and morphologies of the model galaxies used in the
simulations, we start with the pixel-by-pixel profiles of the $z\sim4$
$B$-dropout sample from the HUDF (Bouwens et al.\ 2007) and scale
their sizes as $(1+z)^{-1}$ (for fixed luminosity) to match the
observed size-redshift trends (Bouwens et al.\ 2006; see also Ferguson
et al.\ 2004 and Bouwens et al.\ 2004).  We use our well-tested
``cloning'' software (Bouwens et al.\ 1998a,b; Bouwens et al.\ 2003;
Bouwens et al.\ 2006, 2007) to perform these simulations.  For the
model $UV$ LF at $z$$\,\sim\,$2-6, we adopt the Schechter parameterizations
determined by Reddy \& Steidel (2009) at $z\sim2.5$ and by Bouwens et
al.\ (2007) at $z\sim4-6$.

In performing these simulations, we account for the modest correlation
between the $UV$-continuum slopes $\beta$ of galaxies and their
surface brightnesses.  We determined the approximation correlation by
examining 192 luminous ($\sim$1-2 $L_{UV}^{*}$) galaxies in our GOODS
$B$-dropout selection and comparing their observed $\beta$'s with
their sizes (half-light radii).  Any correlation is potentially
important, since it could substantially lower the efficiency with
which we can select galaxies with very red $UV$-continuum slopes
$\beta$ (e.g., Figure~\ref{fig:seleffect}).  We find that sources with
$\beta\gtrsim-1$ are $\sim$15\% larger than galaxies with
$-2\lesssim\beta\lesssim-1$ and $\sim$40\% larger than galaxies
$\beta\lesssim-2$ -- though there is significant object-to-object
scatter (the correlation coefficient between size (half-light radius)
and $\beta$ is just $\sim$0.3).  Fortunately, this correlation seems
to only have a modest effect on these selection volumes, decreasing it
by only $\sim10$\% for the red galaxies and increasing it by only
$\sim$10\% for the blue galaxies.

We experimented with a range of model $UV$-continuum slope $\beta$
distributions to determine the effect of object selection and
photometric scatter on the observed distribution of $UV$-continuum
slopes $\beta$.  These experiments were perfomed as a function of
magnitude and the input color distribution (with mean $UV$-continuum
slopes ranging from $-1.5$ and $-2.2$ and the input color distribution
taken to be Gaussian).  In general, we found that the recovered
distribution of $UV$-continuum slopes $\beta$ (after selection) is
\textit{bluer} than the input distributions (by $\Delta\beta\sim0.1$)
at lower luminosities for all of our dropout samples.  As expected, we
found that noise in the observations broadened the distribution of
$UV$-continuum slopes $\beta$ somewhat over that present in the input
distribution.  Figure~\ref{fig:seleffect} illustrates how
observational selection and noise modifies the input distribution of
slopes for a $B$-dropout selection over the HUDF.

We used the simulations described above to quantify changes in this
distribution as simple shifts in the mean $UV$-continuum slope and
$1\sigma$ scatter.  We used the estimated shifts to correct the
observed distribution of $UV$-continuum slopes $\beta$ for these
effects.  This corrected distribution of $UV$-continuum slopes is
plotted in Figure~\ref{fig:colmag} as open red squares, with the
$1\sigma$ scatter shown with the red error bars.

To verify that the corrections we apply in this section are accurate,
we compared the mean $UV$-continuum slope $\beta$ we estimate near the
faint-end ($z_{850,AB}$$\sim$26.5-27) of our shallower GOODS
$B$-dropout selections (after correction) with those derived from our
much higher S/N HUDF $B$-dropout selections in the same magnitude
range.  We find that they are in excellent agreement, i.e.,
$-1.81\pm0.04$ at $-19.2$ AB mag for the GOODS fields
vs. $-1.73\pm0.07$ for the HUDF.  This suggests that the corrections
we apply in this section are reasonably accurate ($\Delta \beta
\lesssim0.1$).

\begin{figure}
\epsscale{1.13} \plotone{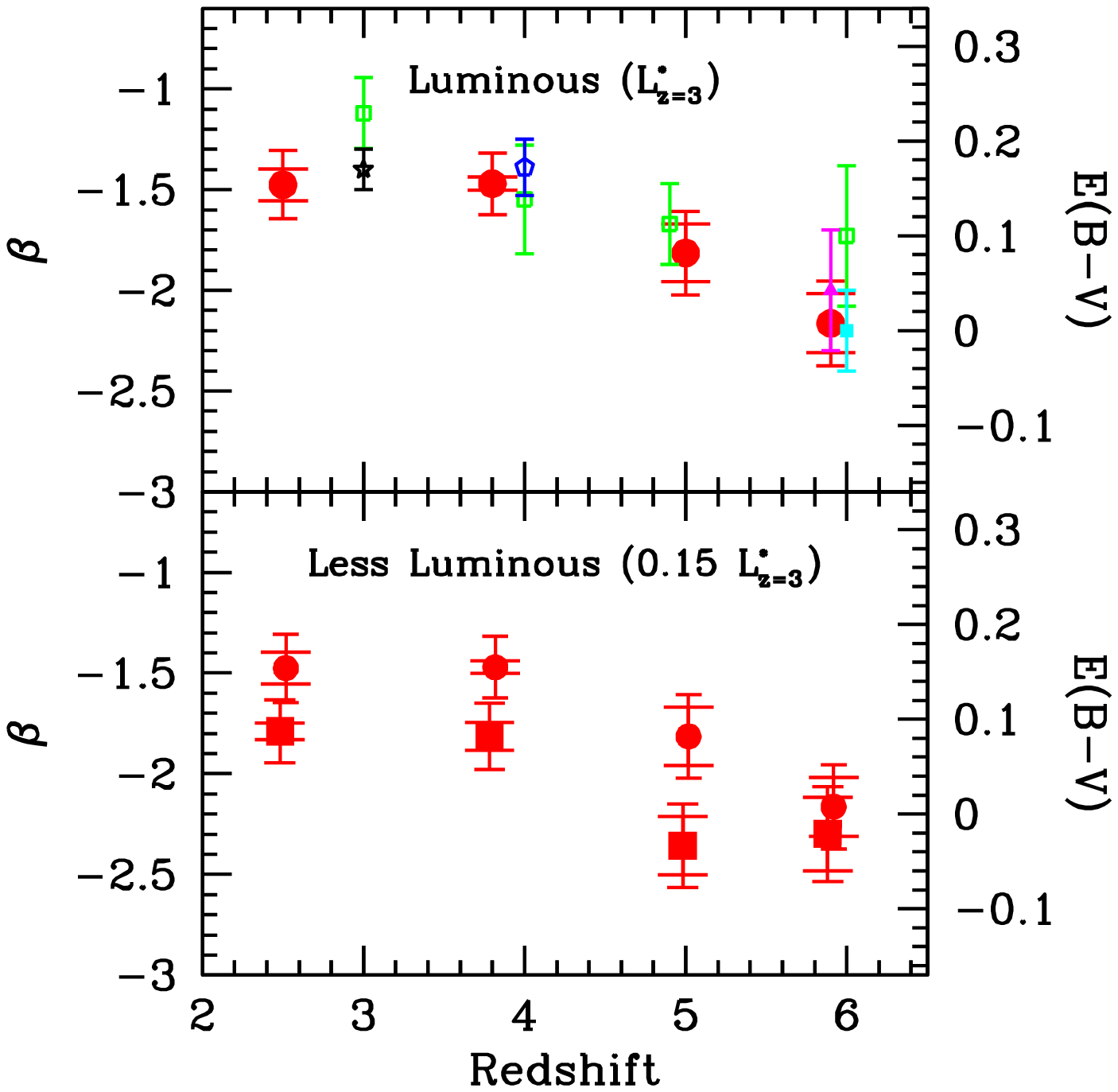}
\caption{(\textit{top}) Determinations of the mean $UV$-continuum
  slope $\beta$ for galaxies with a $UV$ luminosity of $L_{z=3}^{*}$
  ($M_{UV}\sim-21$) as a function of redshift (solid red circles).
  The error bars in $\beta$ are $\geq0.15$ and include possible
  systematic errors in both the measured colors and corrections for
  selection effects.  The error bars in $\beta$ (with the wider
  horizontal bars) are for the random errors.  The random errors are
  generally much smaller than those including the estimated systematic
  errors (shown with the narrower horizontal bars).  The right axes
  give the equivalent $E(B-V)$ extinctions (Calzetti et al.\ 2000 law)
  for a given $\beta$ if the input spectrum is a young star-forming
  galaxy ($\sim$100 Myr of constant star formation).  Published
  determinations of the mean $UV$-continuum slope $\beta$ at $z\sim3$
  (\textit{black star}: Adelberger \& Steidel 2000), at $z\sim4$
  (\textit{blue open pentagon}: Ouchi et al.\ 2004), at $z\sim6$
  (\textit{solid cyan square}: Stanway et al.\ 2005 and \textit{solid
    magenta triangle}: Bouwens et al.\ 2006), and at $z\sim4-6$
  (\textit{green open squares}: Hathi et al.\ 2008) are also plotted.
  $L_{z=3}^{*}$ galaxies are found to have bluer colors at
  $z$$\,\sim\,$5-6 than they do at $z$$\,\sim\,$2-4.
  (\textit{bottom}) Determinations of the mean $UV$-continuum slope
  $\beta$ for galaxies with a $UV$ luminosity of $0.15L_{z=3}^{*}$
  ($M_{UV}\sim-19$) as a function of redshift (solid red squares).
  For comparison, our determinations of the mean $UV$-continuum slope
  $\beta$ at $L_{z=3}^{*}$ (from top panel) are included in this panel
  as solid red circles (and offset somewhat from the lower luminosity
  determinations in redshift for clarity).  For each of our dropout
  samples, the lower luminosity galaxies are found to be
  bluer.\label{fig:mcolor}}
\end{figure}

\begin{deluxetable}{ccc}
\tablewidth{2.5in}
\tablecaption{The mean $UV$-continuum slope $\beta$ and $1\sigma$
  scatter of galaxies, as a function of $UV$
  luminosity.\tablenotemark{a,b}\label{tab:uvslope}} \tablehead{
  \colhead{} & \multicolumn{2}{c}{$UV$-continuum slope $\beta$}
  \\ \colhead{$<M_{UV,AB}>$} & \colhead{Mean\tablenotemark{c}} &
  \colhead{$1\sigma$ Scatter\tablenotemark{d}}} \startdata
\multicolumn{3}{c}{$z\sim2.5$ $U$-dropouts} \\ 
$-$21.73 & $-$1.18$\pm$0.17$\pm$0.15 & 0.32\\
$-$20.73 & $-$1.58$\pm$0.10$\pm$0.15 & 0.27\\
$-$19.73 & $-$1.54$\pm$0.06$\pm$0.15 & 0.34\\
$-$18.73 & $-$1.88$\pm$0.05$\pm$0.15 & 0.36\\
\multicolumn{3}{c}{$z\sim4$   $B$-dropouts} \\ 
$-$22.22 & $-$1.32$\pm$0.12$\pm$0.15 & 0.51\\
$-$21.72 & $-$1.33$\pm$0.07$\pm$0.15 & 0.31\\
$-$21.22 & $-$1.44$\pm$0.04$\pm$0.15 & 0.44\\
$-$20.72 & $-$1.51$\pm$0.03$\pm$0.15 & 0.40\\
$-$20.22 & $-$1.58$\pm$0.10$\pm$0.15 & 0.34\\
$-$19.72 & $-$1.68$\pm$0.09$\pm$0.15 & 0.35\\
$-$19.22 & $-$1.73$\pm$0.07$\pm$0.15 & 0.55\\
$-$18.72 & $-$1.93$\pm$0.06$\pm$0.15 & 0.31\\
$-$18.22 & $-$1.98$\pm$0.05$\pm$0.15 & 0.54\tablenotemark{e}\\ 
$-$17.72 & $-$2.03$\pm$0.04$\pm$0.15 & 0.29\tablenotemark{e}\\ 
$-$17.22 & $-$1.91$\pm$0.05$\pm$0.15 & 0.26\tablenotemark{e}\\ 
\multicolumn{3}{c}{$z\sim5$ $V$-dropouts} \\ 
$-$21.90 & $-$1.81$\pm$0.22$\pm$0.15 & 0.45\\
$-$20.90 & $-$1.76$\pm$0.10$\pm$0.15 & 0.64\\
$-$19.90 & $-$2.17$\pm$0.08$\pm$0.15 & 0.66\tablenotemark{e}\\ 
$-$18.90 & $-$2.64$\pm$0.16$\pm$0.15 & 1.02\tablenotemark{e}\\ 
\multicolumn{3}{c}{$z\sim6$ $i$-dropouts} \\ 
$-$20.76 & $-$2.16$\pm$0.15$\pm$0.15 & 0.41\\
$-$18.76 & $-$2.32$\pm$0.19$\pm$0.15 & 0.91\tablenotemark{e}\\ 
\enddata
\tablenotetext{a}{The mean $UV$-continuum slopes $\beta$ (1600\AA$\,\,$to 2300\AA) presented here are also shown on Figure~\ref{fig:colmag} as the red squares.}
\tablenotetext{b}{The
  slopes presented here have been corrected for selection effects and
  measurement errors (see \S3.7).  The tabulated $UV$ luminosities are
  the geometric mean of the $UV$ luminosity measured in the two bands
  used to establish the $UV$ slope (see Table~\ref{tab:pivotbands}).
} 
\tablenotetext{c}{Both random and systematic errors are quoted
  (presented first and second, respectively).  In \S3.8, we estimate
  the likely size of the systematic errors.}
\tablenotetext{d}{The
  $1\sigma$ scatter presented here has been corrected for photometric
  scatter using the simulations described in \S3.7 (see also
  Figure~\ref{fig:seleffect}) and therefore should reflect the
  intrinsic $1\sigma$ scatter in the $UV$-continuum slope $\beta$
  distribution.  Typical uncertainties are $\sim$0.05-0.10.}
\tablenotetext{e}{Because the observed scatter in the $UV$-continuum
  slopes for the faintest sources is dominated by the photometric
  errors, it is very difficult to estimate the intrinsic
  $1\sigma$ scatter in the $\beta$ distribution, and therefore the
  uncertainties on our estimates of the intrinsic scatter in the
  $\beta$ distribution are large, i.e., $\gtrsim0.2-0.4$.}
\end{deluxetable}

\subsection{Uncertainties and Model Dependencies} 

Before presenting the distribution of $UV$-continuum slopes $\beta$
derived for each of the dropout selections, it is worthwhile to ask
ourselves how our determinations may be affected by specific
assumptions we make.  We have already discussed the effect that errors
in our photometry (and zeropoints) would have on the $UV$-continuum
slope determinations (\S3.3).  A $\sim0.05$ mag error in the derived
colors would result in a $\sim0.15$ change in the derived
$UV$-continuum slope $\beta$.  In an attempt to minimize such errors,
we have exercised great care in obtaining a consistent set of colors
across the optical and near-IR passbands for which we have data.

We would expect a similar error in the $UV$-continuum slope $\beta$
from the fiducial SEDs we use to convert from the observed colors to
$UV$-continuum slope $\beta$.  The uncertainties on the derived slope
from this conversion are $\lesssim$0.1 for our $z\sim2.5$, $z\sim5$,
and $z\sim6$ and $\lesssim$0.2 for our $z\sim4$ sample (see Appendix
A).

The $UV$-continuum slopes we derive for our selections also show some
dependence upon the model redshift distributions.  As discussed in
Appendix A, small differences ($\Delta z \sim 0.1$) between the mean
redshifts of dropout galaxies in our models and that present in
reality have a small effect on the derived $UV$-continuum slopes
$\beta$.  Errors of $\Delta z\sim0.1$ in the mean redshifts of these
selections shift the derived $\beta$'s by 0.02, 0.02, 0.01, and 0.05
for our $z\sim2.5$ $U$, $z\sim4$ $B$, $z\sim5$ $V$, and $z\sim6$ $i$
dropout selections, respectively.  These are much smaller
uncertainties than the $\Delta \beta$ $\sim0.15$ errors that would
result from small systematics in the photometry (or uncertainties in
the conversion from observed colors to $UV$-continuum slopes).

Similarly, the presence or absence of Ly$\alpha$ emission (at 1216\AA)
in the spectra of sources in our selections is not expected to have a
large effect on the measured $UV$-continuum slopes $\beta$ themselves.
This is because the measurements are made in passbands which are only
sensitive to light redward of 1216\AA.  The only exception to this is
for our $z\sim6$ $i$-dropout selection, but even there for typical
Ly$\alpha$ EWs ($\sim$50\AA$\,\,$rest-frame: Dow-Hygelund et al.\ 2007;
Vanzella et al.\ 2006; Stanway et al.\ 2007; Vanzella et al.\ 2009),
the $J_{110}-H_{160}$ colors would only change by $\sim$0.07 -- which
would make the derived $\beta$'s $\sim0.17$ steeper.

Nonetheless, the amount of flux in Ly$\alpha$ has an effect on the
redshift distribution of the sources selected with our dropout
criteria.  For galaxies with Ly$\alpha$ EWs towards the upper end of
the observed range ($\sim$50\AA$\,\,$rest-frame: Shapley et al.\ 2003;
Dow-Hygelund et al.\ 2007; Vanzella et al.\ 2006; Stanway et
al.\ 2007; Vanzella et al.\ 2009), the mean redshift of our dropout
selections increases by $\Delta z\sim0.3$.  This changes the derived
$\beta$'s by 0.06, 0.06, 0.03, and 0.15, respectively, for our four
selections (Appendix A).  However, since only a small fraction
($\lesssim50$\%) of the star-forming galaxies at $z$$\,\sim\,$2-6
appear to show Ly$\alpha$ emission at these levels (Shapley et
al.\ 2003; Dow-Hygelund et al.\ 2007; Vanzella et al.\ 2006; Stanway
et al.\ 2007), the effect of the quoted uncertainties on $\beta$ is
likely much smaller than this (i.e., $\lesssim$0.075).  Again, this is
much smaller than the errors we would expect to result from small
systematics in the photometry (or conversions to $UV$-continuum slopes
$\beta$).

Lastly, one might ask whether interstellar absorption features may
have an effect on the $UV$-continuum slopes $\beta$ estimated from the
broadband photometry.  Fortunately, these absorption lines appear not
to have a big effect on the measured slopes (i.e., $\Delta \beta
\lesssim 0.1$) as demonstrated by Meurer et al.\ (1999: see \S3.4 and
Figure 3 from that work) given the wide wavelength range of the
broadband filters used to estimate the slopes and the fact that most
of these absorption lines occur at $\lesssim$1700\AA.

\subsection{Results}

In Table~\ref{tab:uvslope}, we present the mean $UV$-continuum slopes
and $1\sigma$ scatter observed for our four dropout samples after
correction for selection effects and photometric scatter.  This
distribution is the same as that presented in
Figure~\ref{fig:colmag}).  We assume a minimum systematic error in the
mean $UV$-continuum slope $\beta$ of 0.15 as a result of possible
($\sim0.05$ mag) systematics in the measured colors (see \S3.3) and
small model dependencies in the conversion from the observed colors to
$UV$-continuum slope $\beta$ (see Appendix A).

There are clear trends that seem to be present in the distributions of
$UV$-continuum slopes $\beta$ as a function of redshift and $UV$
luminosity.  The first trend is a correlation between the mean
$UV$-continuum slope and $UV$ luminosity, in the sense that galaxies
become bluer at lower $UV$ luminosities.  This trend is particularly
significant for our $z\sim2.5$ $U$-dropout and $z\sim4$ $B$-dropout
selections, and fitting the mean $\beta$ vs. $UV$ luminosity to a
line, we find
$\beta=(-0.20\pm0.04)(M_{UV,AB}+21)-(1.40\pm0.07\pm0.15)$ for our
$z\sim2.5$ $U$-dropout sample and
$\beta=(-0.15\pm0.01)(M_{UV,AB}+21)-(1.48\pm0.02\pm0.15)$ for our
$z\sim4$ $B$-dropout sample (\textit{the fit is shown on
  Figure~\ref{fig:colmag} with the red lines}).\footnote{While we find
  a strong correlation between the $UV$-continuum slope $\beta$ and
  the observed $UV$ magnitude at $z\sim2.5$ over the range
  $V_{606,AB}\,\sim\,23$-27, it appears that the $UV$-continuum slope
  $\beta$ shows a weaker dependence on magnitude at brighter
  magnitudes, i.e., $V_{606,AB}\,\sim\,23$-25.5.  Adelberger \&
  Steidel (2000) and Reddy et al.\ (2008) find no correlation between
  these quantities over this magnitude range, and a t-test applied to
  our $z\sim2.5$ $U$-dropout sample shows a correlation at only 75\%
  confidence over this magnitude range.}  The $\Delta \beta$ errors of
$\pm0.15$ given in the equations above are our estimates of the
systematic errors (see discussion in previous paragraph).  We find no
significant trend in the width of the $UV$-continuum slope $\beta$
distribution as a function of luminosity.

We also find a clear correlation between the mean $UV$-continuum slope
$\beta$ and redshift, in the sense that higher redshift galaxies are
bluer.  This is most evident in Figure~\ref{fig:mcolor} where the mean
$UV$-continuum slope $\beta$ is shown at a luminosity of $L_{z=3}^{*}$
(i.e., $M_{UV}^{*}\sim-21$) and $0.15L_{z=3}^{*}$ (i.e.,
$M_{UV}^{*}\sim-19$) for each of our dropout samples.  At
$z$$\,\sim\,$2-4, the mean $UV$-continuum slope is $\sim-$1.5 while at
$z\gtrsim5$, it is $\lesssim-1.8$.  No change is evident in the width
of the $UV$-continuum slope distribution -- though this width is
difficult to quantify for our highest redshift samples due to the
small number of sources in these high redshift samples and the large
photometric errors of the sources that are available.

\section{The $UV$-continuum slope Distribution: Direct Implications}

In \S3, the $UV$-continuum slope $\beta$ distribution was derived as a
function of both redshift and luminosity.  Being able to establish
this distribution as a function of these two quantities is an
important empirical result and has a number of noteworthy
implications, which we will detail in this section.  However, before
describing these implications, it is prudent to compare the present
$UV$-continuum slope $\beta$ determinations with those in the
literature to put them in context (see \S4.1).

One of the most significant implications of these results is for the
completeness of dropout selections at $z\sim2-6$, which we discuss in
\S4.2.  Being able to establish these distributions is also important
for a determination of the $z\sim2-6$ LFs (\S4.3).  In \S4.4, we
consider the question of how variations in the $UV$-continuum slope
$\beta$ likely arise, and we argue that changes in the dust extinction
likely dominate the observed variations.  Finally, in \S4.5, we
attempt to connect the sequence we find in the $UV$-continuum slope
versus luminosity to similar trends found at lower redshift (and over
other wavelength baselines).

\subsection{Comparison to previous determinations of the $UV$-continuum slope}

In \S3, we derived the distribution of $UV$-continuum slopes $\beta$
over a wide range in redshift and $UV$ luminosity using a very
systematic approach while taking advantage of a wide-variety of both
deep and wide-area HST data.  We presented evidence that the mean
$UV$-continuum slope $\beta$ is bluer at $z$$\,\sim\,$5-6 than it is at
$z$$\,\sim\,$2-4 and that this slope is also bluer at lower $UV$
luminosities at $z$$\,\sim\,$2-4.

Previously, there had been a variety of different attempts to measure
these slopes at specific redshifts or luminosities (typically $\sim
L_{z=3}^{*}$), e.g., Steidel et al.\ (1999), Meurer et al.\ (1999),
Adelberger \& Steidel (2000), Ouchi et al.\ (2004), Stanway et
al.\ (2005), Bouwens et al.\ (2006), and Hathi et al.\ (2008).  The
top panel of Figure~\ref{fig:mcolor} provides a summary of many of
these previous measurements.  Most of the early high-redshift work
focused on $z$$\,\sim\,$2-3 and was based upon $U$ dropout selections
from the HDF and large ground-based LBG searches (e.g., Steidel et
al.\ 1999; Adelberger \& Steidel 2000; Meurer et al.\ 1999).  In those
papers, the $UV$-continuum slope $\beta$ was found to have a mean
value of $\sim-$1.4 and $1\sigma$ dispersion of $\sim$0.5-0.6 at $\sim
L_{UV}^{*}$.  No significant correlation of $UV$-continuum slope
$\beta$ with $UV$ magnitude was found to $-20$ AB mag (Adelberger \&
Steidel 2000; Reddy et al.\ 2008), though there would appear to be
some evidence in the $\beta$ distributions presented by Meurer et
al.\ (1999: e.g., Figure 5 from that paper) that the $\beta$
distribution becomes a little bluer at lower $UV$ luminosities (i.e.,
$-$18.5 AB mag).\footnote{Unfortunately, Meurer et al.\ (1999) do not
  provide a lot of discussion on the possible correlation of $\beta$
  with observed $UV$ luminosity (despite the existence of a likely
  trend) and instead emphasizes the correlation of $\beta$ with
  \textit{dust-corrected} $UV$ luminosity.}

Ouchi et al.\ (2004) extended these studies to higher redshift by
presenting a determination of the $UV$-continuum slope distribution at
$z\sim4$ based on a large selection of $B$ dropout galaxies from the
Subaru Deep Field and Subaru XMM/Newton Deep Field.  Ouchi et
al.\ (2004) found that the $UV$-continuum slope distribution at
$z\sim4$ was consistent with that determined at $z\sim3$ and that
there was only a marginal trend (not statistically significant)
towards bluer slopes at lower luminosities.  Papovich et al.\ (2004),
by contrast, working with a selection of $z\sim4$ $B$ dropouts from
the GOODS fields, found that galaxies at $z\sim4$ were slightly bluer
in their $UV$ colors than at $z\sim2.5$ in the Hubble Deep Field North
and South.  At somewhat fainter magnitudes, Beckwith et al.\ (2006)
remarked that the $UV$ colors of $z\sim4$ $B$ dropouts in the HUDF
were very blue in general (with $\beta$'s of $\sim$$-2$) and hence
suggested very little dust extinction.  The somewhat bluer
$UV$-continuum slopes found by Beckwith et al.\ (2006) than by
e.g. Ouchi et al.\ (2004) is not particularly surprising given the
substantial differences in the mean luminosities of the two samples.
Incidentally, the trend towards bluer $UV$-continuum slopes at lower
luminosities reported on here is evident in Figure 18 of Beckwith et
al.\ (2006) although it was not specifically noted.

At even higher redshifts, Lehnert \& Bremer (2003) found that $z\sim5$
$V$ dropouts had $UV$-continuum slopes $\beta$ very close to $-2$
while later Stanway et al.\ (2005) and Bouwens et al.\ (2006) found
$UV$-continuum slopes of $-2.2\pm0.2$ and $-2.0\pm0.3$, respectively,
at $z\sim6$ from a selection of $i$-dropouts in the HUDF.  Hathi et
al.\ (2008) used a small sample of galaxies at $z\sim4-6$ from the
HUDF and larger sample of $U$-dropouts at $z\sim3$ in an attempt to
quantify the change in the $UV$-continuum slope $\beta$ as a function
of redshift.  Hathi et al.\ (2008) found that the $UV$-continuum slope
$\beta$ showed only a slight change from $z\sim3$ (where
$\beta\sim-1.5$) to $z$$\,\sim\,$5-6 (where $\beta\sim-1.8$).

Broadly, the picture that has emerged from these studies is that
star-forming galaxies become bluer towards higher redshifts and to
lower luminosities, but it has been difficult, given possible
systematics between the different studies, to quantify the size of the
changes.  We confirm this overall picture, finding that the mean
$UV$-continuum slope is $\sim1.0\pm0.4$ bluer at $z\sim6$ than it is
at $z\sim2.5$ and $\sim0.5\pm0.1$ bluer at $-18$ AB mag
($\sim0.1L_{z=3}^{*}$) than it is at $-21$ AB mag ($\sim
L_{z=3}^{*}$).  Because of the larger samples and use of a consistent
approach in deriving these slopes at all redshifts and luminosities,
the differential evolution we measure is more robust than in previous
studies.

Looking more specifically at the $UV$-continuum slope $\beta$
measurements we have derived at various redshifts and comparing those
measurements with those obtained in previous studies, we find very
good agreement in general, in almost all cases within the quoted
errors.  The only significant exception to this is the Hathi et
al.\ (2008) measurements of the $UV$-continuum slope $\beta$ at
$z\sim6$ where a mean $\beta$ of $-1.73\pm0.35$ was found.  This
appears to result from the relation Hathi et al.\ (2008) use to
compute the $UV$-continuum slope $\beta$ at $z\sim6$ (i.e.,
$\beta=2.56(J-H)-2$).  This relation does not account for the fact
that the NICMOS $J_{110}$ band extends down to 8000\AA$\,\,$and
therefore $z\sim6$ $i$-dropouts partially drop out in the $J_{110}$
band (i.e., are fainter in the $J_{110}$ band and hence have redder
$J_{110}-H_{160}$ colors).  Comparing $\beta=2.56(J-H)-2$ with
Eq.~\ref{eq:ibeta} from Appendix A, we can readily see why the mean
$UV$-continuum slopes we derive are much steeper (by $\sim$0.3).  We
note that an additional $-0.1$ mag shift in $\beta$ relative to
previous $z\sim6$ measurements of $\beta$ (e.g., Bouwens et al.\ 2006;
Hathi et al.\ 2008) comes from small offsets we make to the
$J_{110}$-band photometry (\S3.3).

\begin{figure*}
\epsscale{1.05}
\plotone{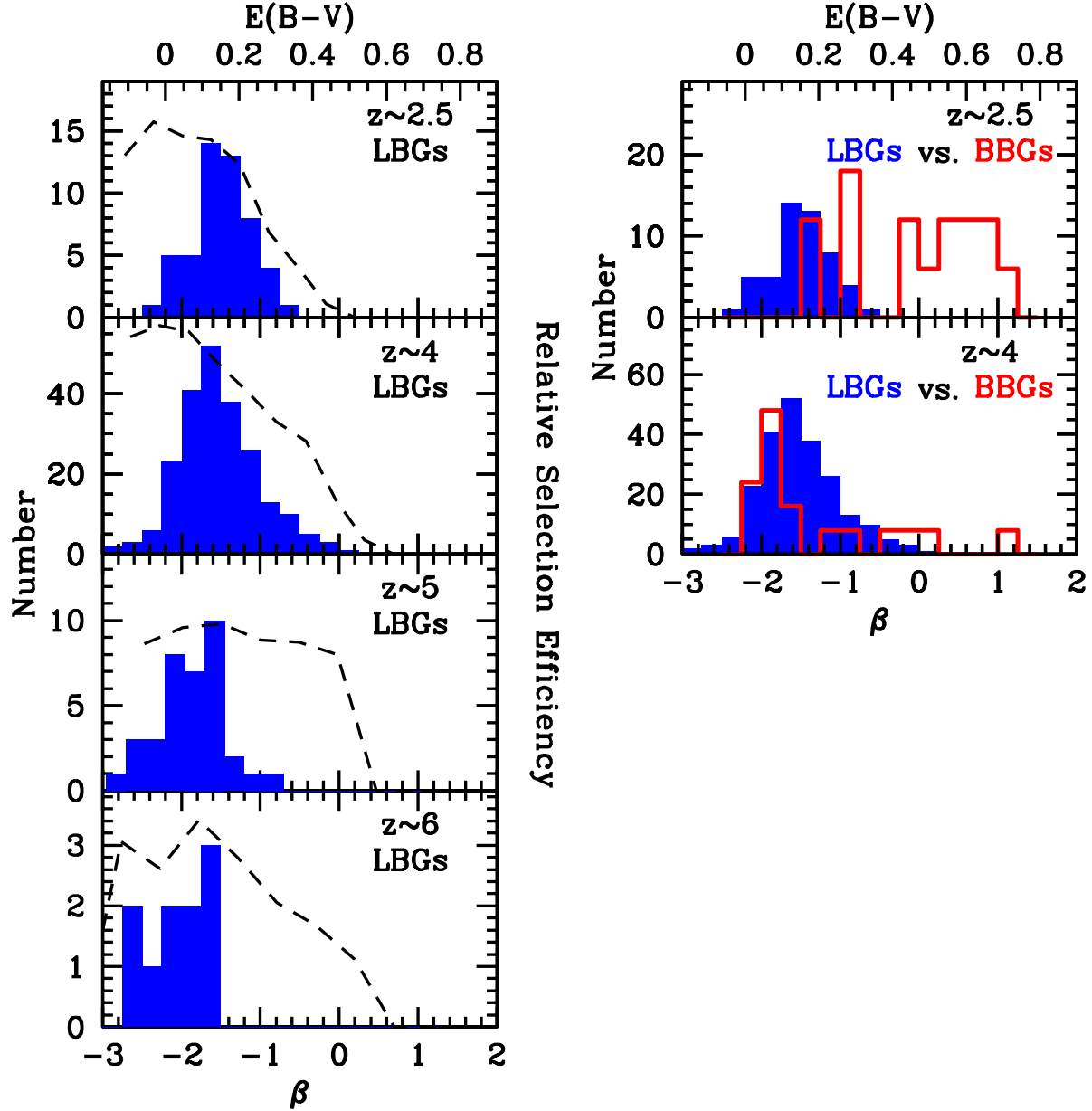}
\caption{(\textit{left}) Distribution of $UV$-continuum slopes $\beta$
  observed at $z$$\,\sim\,$2-6 for the present Lyman-Break Galaxy
  selections (solid blue histograms).  We only include the observed
  $\beta$ values for Lyman-Break galaxies $>$1 mag brighter than the
  selection limit to avoid being dominated by selection effects.  On
  the top axes are the $E(B-V)$ extinctions (Calzetti et al.\ 2000
  law) that are equivalent to a given $\beta$ if the base spectrum is
  a young star-forming galaxy ($\sim100$ Myr of constant star
  formation).  The selection volume (\S3.7) available for galaxies
  with $UV$-continuum slope $\beta$ and magnitudes $\gtrsim1$ mag
  brighter than the selection limit is shown with the thick dashed
  lines in each panel (the normalization is arbitrary).  While it is
  more difficult to select galaxies with redder $UV$-continuum slopes
  $\beta$ with a Lyman Break criterion, our simulations (\S3.7)
  indicate that our dropout selections should be capable of
  identifying modest numbers of galaxies with $\beta$'s as red as 0.5.
  Indeed, such galaxies (with $\beta$'s redder than $-1$) are present
  in our $z$$\,\sim\,$2-4 LBG selections albeit in small numbers.
  However, they are conspicuously absent in our $z\gtrsim5$
  selections.  This suggests that such sources are exceedingly rare at
  $z\gtrsim5$.  Conversely, the modest number of galaxies with red
  $UV$-continuum slopes $\beta$'s seen at $z\sim4$ seem consistent
  with the small samples of $z\sim4$ ULIRGs that have been found in
  the observations (e.g., Daddi et al.\ 2009).  The lack of very blue
  $\beta\lesssim-3$ galaxies is also consistent with expectations from
  models and suggest that the $\beta$ distributions derived here are
  reliable.  (\textit{right}) Distribution of $UV$-continuum slopes
  observed for Lyman Break Galaxy selections at $z$$\,\sim\,$2-4
  (solid blue histogram) and for Balmer-Break Galaxy selections at
  $z$$\,\sim\,$2-4 (red histogram: Brammer \& van Dokkum 2007).  The
  $E(B-V)$ extinctions given on the top axis are as in the left panel
  (but are not expected to be particularly representative for the
  somewhat older BBGs).  Both $\beta$ distributions plotted are as
  observed, with no correction for selection or measurement effects.
  Balmer-Break Galaxy (BBG) selections provide us with an independent
  measure of the $UV$-continuum slope distribution since they do not
  depend upon their $UV$-continuum slope or the brightness of these
  sources in the $UV$-continuum.  The fact that the distribution of
  $UV$-continuum slopes $\beta$ is becoming substantially bluer for
  both selections (and in particular the Brammer \& van Dokkum 2007
  BBG selection at $z\sim4$) suggests that Lyman-Break selections
  (sensitive to galaxies with $\beta$'s bluer than 0.5: see
  Figures~\ref{fig:sel} and \ref{fig:seleffect}) should be largely
  complete at $z\gtrsim5$ (see \S4.2).  \label{fig:complete}}
\end{figure*}

\subsection{Do LBG selections miss a substantial population of red 
star-forming galaxies at $z\gtrsim4$?}

The Lyman Break galaxy selection technique identifies galaxies at high
redshift through simple color criteria.  This technique is well
established to be an efficient and robust method for identifying
star-forming galaxies over a wide range in redshift $z$$\,\sim\,$2-6
(Steidel et al.\ 1996; Williams et al.\ 1996; Bunker et al.\ 2003;
Vanzella et al.\ 2006; Dow-Hygelund et al.\ 2007; Vanzella et
al.\ 2009).  This technique provides a very complete census of $UV$
light at high redshift -- simply by virtue of the selection wavelength
itself.  However, it is less efficient for probing the total stellar
mass or even the total SFR at high redshift (e.g., van Dokkum et
al.\ 2006).  This is because galaxies with the highest stellar masses
or SFRs are often either old or dust obscured -- making them fainter
in the $UV$ and thus more difficult to select with the LBG technique.

Our interest here is in examining the systemic completeness of LBG
selections.  We want to determine whether there is a set of galaxies
at high redshift that we miss altogether by virtue of the LBG
selection technique.  Note that this is a very different question from
determining whether there is a class of galaxy at high redshift that
we select \textit{less efficiently} because they are faint in the UV
(e.g., because of dust or age).  In general, we would expect sources
to miss our LBG selections if one of the two LBG color criteria failed
to hold, i.e., if (1) the sources did not show a strong Lyman break or
(2) the sources were too red in their $UV$-continuum slope to satisfy
the LBG selection.  We would expect criterion (1) to always hold for a
sufficiently high redshift source, as a result of line blanketing by
the Ly$\alpha$ forest.  However, we might expect criterion (2) to fail
if the $UV$-continuum slopes $\beta$ of the sources were sufficiently
red.

We can attempt to address this question by looking at how many
galaxies have $UV$-continuum slopes $\beta$ that lie close to the
selection limits.  If our samples contain a large number of such
galaxies, it would suggest that our LBG selections suffer from
significant incompleteness near these limits.  Figure~\ref{fig:sel}
shows the selection criteria for each of LBG selections and
Figure~\ref{fig:complete} presents the effective search volume (\S3.7)
for galaxies (\textit{dashed black lines} on the left-hand panels) as
a function of $UV$-continuum slope $\beta$ (see also
Figure~\ref{fig:seleffect}).  It is quite clear from these figures
that our selection criteria is effective in identifying star-forming
galaxies at $z$$\,\sim\,$2-6 with $\beta$'s bluer than $\sim0.5$, and it is
striking how much redder this limit is than the $UV$-continuum slopes
derived for galaxies in our dropout samples at $z$$\,\sim\,$2-6
(\textit{solid blue histograms} in Figure~\ref{fig:complete}).  The
situation is particularly conspicuous for galaxies with $UV$-continuum
slopes $\beta$ redder than $-1$ at $z\gtrsim5$.  While our simulations
(\S3.7) suggest that galaxies with these colors should show up in our
selections if they existed (\textit{dashed lines} in
Figure~\ref{fig:complete}), essentially none are found.  This suggests
that star-forming galaxies with red $UV$-continuum slopes $\beta$ are
extremely rare.

\textit{Complementary Balmer Break Selections:} Another way we can
investigate the issue of possible systemic incompleteness in LBG
selections is through other selection techniques.  One such technique
identifies galaxies based upon their rest-frame optical light and
searches for a prominent Balmer Break at
$\sim$3700\AA$\,\,$rest-frame.  Such selections are not surprisingly
called Balmer Break Galaxy (BBG) selections and typically identify the
oldest and most massive galaxies.  Since such galaxies tend to be more
chemically evolved, they are frequently more dusty.  Considering such
selections therefore permit us to evaluate the extent to which our LBG
selections may miss redder and more dust-obscured starbursts.  Brammer
\& van Dokkum (2007) performed such a selection at 2$<$$z$$<$3 and
3$<$$z$$<$4.5 (see right-hand panels on Figure~\ref{fig:complete})
based upon the Faint Infrared Extragalactic Survey (FIRES) data
(Labb{\'e} et al.\ 2003; F{\"o}rster Schreiber et al.\ 2006) and
derived $UV$-continuum slopes for sources in their selections from SED
fits to the photometry.  They found that $\sim$67\% of the galaxies in
their $z$$\,\sim\,$2-3 selection had $UV$-continuum slopes bluer than
0.5, but almost all ($\gtrsim90$\%) of the galaxies in their
$z\sim$3-4.5 sample had such blue slopes.  This is highly encouraging
for $z\sim4$ LBG selections given the independent nature of Balmer
Break selections -- suggesting that $z\sim4$ LBG selections may be
largely complete.

By contrast, the fact that $\sim$33\% of the galaxies in the Brammer
\& van Dokkum (2007) $z$$\,\sim\,$2-3 BBG sample have $\beta$'s redder
than 0.5 suggests that completeness could be somewhat of a concern for
LBG selections at $z\sim2.5$, and in fact it is well known that at
$z$$\,\sim\,$2-3, there is a substantial population of dust-obscured
galaxies undergoing vigorous star formation (e.g., Hughes et
al.\ 1998; Barger et al.\ 1999; Chapman et al.\ 2005; Labb{\'e} et
al.\ 2005; Papovich et al.\ 2006; Pope et al.\ 2006).

Searches for Balmer Break galaxies at $z\gtrsim5$ have also been
conducted (e.g., Dunlop et al.\ 2007; Wiklind et al.\ 2008: see also
Rodighiero et al.\ 2007 and Mancini et al.\ 2009).  As with the
Brammer \& van Dokkum (2007) study, these searches would seem to be
relevant to the present discussion we are having about the
completeness of LBG selections for star-forming galaxies at high
redshift.  Unfortunately, there is a wide diversity of search results
in this area that make it difficult to draw clear conclusions.
Wiklind et al.\ (2008) report 11 plausible Balmer Break galaxies at
$z$$\,\sim\,$5-7 over the CDF South GOODS field, 7 of which are detected at
$24\mu$ with MIPS, while Dunlop et al.\ (2007) find no Balmer Break
galaxies over the same redshift range, in their analysis of the same
field.  While the MIPS detections for 7 of the 11 $z$$\,\sim\,$5-7 Wiklind
et al.\ (2008) BBG candidates may be interpreted as due to obscured
AGN, we believe a much more likely explanation is due to PAH emission,
as Chary et al.\ (2007b) argue for HUDF-JD2 (Mobasher et al.\ 2005).
The other BBG candidates from Wiklind et al.\ (2008) may have
redshifts of $z$$\,\sim\,$5-7, but may also have lower redshifts.

\textit{Implications:} Putting together the present LBG selection
results at $z$$\,\sim\,$2-6 with those from Balmer Break selections over the
same range and other results in the literature, we find clear evidence
that the galaxy population at high redshift is increasingly blue as
one moves out to high redshift, and therefore high-redshift LBG
selections seem likely to be increasingly complete (and consequently
suffer much less from systematic incompleteness).  These trends are
very clear in LBG selections at $z$$\,\sim\,$2-6, and since there is no
reason to suppose that these trends come from various selection biases
(not only do the sources have $\beta$'s very far from the selection
limits, but the effect of observational selection are corrected for),
it seems reasonable to take the observed evolution at face value.
Supporting evidence comes from a similar evolution seen in
complementary Balmer Break selections (at least according to Brammer
\& van Dokkum 2007).  Previously, Bouwens et al.\ (2007: \S4.1) have
discussed this in the context of $z\sim4$ $B$-dropout selections.

Obviously we would expect galaxies at very high redshifts ($z$$\,\sim\,$5-6)
to have much bluer $UV$-continuum slopes than at $z\sim2-3$, because
of the much smaller time baseline over which to produce metals and
dust, as well as the shorter dynamical time scales of galaxies at
$z\gtrsim5$ (and hence much reduced ages).  We would also expect this
result based upon the evolution of the $UV$ LF.  Since the
characteristic luminosity of galaxies in the UV becomes progressively
smaller as we move to higher redshifts (Dickinson et al.\ 2004;
Shimasaku et al.\ 2005; Bouwens et al.\ 2006; Yoshida et al. 2006;
Bouwens et al.\ 2007; Oesch et al.\ 2009), we would expect to find
fewer and fewer galaxies at high redshift with very large SFRs.  Since
dust extinction is well correlated with the SFR (e.g., Wang \& Heckman
1996; Hopkins et al.\ 2001; Martin et al.\ 2005; Reddy et al.\ 2006;
Buat et al.\ 2007; Zheng et al.\ 2007), we would not expect to find
many galaxies at $z\gtrsim5$ with substantial dust extinction.
Finally there is the evolution seen in the relationship between SFR
and dust extinction.  From $z\sim0$ to $z\sim0.7$ to $z\sim2$, it has
been found that the effective dust extinction for a given star
formation rate decreases quite strongly as we move out to higher
redshifts (Reddy et al.\ 2006; Buat et al.\ 2007; Burgarella et
al.\ 2007).  Each of these considerations suggest that galaxies at
high-redshift should be almost uniformally very blue and LBG
selections very complete.

Nonetheless, we know there is a substantial population of luminous,
dust obscured galaxies at $z\sim2-3$ (e.g., Hughes et al.\ 1998;
Barger et al.\ 1999; Chapman et al.\ 2005; Labb{\'e} et al.\ 2005;
Papovich et al.\ 2006; Pope et al.\ 2006) and extending out to
redshifts as high as $z\sim4.5$ (e.g., Capak et al.\ 2008; Daddi et
al.\ 2009; Wang et al.\ 2009; Dannerbauer et al.\ 2008).  While some
authors (e.g., Daddi et al.\ 2009) make the suggestion that these
populations might be quite significant and contribute substantially to
the overall SFR density at $z\gtrsim4$, we would argue that the role
of these galaxies is likely much more modest in scope and they do not
provide a large fraction of the SFR density, particularly at
$z\gtrsim4$.  We discuss and quantify this point in \S6.2 (see also
Table~\ref{tab:effdust}, Table~\ref{tab:sfrdensulirg},
Figure~\ref{fig:sfrdensulirg}, and Figure~\ref{fig:wrelfrac}).

\subsection{Relevance for LF Determinations}

As shown in Figure~\ref{fig:sel}, Figure~\ref{fig:seleffect} and
Figure~\ref{fig:complete} (and as discussed in the previous section),
the effective selection volume for star-forming galaxies at
$z$$\,\sim\,$2-6 is very sensitive to its $UV$-continuum slope $\beta$
and $UV$ luminosity and can vary by factors of $\gtrsim2\times$ (see
also discussion in Sawicki \& Thompson 2006a or Beckwith et
al.\ 2006).  Quantifying the distribution of $UV$-continuum slopes for
star-forming galaxies (as a function of both luminosity and redshift)
is therefore an important first step in determining the $UV$ LFs at
high redshift.

Given this impact, it has been somewhat surprising that different
teams have made very different assumptions about the distribution of
$UV$-continuum slopes in their determinations of the $UV$ LFs.  Some
teams have adopted $UV$-continuum slopes of $\beta\sim-1.4$ (Sawicki
\& Thompson 2006a; Yoshida et al.\ 2006), other teams have assumed
$UV$ continuum slopes of $\beta\sim-2.0$ (Beckwith et al.\ 2006), and
yet other teams have adopted luminosity-dependent $UV$-continuum
slopes (Bouwens et al.\ 2007).  Such differences have only added to
the large dispersion seen in LF determinations at high redshift (see
Figures 10, 11, and 13 from Bouwens et al.\ 2007).

As a result of the present quantification of $UV$-continuum slopes
$\beta$ at high redshift (vs. redshift and luminosity), we now have a
relatively uniform set of assumptions they can be used for quantifying
the $UV$ LFs at $z$$\,\sim\,$2-6 and in extending such determinations to
$z\gtrsim7$.

\begin{figure}
\epsscale{1.13}
\plotone{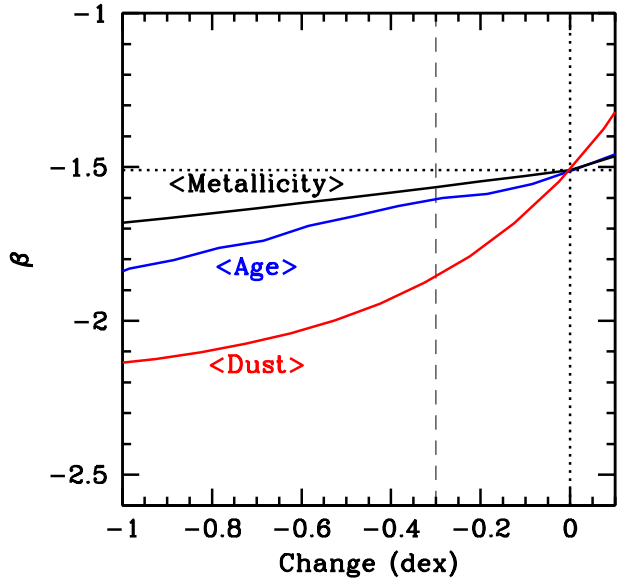}
\caption{$UV$-continuum slope $\beta$ (1600\AA$\,\,$ to 2300\AA$\,\,$
  rest-frame baseline) predicted for specific fractional changes in
  the mean age, metallicity, or dust extinction of high-redshift
  star-forming galaxies from our fiducial stellar population model.
  For our fiducial model (where $\beta=-1.47$), we assume $t = 70$
  Myr, $\tau=10$ Myr, $[Z/Z_{\odot}]=-0.7$, $E(B-V)=0.15$, and a
  Salpeter IMF (where the star formation history is parametrized as
  $e^{-t/\tau}$) from the Papovich et al.\ (2001) fits to $z\sim2.5$
  $U$-dropouts from the WFPC2 HDF North.  In modifying our fiducial
  model to have younger ages, we make changes to both $t$ and $\tau$.
  A factor of 2 (0.3 dex) change (dahsed line) in the mean dust
  content, age, or metallicity of the star-forming galaxy population
  at high redshift shifts the $UV$-continuum slope $\beta$ by
  $\sim0.35$, $\sim0.1$, and $\sim0.05$, respectively.  The effect of
  the stellar IMF on $\beta$ is not shown here, but results in only
  modest changes in $\beta$, i.e., $\Delta\beta \lesssim0.1$ for a
  $\sim0.5$ shift in the slope of the IMF.  It seems clear that
  changes in the mean dust content of galaxies at high-redshift has
  the biggest effect on the $UV$-continuum slope $\beta$ -- though
  some change is also likely the result of differences in the average
  age.  Nonetheless, since stellar population modelling of $z\sim4-6$
  dropout galaxies (Stark et al.\ 2009) suggest that age shows little
  dependence on redshift or luminosity (at least in the median), it
  seems likely that age only plays a minor role ($\Delta
  \beta\lesssim0.2$) in driving the observed trends
  (\S4.4).\label{fig:agemetaldust}}
\end{figure}

\subsection{Interpreting variations in the $UV$-continuum slope $\beta$}

Having used the available observations to derive the distribution of
$UV$-continuum slopes $\beta$ for high-redshift galaxies as a function
of $UV$ luminosity and a range in redshift (Table~\ref{tab:uvslope}
and Figure~\ref{fig:colmag}), we might ask ourselves what this teaches
us about the physical properties of high-redshift galaxies.  Since the
$UV$-continuum slope $\beta$ is purely an observational measure of the
shape of the spectrum of high-redshift galaxies and can be affected by
a variety of different physical conditions (including dust, age,
metallicity, etc.), we cannot use the above measurements to make any
unambiguous observational inferences about the nature of high-redshift
galaxies.

Nevertheless, we will argue that the simplest way to explain most of
observed differences in $\beta$ is through changes in the dust content
of galaxies.  While there are undoubtably variations in other
quantities such as the age, metallicity, or IMF of the stars that
affect these slopes, we will argue that variations in the dust content
of galaxies likely have the largest effect on the observed
$UV$-continuum slopes and we can use the observed variations in these
slopes to make inferences about changes in the effective dust
extinction as a function of galaxy luminosity and redshift.

We note that we would not expect the presence or luminosity of an AGN
to have a big effect on these slopes, given that the observed
incidence of such sources at $z$$\,\sim\,$2-4 is just $\sim3$\% for
moderately bright galaxies at $z\sim2-3$ (e.g., Nandra et al.\ 2002)
and there is little evidence they are more frequent or important at
higher redshift or lower luminosities (e.g., Lehmer et al.\ 2005;
Ouchi et al.\ 2008).

To ascertain which of the aforementioned factors (e.g., the overall
dust content and properties, the age, metallicity, or IMF of a stellar
population) are likely to be the most important for interpreting the
variations we see in the $UV$-continuum slopes $\beta$, we consider
the effect that changes in many of the above quantities would have on
the $UV$-continuum slope $\beta$.  For simplicity, we model the star
formation history of galaxies with a simple $\tau$ model $e^{-t/\tau}$
with $\tau = 10$ Myr (e.g., as in Papovich et al.\ 2001) and $t$ equal
to the age. We also take the metallicity to be fixed and not evolve
over this entire history.  Finally, we implement the dust extinction
by applying the Calzetti et al.\ (2000) law to the SED resulting from
the stellar population models (Bruzual \& Charlot 2003).  Using the
Papovich et al.\ (2001) modelling of $z\sim2.5$ $U$-dropouts in the
HDF North as a guide, we adopt $t = 70$ Myr, Salpeter IMF,
$[Z/Z_{\odot}]=-0.7$, and $E(B-V)=0.15$ as our fiducial parameters.

We then change the age, metallicity, and dust content of this model by
various factors and calculate the effect it would have on the
$UV$-continuum slope $\beta$ predicted for a galaxy, as measured over
the baseline 1600\AA$\,\,$to 2300\AA.  The results are shown in
Figure~\ref{fig:agemetaldust}.  From this figure, it is clear that the
largest changes in the $UV$-continuum slope come from variations in
the dust content of galaxies and the other parameters have a smaller
effect.  For factor of $\sim$2 changes in the age, metallicity, and
dust, we estimate that the $UV$-continuum slope $\beta$ change by
$\sim$0.05, $\sim$0.1, $\sim$0.35, respectively.

Of course, we also see that age has a modest effect on the observed
$UV$-continuum slope $\beta$.  This is well documented in the
literature (e.g., Bell 2002; Kong et al.\ 2004; Cortese et al.\ 2006;
Panuzzo et al.\ 2007).  Despite this sensitivity of $\beta$ to age, no
significant change are found in the median ages of star-forming
galaxies from $z\sim6$ to $z\sim4$ (Stark et al.\ 2009), and little
changes are found in the median ages as a function of luminosity
(there is a hint that lower luminosity galaxies may be
$\sim$1.5$\times$ younger).  Galaxies are found to have median ages of
$\sim$150 Myr.  This suggests that the effect that systematic changes
in the age of the stellar populations on the mean $\beta$ is at most
modest ($\lesssim0.2$).  Moreover, even if we ignore these
observational results, it seems unlikely that the average ages of
star-forming galaxies would increase more rapidly than some multiple
of the dynamical time in a galaxy.  From theory (e.g., Mo et
al.\ 1998), the dynamical times scale as $\rho^{-1/2}\sim(1+z)^{-3/2}$
-- which is a factor of $\sim3$ (0.5 dex) from $z\sim6$ to $z\sim2.5$.
For such a change in age, the change in $\beta$ would be $\Delta
\beta=0.2$, which is small compared to the observed differences --
where $\Delta \beta$ is 0.5-1.0.

Metallicity has an even smaller effect (by a factor of $\sim$8) on the
observed $UV$-continuum slopes $\beta$ than either dust or age do, so
we would require very large variations in the metallicity to affect
the $UV$-continuum slope $\beta$ in any sizeable way.  However, since
any substantial change in metallicity would almost certainly be
accompanied by a similar change in the dust content (given the
correlation between these two quantities), we would again be left with
a situation where the changes in $\beta$ resulting from metallicity
would be completely overwhelmed by changes in $\beta$ resulting from
dust.

Changing the IMF of galaxies only appears to have a modest effect on
the $UV$-continuum slope $\beta$.  For example, changing the slope of
the IMF by $\sim$0.5 only changes $\beta$ by $\lesssim$0.1.  Moreover,
for a steep enough $UV$-continuum slope and a top heavy enough IMF,
there is very red nebular continuum emission (resulting from the
ionizing radiation: Schaerer 2002; Venkatesan et al. 2003; Schaerer
2003; Zackrisson et al.\ 2008; Schaerer \& de Barros 2009) that more
than offsets the very blue spectrum from the stars themselves (Figure
1 from Schaerer \& Pello 2005).

In summary, we expect that the observed variations in the mean
$UV$-continuum slope $\beta$ to be largely the result of changes in
the mean dust extinction.

\subsection{Connection to trends found in star-forming galaxies at $z\sim0-3$:}

One of the most salient trends we found in the $UV$-continuum slopes
$\beta$ derived for $z\sim2.5$ $U$-dropout and $z\sim4$ $B$-dropout
galaxies was the presence of a strong correlation between the
$UV$-continuum slope $\beta$ and $UV$ luminosity (see also Meurer et
al.\ 1999).  As detailed in \S3.9, the mean $UV$-continuum slope
$\beta$ for $z\sim2.5$ and $z\sim4$ galaxies decreases (becomes bluer)
by 0.20$\pm$0.04 and 0.15$\pm$0.01, respectively, for each magnitude
we reach fainter in luminosity.

It seems reasonable to imagine that such trends might also be present
in even lower redshift samples or in other colors for $z$$\,\sim\,$2-4
samples.  In fact, a strong correlation between $UV$-optical colors
and rest-frame optical magnitude has been found for star-forming
galaxies in the ``blue'' cloud (e.g., Papovich et al.\ 2001; Baldry et
al.\ 2004; Papovich et al.\ 2004; Wyder et al.\ 2007; Labb{\'e} et
al.\ 2007), in the sense that more luminous galaxies are redder.  This
is very similar to the relationship we find between the $UV$-continuum
slope $\beta$ and $UV$ luminosity.  Looking at the comparison more
quantitatively, the slope of the color-magnitude relationship
Labb{\'e} et al.\ (2007) measure, for example, is equivalent to $d
\beta$/$dM_{UV}$ of $\sim-0.27$.  This is similar to the
$-0.20\pm0.04$ slope we estimate at $z\sim2.5$ for our $U$-dropout
selection (see \S3.9).

Finally, it is worthwhile to remark on the physical origin of this
slope -- which could plausibly be explained by changes in either the
mean age or the dust content of galaxies as a function of luminosity.
In the previous section, we argued that the simplest way of
accommodating the sizeable change ($\Delta \beta \sim 0.6$) in the
$UV$-continuum slope $\beta$ with luminosity was through a change in
the dust content of the galaxy population.  Explaining the change in
$\beta$ with age would require very large changes (factor of $\sim$10
changes in the mean galaxy age) -- which seem contrary to the modest
changes in age noted by Stark et al.\ (2009) as a function of
luminosity.  

Labb{\'e} et al.\ (2007) also argue that the slope in the
color-magnitude relationship is predominantly the result of a
variation in the dust content.  Labb{\'e} et al.\ (2007) draw this
conclusion based upon an examination of the slope of the
color-magnitude relationship vs. wavelength and through a detailed
examination of the $z\sim0$ Nearby Galaxy Field Sample (Jansen et
al.\ 2000).  This suggests that the color-magnitude relationship we
observe for star-forming galaxies may simply be another manifestation
of the well-known mass-metallicity relationship observed at $z\sim0-3$
(e.g., Tremonti et al.\ 2004; Erb et al.\ 2006a; Maiolino et
al.\ 2008).

\section{Inferred Dust Extinction}

In \S4.4-\S4.5, we argued that the most likely interpretation of the
systematic changes in the mean $UV$-continuum slope $\beta$ (as a
function of redshift or luminosity) is a change in the dust content of
galaxies.  Given the sizeable (factor of $\sim$5) dust corrections
inferred at $z\sim2-3$ for luminous galaxies (e.g., Reddy et
al.\ 2006; Meurer et al.\ 1999; Erb et al.\ 2006b; Reddy \& Steidel
2004), any changes in the estimated dust corrections would have a
substantial effect on the SFR densities inferred at earlier redshifts
and hence its evolution across cosmic time.

In this section, we use the measured $UV$-continuum slope $\beta$
distribution at $z$$\,\sim\,$2-6 to estimate the dust corrections for
$UV$-bright galaxies at $z$$\,\sim\,$2-6.  We begin by describing the
formula we use to estimate dust extinction using the measured
$UV$-continuum slopes (\S5.1).  Then, we discuss the likely
significance of the observed trends in $UV$-continuum slope and dust
extinction, versus luminosity (\S5.2) and redshift (\S5.3).  In \S5.4,
we combine the derived trends in dust extinction with the observed
$UV$ LFs to calculate dust corrections, to various limiting
luminosities.  We then use the typical dust corrections for specific
luminosity galaxies to estimate the bolometric luminosities of
specific luminosity galaxies in the $UV$ (\S5.5).  Finally, in \S5.6,
we discuss the few well-cited cases of dusty high-redshift galaxies
and explain why they do not alter the simple picture we lay out in
this section.

\subsection{Inferred Dust Extinction} 

To estimate the effective dust extinction for galaxies in our various
samples and as a function of redshift, we rely upon the correlation
found between dust extinction and the $UV$-continuum slope $\beta$ at
$z\sim0$ (Meurer et al.\ 1999):
\begin{equation}
A_{1600} = 4.43 + 1.99\beta'
\label{eq:mhc}
\end{equation}
where the dust extinction $A_{1600}$ here is specified at rest-frame
1600\AA$\,\,$and where $\beta'$ is the $UV$-continuum slope measured
from 1300\AA$\,\,$to 2600\AA.  This relationship has been shown to be
a reasonable predictor of the actual dust extinction in galaxies at
$z\sim0$, $z\sim1$, $z\sim2$ for all but the youngest (Reddy et
al.\ 2006; Siana et al.\ 2008; Siana et al.\ 2009) and most obscured
starburst galaxies (e.g., Meurer et al.\ 1995; Meurer et al.\ 1999;
Burgarella et al.\ 2005; Laird et al.\ 2005; Reddy et al.\ 2006), and
therefore it is reasonable for us to use this equation to estimate
extinction in each of our high-redshift dropout samples.  This
technique has already been employed in several previous studies
estimating the SFR density at $z$$\,\sim\,$2-6 (e.g., Adelberger \&
Steidel 2000; Meurer et al.\ 1999; Bouwens et al.\ 2006; Stark et
al.\ 2007).

Of course, there are many reasons for suspecting that Eq~\ref{eq:mhc}
may not work very well at early times, both because of the very
different ages and metallicities of stellar populations and because
the dust itself may have very different characteristics.  After all,
the standard mechanism for forming dust in the envelopes of AGB stars
should not work at $z\gtrsim5$ because the universe is not old enough
at these redshifts to have produced AGB stars.  Dust in $z\gtrsim5$
galaxies has therefore been suggested to form by another mechanism,
most frequently in SNe ejecta (e.g., Maiolino et al.\ 2004; Maiolino
2006).  Eq.~\ref{eq:mhc} is also known to fail at $z\sim3$ for a few
lensed sources (e.g., MS1512-cB58 [Siana et al.\ 2008] or the Cosmic
Eye [Siana et al.\ 2009]) and for very young ($<$100 Myr) star-forming
galaxies at $z\sim2$ (Reddy et al.\ 2006).

To use Eq.~\ref{eq:mhc}, we need to convert the $UV$-continuum slopes
$\beta$ we have derived (which have a baseline 1600\AA$\,\,$to
2300\AA) to that appropriate for Eq.~\ref{eq:mhc} (where $\beta$
assumes a slightly more extended baseline 1300\AA$\,\,$to 2600\AA).
We therefore explored the value of the $UV$-continuum slope over these
two different baselines for a wide variety of star formation histories
and values of the dust extinction.  We found that they are very
similar in general, with an absolute difference generally
$\Delta\beta$$<$0.2.  While there is clearly latitude in the precise
relationship we use to convert the measured $UV$-continuum slopes
$\beta$ to this new wavelength, perhaps the least arbitrary set of
SEDs to use to perform this conversion are those we use in Appendix A
to estimate $\beta$ from the observed colors.  For those SEDs (which
assume a $e^{-t/\tau}$ star formation history, $t=10$ Myr, $\tau=70$
Myr, $[Z/Z_{\odot}]=-0.7$, Calzetti et al.\ 2000 extinction law, and a
Salpeter IMF), we derive the following relationship $\beta' = -2.31 +
1.11 (\beta + 2.3)$.  While obviously there are some uncertainties in
using this conversion, there are at least as many uncertainties in
using Eq.~\ref{eq:mhc} to estimate the dust obscuration in high
redshift galaxies, and so we will ignore these uncertainties in the
subsequent discussion.

\begin{deluxetable}{cccc}
\tablecaption{The effective dust extinction (at $\sim$1600\AA)
  estimated for the LBG population integrated down to various $UV$
  luminosities (see also Figure~\ref{fig:dustz}).\label{tab:effdust}} \tablehead{\colhead{} &
  \multicolumn{3}{c}{Effective Extinction} \\ \colhead{Sample} &
  \colhead{$>0.3L_{z=3}^{*}$} & \colhead{$>0.04L_{z=3}^{*}$} &
  \colhead{$>0$}} \startdata \multicolumn{4}{c}{Using Meurer et
  al.\ (1999) Relationship\tablenotemark{a,b}} \\ 
$U$-dropouts & 6.0$_{-1.4}^{+1.8}$$_{-1.6}^{+2.1}$\tablenotemark{c} & 3.8$_{-0.6}^{+0.8}$$_{-1.0}^{+1.4}$ & 2.8$_{-0.4}^{+0.5}$$_{-0.7}^{+1.0}$ \\
$B$-dropouts & 5.8$_{-0.7}^{+0.8}$$_{-1.5}^{+2.1}$ & 3.8$_{-0.5}^{+0.5}$$_{-1.0}^{+1.3}$ & 2.8$_{-0.3}^{+0.3}$$_{-0.7}^{+1.0}$ \\
$V$-dropouts & 2.7$_{-0.5}^{+0.7}$$_{-0.7}^{+1.0}$ & 1.6$_{-0.2}^{+0.3}$$_{-0.4}^{+0.6}$ & 1.4$_{-0.2}^{+0.2}$$_{-0.4}^{+0.5}$ \\
$i$-dropouts & 1.6$_{-0.3}^{+0.4}$$_{-0.4}^{+0.6}$ & 1.3$_{-0.3}^{+0.3}$$_{-0.4}^{+0.5}$ & 1.2$_{-0.2}^{+0.2}$$_{-0.3}^{+0.4}$ \\
\enddata
\tablenotetext{a}{The effective dust extinctions given here are the
  multiplicative factors needed to correct the observed UV luminosity
  densities at $\sim$1600\AA$\,\,$to their intrinsic values, after
  integrating to specific limiting luminosities (with the limits
  specified in the columns).  These extinctions are estimated based
  upon the distribution of $UV$-continuum slopes observed
  (Table~\ref{tab:uvslope}) and the correlation between $UV$-continuum
  slope and dust obscuration observed at $z\sim0-2$ (Eq.~\ref{eq:mhc}:
  Meurer et al.\ 1999).  Notice that the dust extinctions are much lower
  when integrated to very low luminosities.}
  \tablenotetext{b}{Both random and systematic errors are quoted
  (presented first and second, respectively).}
  \tablenotetext{c}{This estimate is in good agreement with the
    estimates at $z\sim2$ by Reddy et al.\ (2006) and Erb et
    al.\ (2006b) from various multiwavelength data (\S5.4).}
\end{deluxetable}

\subsection{Dependence of Dust extinction on $UV$ Luminosity}

We will first discuss the dependence of the $UV$-continuum slope and
inferred dust extinction on the \textit{$UV$ luminosity}.  We find
that the $UV$-continuum slope $\beta$ shows a strong dependence on the
$UV$ luminosity at both $z\sim2.5$ and $z\sim4$.  This suggests that
dust extinction is positively correlated with the observed $UV$
luminosity, in the sense that more luminous galaxies are more dust
obscured.

At face value, this trend would appear to be contrary to what
Adelberger \& Steidel (2000) and Reddy et al.\ (2008) found in their
examination of $z\sim3$ $U$-dropouts -- where no correlation was found
between the $UV$-continuum slope $\beta$ and the observed $UV$
luminosity over the range ${\cal R}=22$ ($M_{UV,AB}\sim-23.5$) to
${\cal R}=25.5$ ($M_{UV,AB}\sim-20$).  This is a somewhat brighter
range than we probe with good statistics, suggesting that dust
extinction (and the $UV$-continuum slope $\beta$) may be a weaker
function of the observed $UV$ magnitude at $M_{UV,AB}\lesssim-20$
(${\cal R}\lesssim25.5$) than it is over the whole range probed here.
Indeed, using this same magnitude baseline ($V_{606,AB}<25.5$) and a t
test, we only find the correlation between $\beta$ and $UV$ luminosity
to be significant at 75\% confidence from our own $z\sim2.5$
$U$-dropout sample.  Thus, while the correlation would appear to be
weaker for the more luminous sources, the existence of a correlation
between $\beta$ and observed $UV$ magnitude over the entire magnitude
range probed here is very significant (i.e., $>$4$\sigma$ result).

\begin{figure}
\epsscale{1.13}
\plotone{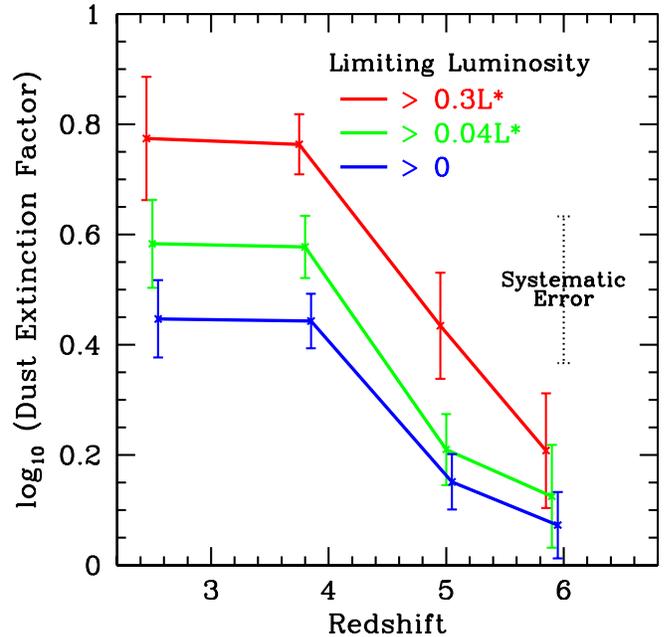}
\caption{The effective dust extinction (at $\sim$1600\AA) estimated
  for LBGs vs. redshift integrated down to a luminosity of
  $0.3L_{z=3}^{*}$ (red lines), $0.04L_{z=3}^{*}$ (green lines), and 0
  (blue lines).  The precise redshifts at which our dust extinction
  estimates are presented are shifted slightly to minimize overlap
  between the different limiting luminosities.  The error bars shown
  here are for the random errors.  The dotted error bar included on
  the right here is the systematic error that likely applies to each
  dust extinction estimate (based upon our conservative estimate that
  all $\beta$ estimates in this paper are uncertain at the $\Delta
  \beta \sim \pm 0.15$ level).  This systematic error is unlikely to
  have any effect on the trends and differences seen here, but could
  change the overall scaling.  The dust extinctions presented here are
  also given in Table~\ref{tab:effdust}.  These extinctions are
  estimated based upon the distribution of $UV$-continuum slopes
  observed (Table~\ref{tab:uvslope}) and the correlation between
  $UV$-continuum slope and dust obscuration observed at $z\sim0-2$
  (Eq.~\ref{eq:mhc}: Meurer et al.\ 1999).  Notice that the dust
  extinctions are much lower when integrated to very low
  luminosities.\label{fig:dustz}}
\end{figure}

In order to understand the differences between our results and those
of Adelberger \& Steidel (2000), we must appreciate that galaxies
exhibit very different behaviors as a function of $UV$ luminosity,
depending upon whether we are dealing with galaxies at higher or lower
luminosities.  For the lower luminosity regime, a large fraction of
the sources should have very low star formation rates.  In light of
the well-known correlation between \textit{the star formation rate}
and dust extinction (e.g., Wang \& Heckman 1996; Hopkins et al.\ 2001;
Martin et al.\ 2005; Reddy et al.\ 2006; Zheng et al.\ 2007), these
galaxies would have low dust extinction (less than $\lesssim0.2$ mag),
and therefore the $UV$ luminosity should be approximately proportional
to the star formation rate.  Consequently, we might expect the dust
extinction (and $UV$-continuum slope) to be approximately proportional
to $UV$ luminosity at low luminosities.  This is what we observe.

In the higher luminosity regime, we might expect the SFR to be much
larger in general.  We would also expect the dust extinction in these
galaxies to also be larger, given the correlation of dust extinction
with the SFR.  As such, any increase in the SFR of a galaxy would be
largely offset by a similar increase in the dust extinction --
resulting in a very weak correlation between the $UV$ light output
from a galaxy and its SFR (or dust extinction).  This is exactly what
Adelberger \& Steidel (2000) and Reddy et al.\ (2008) found in their
sample of luminous ($>0.3L_{z=3}^{*}$) $z\sim3$ $U$-dropout galaxies.
Of course, we would also expect galaxies with very large extinctions
to be present in galaxy samples at lower $UV$ luminosities, albeit as
a small fraction.  We can infer this latter fact from the volume
density of such systems: $>10^{12}$ $L_{\odot}$ IR ultra-luminous
galaxies (e.g., from the $z\sim2$ IR LF of Caputi et al.\ 2007) have a
much smaller volume density than $UV$ faint sources (by factors of
$>$50: see discussion in \S8 of Reddy \& Steidel 2009).  Consequently,
we would expect the contribution from such $IR$ ultra luminous
galaxies to the overall SFR density in $UV$ faint samples to be
relatively small.

\subsection{Dependence of Dust Extinction on Redshift}

Very high-redshift ($z\gtrsim5$) galaxies are also much bluer in
general than $z$$\,\sim\,$2-4 galaxies of the same luminosity.  This
suggests $z\gtrsim5$ galaxies are less dust obscured as well.  This
smaller dust obscuration is probably a consequence of an evolution in
the relationship between dust extinction and the star formation rate,
such that for a given star formation rate the observed dust
obscuration increases as a function of cosmic time.  This would cause
higher redshift galaxies of a given $UV$ luminosity to have bluer
$UV$-continuum slopes.  Such an evolution in extinction-SFR
relationship is observed from $z\sim2$ to $z\sim0$ (Reddy et
al.\ 2006; Buat et al.\ 2007; Burgarella et al.\ 2007), and it is
expected that this relationship continues to evolve from $z\sim6$.
This is presumably a result of the gradual build-up in metals in the
universe with cosmic time.

\subsection{Average Dust Extinction for LBGs to Specific Luminosity Limits}

It is interesting to estimate what the observed distribution of
$UV$-continuum slopes $\beta$ would imply for the dust extinction, if
Eq.~\ref{eq:mhc} holds.  For this calculation, we assume that we can
represent the distribution of $UV$-continuum slopes $\beta$ as a
simple gaussian with mean and $1\sigma$ scatter as derived earlier in
our paper (Table~\ref{tab:uvslope}: see \S3.7).  We then fold this
distribution through Eq.~\ref{eq:mhc} to calculate the effective dust
extinction integrated down to various limiting luminosities.  We adopt
the $UV$ luminosity function of Bouwens et al.\ (2007) at $z\sim4$,
$z\sim5$, and $z\sim6$ and that of Reddy \& Steidel (2009) at $z\sim3$
for this calculation.

The effective dust extinction is presented in Table~\ref{tab:effdust}
and Figure~\ref{fig:dustz}.  To a limiting luminosity of 0.3
$L_{z=3}^{*}$, we infer a dust extinction of 6.0$\pm$2.5$\times$ (or
1.6 mag) at $z\sim2.5$, very close to the 5.4$\pm$0.9$\times$ dust
attention Meurer et al.\ (1999) estimated previously from the $UV$
continuum slopes of HDF-North $U$-dropouts.  It is encouraging that
this extinction estimate is very similar to the values estimated at
$z\sim2$ using a wide variety of multiwavelength data and SFR
calibrators (e.g., Reddy \& Steidel 2004; Reddy et al.\ 2006; Erb et
al.\ 2006b).  However, these extinction estimates are a few times
higher than those inferred by Carilli et al.\ (2008) at $z\sim3$ by
stacking the Very Large Array (VLA) radio observations of
$>0.2L_{z=3}^{*}$ $U$-dropouts in the COSMOS field (where a factor of
1.8$\pm$0.4 extinction was inferred).  This may indicate that the
Meurer et al.\ (1999) relationship substantially overestimates the
dust extinction at $z>2$, or it may indicate that the radio luminosity
for a given star formation rate is systematically lower at high
redshift (Carilli et al.\ 2008).  At this time, it is not obvious how
this issue will be resolved, though it seems clear that dust
extinction will be less of a concern as we reach back further in
cosmic time (e.g., see \S6.2).

\begin{figure}
\epsscale{1.13}
\plotone{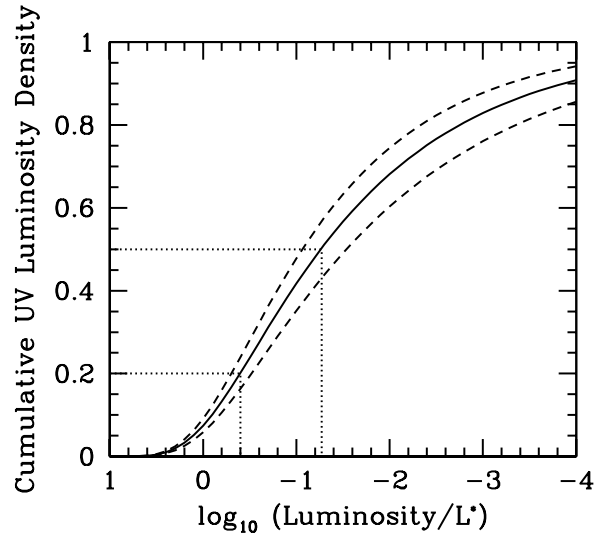}
\caption{Fraction of the total $UV$ luminosity density that arises
  from galaxies brighter than a given luminosity (solid line).  This
  calculation assumes the faint-end slope $\alpha$ of $-1.73$
  preferred in recent determinations of the LF at $z$$\,\sim\,$2-6
  (e.g., Yoshida et al.\ 2006; Bouwens et al.\ 2007; Oesch et
  al.\ 2007; Reddy \& Steidel 2009).  The dashed lines show the
  fraction of the luminosity density brighter than specific
  luminosities for faint-end slopes $\alpha$ of $-1.68$ and $-1.78$
  (which differ at the $1\sigma$ level from the $\alpha=-1.73$ value
  preferred at $z\sim4$: Bouwens et al.\ 2007).  Note that $>50$\% and
  $>80$\% of the luminosity density occurs in galaxies fainter than
  $10^{-1.3} L^{*} \sim 0.05 L^{*}$ and $10^{-0.4} L^{*}\sim0.4L^*$,
  respectively, demonstrating how important it is to understand the
  dust correction down to very low luminosities.\label{fig:relfrac}}
\end{figure}

At higher redshifts and integrated down to lower luminosities, the
effective dust extinction is much less, i.e., factors of $\lesssim$2-3
($\lesssim1$ mag).  Qualitatively, we would expect such a change due
to the much bluer UV continuum slopes $\beta$ found at these
luminosities and redshifts.  Nonetheless it is still quite striking
how much smaller the estimated dust extinction is when including light
from the lowest luminosity galaxies.  In detail, this is because lower
luminosity galaxies dominate the total luminosity density in the $UV$
continuum (see Figure~\ref{fig:relfrac}) and the average dust
extinction in these lower luminosity galaxies is quite low.  Bouwens
et al.\ (2007) estimated that half of the light in the $UV$-continuum
originated in galaxies fainter than 0.06 $L_{z=3}^*$ ($\sim$$-18$ AB
mag: see Figure~\ref{fig:relfrac}).  At $\sim-18$ AB mag, we find that
galaxies have an average $UV$-continuum slope $\beta$ of $\sim-2$ to
$-2.5$.

\begin{figure*}
\epsscale{1.13}
\plotone{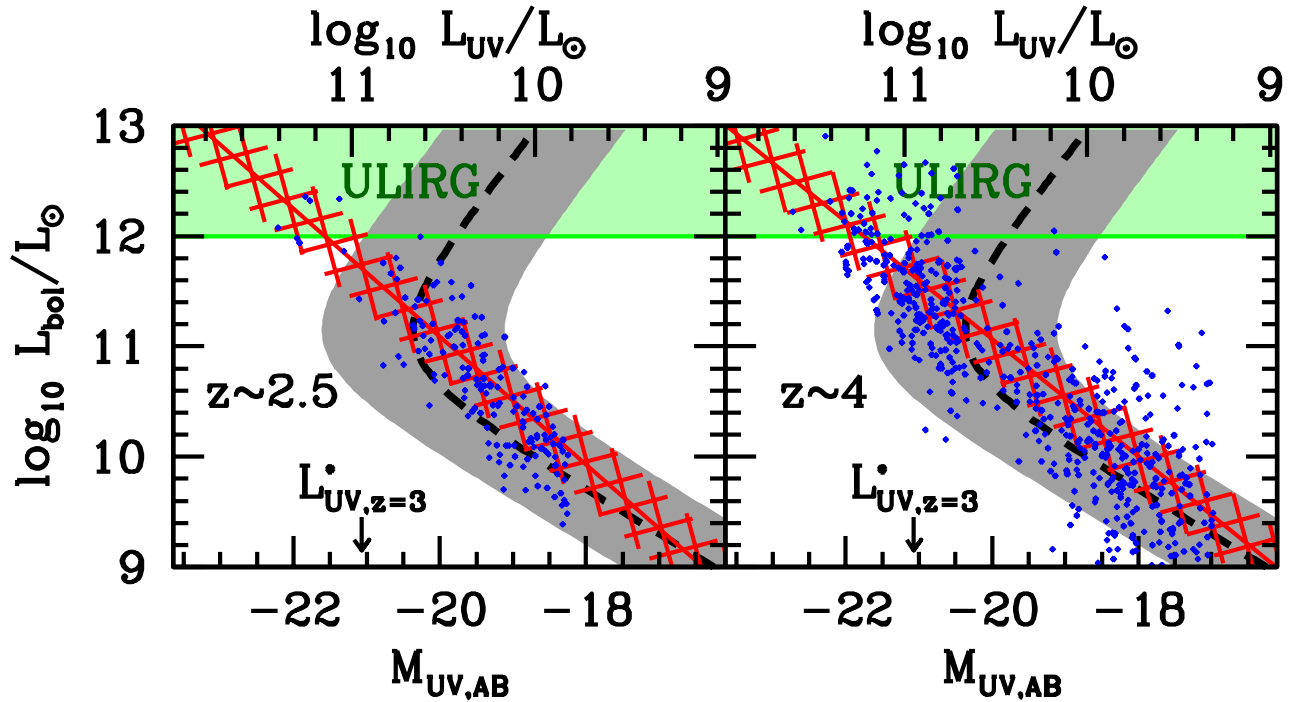}
\caption{Bolometric luminosity versus observed $UV$ luminosity
  $M_{UV,AB}$.  The top axis gives this $UV$ luminosity in units of
  $L_{\odot}$.  The blue points give the $UV$ luminosities and
  equivalent bolometric luminosities (after dust correction) for
  individual sources in our $z\sim2.5$ $U$-dropout and $z\sim4$
  $B$-dropout samples (see Figure~\ref{fig:colmag}).  The dust
  correction utilizes the observed $UV$-continuum slopes $\beta$ and
  the Meurer et al.\ (1999) IRX-$\beta$ relationship.  Note that the
  red line is not a direct fit to the blue points (see \S3.7 and
  Figure~\ref{fig:colmag}).  The red line shows the expected relation
  between bolometric luminosity and $UV$ luminosity using the
  luminosity-dependent dust correction derived in \S5.5 (see
  Eq.~\ref{eq:lbol}).  These luminosity-dependent dust corrections are
  derived utilizing the correlation between $UV$-continuum slope
  $\beta$ and $M_{UV,AB}$ we found for $z\sim2.5$ $U$-dropouts and
  $z\sim4$ $B$-dropouts (see \S3.9) and the Meurer et al.\ (1999)
  relationship.  The hatched red region indicates the expected
  $1\sigma$ scatter in the bolometric luminosities for a given $UV$
  luminosity (based upon the observed scatter of $\sim$0.4 in the
  $UV$-continuum slope at a fixed $UV$ luminosity).  Since it is
  unclear whether the Meurer et al.\ (1999) IRX-$\beta$ relationship
  can be accurately used to estimate extinctions for galaxies with
  $>10^{12}$ $L_{\odot}$ luminosities (e.g., Chapman et al.\ 2005;
  Reddy et al.\ 2006), caution should be used in interpreting the blue
  points or red lines which extend beyond $10^{12}$ $L_{\odot}$
  (\textit{green shaded region}).  The black dashed line shows the
  relationship between $UV$ luminosity and $\beta$ that one would
  derive utilizing the correlation Reddy et al.\ (2006) found between
  bolometric luminosity and dust obscuration (see \S5.5).  The shaded
  grey region corresponds to a $\sim0.4$ dex scatter in the dust
  obscuration.  Notice how the population of bluer, more dust-free
  dropout galaxies in our selections span a wide range in bolometric
  luminosity, extending all the way from $10^9$ $L_{\odot}$ to
  $10^{12}$ $L_{\odot}$ where the LBG population transitions to a more
  dust-obscured ULIRG-like population.  It is quite interesting that
  the $UV$ luminosity where galaxies begin reaching bolometric
  luminosities of $\sim10^{12}$ $L_{\odot}$ -- where dust obscuration
  becomes very important -- is very close to $L_{z=3}^{*}$ (Steidel et
  al.\ 1999: shown on the figure).  The $UV$ LF therefore begins to
  cut off at precisely the same luminosities as dust obscuration is
  becoming increasingly important in attenuating the light.  This
  suggests that the value of $L^{*}$ at $z\sim2-4$ may be set (in
  part) by the luminosities at which dust obscuration becomes
  dominant.  It may also help to explain the mild evolution in the
  $UV$ LF from $z\sim4$ to $z\sim2$ (see \S5.5).\label{fig:colmagill}}
\end{figure*}

These dust corrections are much more modest than have occasionally
been assumed to calculate the total star formation rate density at
early times, e.g., the factor of $\sim8$ adopted by Giavalisco et
al.\ (2004b) or the factor of $\sim$3 adopted by Hopkins (2004) for
their common obscuration correction.  Previously, Bouwens et
al.\ (2006) and Lehnert et al.  (2005) have argued that the dust
extinction was likely less than a factor of $\lesssim2$ at
$z$$\,\sim\,$5-6 and Reddy \& Steidel (2009) argued that when
integrated to lower luminosities the dust extinction was less than a
factor of $\sim$1.5-2 at $z$$\,\sim\,$2-3 (adopting their Luminosity
Dependent Reddening [LDR] model).  The Reddy \& Steidel (2009) LDR
dust extinctions are somewhat lower than what we estimate here,
largely because they assume there is little dust extinction for
galaxies with luminosities fainter than $\sim-18.5$ (0.1
$L_{z=3}^{*}$).  By contrast, the dust extinction we adopt come from
the distribution of $UV$-continuum slopes we find and using
Eq.~\ref{eq:mhc}.  While it is unclear which set of dust corrections
is most accurate, it seems quite clear that the dust extinction must
be very low for galaxies with lower $UV$ luminosities.  This follows
from two observational findings: (1) the much larger volume density
(i.e., $\sim10^{-2}$ Mpc$^{-3}$) of galaxies with low $UV$
luminosities than ULIRG-type sources (i.e., where $10^{-3.5}$
Mpc$^{-3}$) and (2) the correlation between dust extinction and
bolometric luminosity (e.g., Wang \& Heckman 1996; Hopkins et
al.\ 2001; Martin et al.\ 2005; Reddy et al.\ 2006; Buat et al.\ 2007;
Burgarella et al.\ 2007).  As a result of these two findings, we can
conclude that essentially all of the galaxies with lower $UV$
luminosities ($>$99\%) do not have very high bolometric luminosities,
and therefore do not exhibit large dust extinctions.

\subsection{Bolometric Luminosities of Galaxies in our $UV$ selections}

The viability of the Meurer et al.\ (1999) IRX-$\beta$ relationship
(Eq. 1) we used to correct the observed $UV$ luminosities for the dust
extinction is a strong function of the bolometric luminosity of the
sources.  For example, Reddy et al.\ (2006) found that while the
Meurer et al.\ (1999) relationship worked well between $10^{11}
L_{\odot}$ and $10^{12} L_{\odot}$ at $z\sim2$, it failed badly for
galaxies with luminosities $>10^{12} L_{\odot}$ (see also Chapman et
al.\ 2005).

To investigate the validity of the estimates we made in the previous
section, it is interesting therefore to estimate the bolometric
luminosities for galaxies with specific $UV$ luminosities in our LBG
selections.  The results are shown in Figure~\ref{fig:colmagill}.  We
again take advantage of Eq. (1), as well as the relationships between
the $UV$-continuum slope $\beta$ and $UV$ luminosity (\S3.9) to make
this estimate.  This results in the following relationship between the
absolute magnitude of a galaxy in the $UV$ and the bolometric
luminosity $L_{bol}$ at $z\sim2.5$:
\begin{equation}
L_{bol} = 10^{11.67 - 0.58 (M_{UV,AB}+21)} L_{\odot}\label{eq:lbol}
\end{equation}
The relationship at $z\sim4$ is similar.  We emphasize that this
relationship does not generally hold for galaxies where $L_{bol} >
10^{12} L_{\odot}$ (e.g., Reddy et al.\ 2006; Chapman et al.\ 2005) or
for galaxies which are particularly young (Reddy et al.\ 2006),
because of difficulties in using the $UV$-continuum slope $\beta$ to
correct for dust extinction in these cases.

The above result is shown in Figure~\ref{fig:colmagill} with the red
lines for the $z\sim2.5$ and $z\sim4$ samples.  The hatched region
shows the approximate scatter expected in the bolometric luminosity
for a given $UV$ luminosity as a result of scatter in the
$UV$-continuum slope $\beta$.  Also shown are the bolometric
luminosities for individual dropout galaxies in the two samples, after
correcting their $UV$ luminosities for dust extinction using their
measured $\beta$'s and Eq. (1).

From Figure~\ref{fig:colmagill} and Eq.~\ref{eq:lbol}, it seems clear
that we would expect a substantial fraction of our dropout sample to
have bolometric luminosities less than $10^{12}$ $L_{\odot}$.  As
such, we can plausibly make use of the IRX-$\beta$ relationship to
predict the mean dust extinction in galaxies -- as we have done in the
previous section.  Of course, we might expect a few $>10^{12}$
$L_{\odot}$ dust-obscured galaxies to be present in the normal dropout
samples, with typical $UV$ luminosities and $UV$-continuum slopes
$\beta$ where Eq.~\ref{eq:lbol} may not apply (e.g., Chapman et
al.\ 2005).  However, given that the volume density of such galaxies
is observed to be much lower ($>10$-100$\times$) than that of typical
dropout galaxies, we would expect Eq.~\ref{eq:lbol} to be reasonably
accurate in most cases.

Examining Figure~\ref{fig:colmagill}, it is certainly quite striking
that $z$$\,\sim\,$2-4 LBGs start becoming extraordinarily rare
brightward of $\sim-21$ or $-22$, at precisely the same luminosities
(i.e., $>10^{12}$ $L_{\odot}$) as we expect dust extinction to become
increasingly important in galaxies.  This suggests the abrupt cut-off
in the $UV$ LF at $z$$\,\sim\,$2-4, i.e., $L_{UV}^{*}$, is set by dust
extinction -- or more precisely the luminosity at which dust
obscuration becomes so significant as to offset the increase in energy
output from stars.  We can examine this by using the observed
correlation between dust extinction and bolometric luminosity in the
blue star-forming population at $z\sim2$ (Reddy et al.\ 2006) $\log
L_{bol} = (0.62\pm0.06)\log \frac{L_{IR}}{L_{1600}} + (10.95\pm0.07)$
and applying it to Figure~\ref{fig:colmagill}.  Adopting the observed
scatter of $\Delta \log \frac{F_{IR}}{F_{UV}}\sim0.4$, we show the
envelope of $M_{UV}$ and $L_{bol}$ values expected for a wide range in
$L_{bol}$.  It seems clear from this exercise that we would not expect
galaxies to have $UV$ luminosities much brighter than $-22$ ($\sim$2
$L_{UV}^{*}$) and that dust effectively sets the cut-off in the UV LF.
\textit{This may explain why the value of $L_{UV} ^*$ (e.g., Steidel
  et al.\ 1999; Reddy et al.\ 2008) does not evolve much from $z\sim4$
  to $z\sim2$.}

\subsection{High-Redshift Galaxies from the Literature that would seem to show Substantial Dust Extinction} 

The dust extinctions we infer are substantially less than have been
inferred in some studies of specific high redshift galaxies.  One such
example is the $z=6.56$ galaxy behind Abell 370 (HCM6A: Hu et
al.\ 2002).  This source was found to have a significantly larger flux
in the $4.5\mu$ band than in the $3.6\mu$ band -- which may be due to
substantial $H\alpha$ emission falling within the $4.5\mu$ band (Chary
et al.\ 2005).  If $H\alpha$ is in fact the explanation, the inferred
star formation rate for this source from this $H\alpha$ emission is
$\sim10\times$ higher than would be inferred from the UV or Ly$\alpha$
emission alone.  Detailed stellar population modelling on the optical,
near-IR, and IRAC imaging data of HCM6A suggest an extinction factor
of $\sim11\times$ ($A_{UV}=2.6$ mag) which is much larger than what we
infer on average for $z\sim6$ $i$-dropouts (Chary et al.\ 2005;
Schaerer \& Pello 2005; see Table~\ref{tab:effdust}).  However, in
subsequent and much deeper IRAC observations of HCM6A (K. Lai et
al.\ 2009, in prep), no significant difference between the fluxes of
HCM6A in the 3.6$\mu$ and 4.5$\mu$ bands is found, leaving us with no
reason to believe this source shows significant $H\alpha$ emission and
hence no reason to suspect that its UV light is significantly
obscured.  Direct observations of this source with the MAMBO-2
bolometric array at $1.2mm$ also set useful upper limits on its dust
obscuration (Boone et al.\ 2007).

Other examples might include the many high redshift QSOs discovered by
the Sloan Digital Sky Survey (SDSS: Fan et al.\ 2006).  The inferred
dust masses of SDSS QSOs can be quite substantial (e.g., J1148+5251:
Bertoldi et al.\ 2003), providing clear evidence that specific high
redshift sources \textit{can} show modest to sizeable dust extinction.
Nonetheless, we would not expect QSOs like J1148+5251 to be
representative of $z\sim6$ galaxies, associated as they are with very
massive halos and thus in very rich environments.  Galaxies in such
environments are known to show much greater evolution than galaxies in
more typical environments (e.g., Tanaka et al.\ 2005) and show much
greater metal enrichment.

\begin{deluxetable}{lccc}
\tablewidth{0pt}
\tabletypesize{\footnotesize}
\tablecaption{Inferred Star Formation Rate Densities from $z\sim2-6$
  LBG samples.\tablenotemark{a,b}\label{tab:sfrdens}}
\tablehead{
\colhead{Dropout} & \colhead{} & \multicolumn{2}{c}{$\textrm{log}_{10}$ SFR density ($M_{\odot}$ Mpc$^{-3}$ yr$^{-1}$)} \\
\colhead{Sample} & \colhead{$<z>$} & \colhead{$L>0.3 L_{z=3}^{*}$} & 
\colhead{$L> 0.04 L_{z=3}^{*}$}}
\startdata
& & \multicolumn{2}{c}{Uncorrected} \\
$U$ & 2.5\tablenotemark{c} & $-$1.70$\pm$0.03 & $-$1.36$\pm$0.03 \\
$B$ & 3.8 & $-$1.81$\pm$0.05 & $-$1.48$\pm$0.05 \\
$V$ & 5.0 & $-$2.15$\pm$0.06 & $-$1.78$\pm$0.06 \\
$i$ & 5.9 & $-$2.31$\pm$0.08 & $-$1.83$\pm$0.08 \\
$z$ & 7.4 & $-$2.58$\pm$0.21 & --- \\
& & \multicolumn{2}{c}{Dust-Corrected} \\
$U$ & 2.5\tablenotemark{c} & $-$0.93$\pm$0.03 & $-$0.78$\pm$0.03 \\
$B$ & 3.8 & $-$1.05$\pm$0.05 & $-$0.90$\pm$0.05 \\
$V$ & 5.0 & $-$1.72$\pm$0.06 & $-$1.57$\pm$0.06 \\
$i$ & 5.9 & $-$2.10$\pm$0.08 & $-$1.70$\pm$0.08 \\
$z$ & 7.4 & $-$2.37$\pm$0.21 & --- \\
\enddata 
\tablenotetext{a}{Based upon LF parameters in Table 2 of Reddy \& Steidel (2009), Table 7 of Bouwens et al.\ (2007:see \S3.5), and Table
  4 of Bouwens et al.\ (2008).}
\tablenotetext{b}{The SFR density here tabulated in terms of the
  Salpeter IMF.  Expressing these results in terms of a Kroupa (2001)
  IMF, one should divide the results given here by a factor of
  $\sim$1.7.}
\tablenotetext{c}{We adopt the Reddy \& Steidel (2009) UV LF at
  $z=3$ for computing the SFR density at $z\sim2.5$.}
\end{deluxetable}

\begin{figure}
\epsscale{1.13}
\plotone{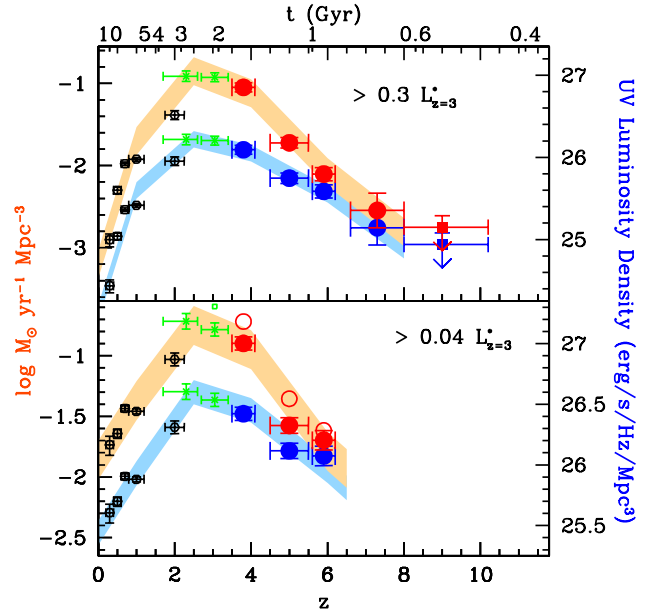}
\caption{The star formation history inferred using the dust
  corrections determined here (upper set of points: orange shaded
  region) and without those dust corrections (lower set of points:
  blue shaded region).  These dust corrections
  (Table~\ref{tab:effdust}) are applied to the luminosity density
  determinations at $z>2$ by Reddy \& Steidel (2009), Bouwens et
  al.\ (2007), and Bouwens et al.\ (2008).  Otherwise, the dust
  corrections of Schiminovich et al.\ (2005) are assumed.  The top
  panel is based upon the total $UV$ luminosity density integrated to
  0.3 $L_{z=3}^{*}$ and the bottom panel is based upon the total $UV$
  luminosity density integrated to $0.04L_{z=3}^{*}$.  The rest-frame
  UV continuum luminosity density is converted to a star formation
  rate density assuming a constant $>10^8$ yr star formation model and
  a Salpeter (1955) IMF (Madau et al.\ 1998).  Conversion to a Kroupa
  (2001) IMF would result in a factor of $\sim$1.7 (0.23 dex) decrease
  in the SFR density estimates given here.  Assuming younger ages for
  star-forming galaxies in high-redshift samples would increase the
  SFR density estimates by a similar factor (Verma et al.\ 2007).
  Shown are the luminosity density determinations by Schiminovich et
  al.\ (2005: \textit{black hexagons}), Reddy \& Steidel (2009:
  \textit{green crosses}), Bouwens et al.\ (2007: \textit{blue/red
    circles}), and Bouwens et al.\ (2008: \textit{blue/red circles and
    $1\sigma$ upper limit}).  The open symbols show the star formation
  rate densities that would be inferred if we applied the same dust
  correction to $>0.04L_{z=3}^{*}$ as we apply to $>0.3L_{z=3}^{*}$.
  It is clear then that ignoring the fact that lower luminosity
  galaxies have much lower values for the dust extinction will result
  in an overestimate of the SFR density.  See also
  Table~\ref{tab:effdust}.  Table 5 and Figure 10 from Reddy \&
  Steidel al.\ (2009) make a similar point.  The equivalent star
  formation rate density (UV luminosity density) is also shown
  assuming no extinction correction (lower set of
  points).\label{fig:sfrdens}}
\end{figure}

\section{Possible Implications for the Star Formation History}

In \S5, we used the observed distribution of $UV$-continuum slopes
$\beta$ to estimate the dust extinction in star-forming galaxies at
$z\sim2-6$.  In the current section, we apply these dust extinctions
to $UV$ LFs derived at $z\sim2-6$ to estimate the SFR density from
typical star-forming galaxies (\S6.1).  In \S6.2, we use recent search
results for ULIRGs to estimate their likely contribuion to SFR density
(since they will not be accurately accounted for by the dust-corrected
LBG results).  In \S6.3, we use these two results to estimate the
bolometric flux from SFR to be output in the UV and IR.  Finally, in
\S6.4, we put together these results to estimate the SFR density from
all star-forming galaxies at $z\sim2-6$.

\subsection{The Inferred Star Formation Rate Density at $z$$\,\sim\,$2-6}

It makes sense for us to use the effective dust extinction inferred in
the previous section (Table~\ref{tab:effdust}) to revise our estimates
the star formation rate density at $z$$\,\sim\,$2-6.  As per our calculation
above, we adopt the $UV$ LFs derived by Reddy \& Steidel (2009) at
$z\sim3$ and Bouwens et al.\ (2007) in calculating the SFR densities.
In order to convert luminosities in the $UV$ into estimates of the SFR
we use the transformation presented by Madau et al.\ (1998: see also
Kennicutt 1998):
\begin{equation}
L_{UV} = \left( \frac{\textrm{SFR}}{M_{\odot} \textrm{yr}^{-1}} \right) 8.0 \times 10^{27} \textrm{ergs}\, \textrm{s}^{-1}\, \textrm{Hz}^{-1}\label{eq:mad}
\end{equation}
where a $0.1$-$125\,M_{\odot}$ Salpeter IMF and a constant star formation rate
of $\gtrsim100$ Myr are assumed.  Conversion to a Kroupa (2001) IMF
would result in factor of $\sim$1.7 smaller SFR estimates.

In view of the young ages ($\sim10$-50 Myr) inferred for many
star-forming galaxies at $z$$\,\sim\,$5-6 (e.g., Yan et al.\ 2006;
Eyles et al.\ 2007; Verma et al.\ 2007), it is clear that adopting a
simple formula like Eq.~\ref{eq:mad} (which require the assumption of
$\gtrsim100$ Myr ages) would result in a systematic underestimate of
the SFR density of the universe at very early times (e.g., Verma et
al.\ 2007; Reddy \& Steidel 2009; Stanway et al.\ 2005).  In any case,
given our uncertain knowledge of how the age distribution of
star-forming galaxies varies as a function of redshift and luminosity,
it is difficult to know precisely how much we must revise our
estimates upward.  Working with a sample of bright $z\sim5$ galaxies,
Verma et al.\ (2007) estimate that the actual SFRs may be twice as
large as what Eq.~\ref{eq:mad} would suggest.  Of course, this effect
may be partially offset by the fact that for these same young
star-forming systems that standard dust corrections (e.g., following
Meurer et al.\ 1999) seem to overpredict the SFRs by factors of
$\gtrsim$2 (Reddy et al.\ 2006; Siana et al.\ 2008).  In any case,
going forward, we will need to obtain tighter constraints on the age
distribution of galaxies as a function of redshift and especially to
very low luminosities (since faint galaxies provide the dominant
contribution to the SFR density).

Our results are presented in Figure~\ref{fig:sfrdens} and
Table~\ref{tab:sfrdens}.  These dust-corrected SFR densities are
somewhat lower than presented by Bouwens et al.\ (2008).  This is
because the average dust extinction including the lowest luminosity
galaxies is less than when only considering $L^*$ galaxies.  We will
discuss these results in more depth in \S6.4 after including the
contribution from ULIRGs.

\subsection{Contribution of ULIRGs to the Star Formation Rate Density at $z$$\,\sim\,$2-6}

Of course, we would expect the extinction estimates above to
underestimate the effective extinction, given that Eq.~\ref{eq:mhc}
does not accurately describe galaxies with very high SFRs, dust
extinctions, and bolometric luminosities (e.g., Elbaz et al.\ 2007),
particularly those systems where $L_{bol}>10^{12} L_{\odot}$.  The SFR
from such galaxies can be better accounted for by a direct
consideration of IR LFs (e.g., Caputi et al.\ 2007).\footnote{While
  there has been some suggestion that galaxies whose IR luminosities
  exceed that predicted by the IRX-$\beta$ relationship fail due to a
  contribution from an obscured AGN (e.g., Daddi et al.\ 2007) and
  therefore the SFR derived from the IRX-$\beta$ is correct after all,
  it has been found (e.g., Murphy et al.\ 2009) that this is only
  partially the explanation, and that part of the IR excess would
  appear to be due to star formation.  We therefore consider it safer
  to include the contribution from $>10^{12} L_{\odot}$ IR-luminous
  galaxies explicitly through the use of IR surveys.}  Reddy \&
Steidel (2009) have already performed such a calculation at $z\sim2$,
integrating the $z\sim2$ IR LF of Caputi et al.\ (2007) down to
bolometric luminosities of $10^{12} L_{\odot}$.\footnote{There has
  been sizeable variation in the literature in the relationships used
  to convert from $24\mu$ fluxes to the bolometric flux in the IR
  (e.g., Bavouzet et al.\ 2008 vs.  that implied by the Chary \& Elbaz
  2001 templates found to work in the local universe), and therefore
  large differences in the reported LFs in the IR (e.g.,
  P{\'e}rez-Gonz{\'a}lez et al.\ 2005 vs. Caputi et al.\ 2007).
  Fortunately, the situation seems to have become more clear from the
  observations, and Caputi et al.\ (2007) adopt an apparently more
  justified linear relationship (Bavouzet et al.\ 2008) to convert
  between the 24$\mu$m flux and the bolometric IR flux.}  Reddy \&
Steidel (2009) infer a SFR density of 0.03 $M_{\odot} \textrm{yr}^{-1}
\textrm{Mpc}^{-3}$ (Kroupa IMF) at $z\sim2$ from this population.
Obviously, there are some modest uncertainties as to the exact
bolometric luminosity down to which one should integrate the Caputi et
al.\ (2007) LF (to correct for the ULIRG population not reasonably
accounted for by the Meurer et al.\ 1999 relation), and so this SFR
density could be larger.  For example, assuming we integrated the
Caputi et al.\ 2007 LF down to $10^{11.8}$ $L_{bol}$ instead, the
ULIRG contribution could increase by $\sim$50\%.

Despite the existence of these possible uncertainties, we adopt as our
baseline values at $z\sim2.5$ the contribution Reddy \& Steidel (2009)
derived from the $z\sim2$ Caputi et a.\ (2007) IR LF.  We therefore
correct our SFR density estimates at $z\sim2.5$ in
Table~\ref{tab:sfrdens} for this population.  The results are given in
the lower section of Table~\ref{tab:sfrdensulirg}.  The obscured
population only has a modest effect on the dust extinction (and SFR
density), increasing it by $\sim$20\% (see also discussion in Reddy \&
Steidel 2009).

\begin{figure}
\epsscale{1.13}
\plotone{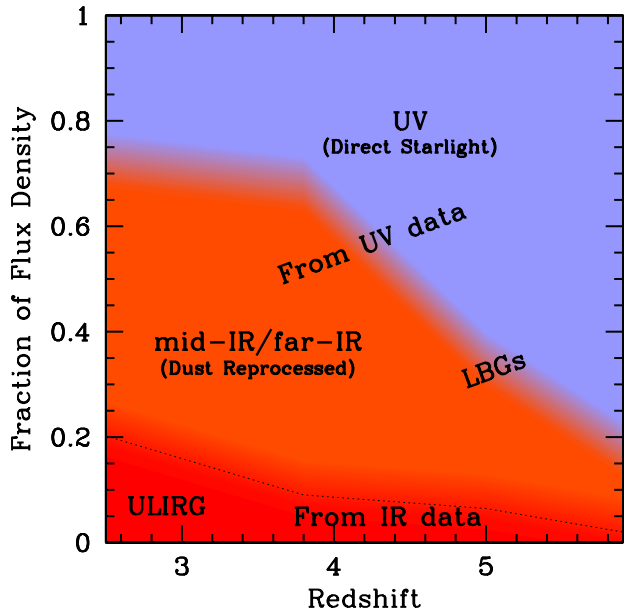}
\caption{Fraction of the total luminosity density from stars at
  $z$$\,\sim\,$2-6 emitted by galaxies in the rest-frame $UV$ and in
  the mid-IR/far-IR (after dust reprocessing).  In calculating the
  energy output in the UV vs. IR, we adopt the extinction estimates in
  Table~\ref{tab:effdust} and assume that the total energy output from
  ULIRGs occurs in the infrared (see \S6.3).  A dotted line is
  included to show the relative contribution of LBGs and ULIRGs to the
  mid-IR/far-IR energy output.  The contribution of star-forming
  galaxies is included all the way down to zero luminosity.  Most of
  the energy output at $z$$\,\sim\,$2-4 would appear to occur in the
  IR, while at $z\gtrsim5$ the energy output occurs mostly in the
  $UV$.  Note that despite most of the energy output at $z\sim2.5-4$
  occurring at IR wavelengths, the SFR densities we derive from LBG
  surveys (after dust corrections) should account for $\gtrsim80$\% of
  the SFR density at $z\gtrsim3$ (see
  Figure~\ref{fig:wrelfrac}).\label{fig:irfrac}}
\end{figure}

Accounting for the effect of such ULIRGs on the SFR density (and
extinction estimates) at $z\gtrsim4$ is a much less certain
enterprise.  Not only is it difficult to probe to even $10^{12}$
$L_{\odot}$ at $z\sim4$ with current instrumentation, but conducting
follow-up observations on specific $z\gtrsim4$ sources (e.g., to
determine redshifts) is also quite challenging at present.  As a
result, only a handful of high quality $z\gtrsim4$ ULIRG candidates
(e.g., Capak et al.\ 2008; Daddi et al.\ 2009; Wang et al.\ 2009;
Schinnerer et al.\ 2008; Dunlop et al.\ 2004) currently exist, and
essentially all of them are at $z\sim4$.  Perhaps the deepest, most
well-defined selection of such sources is found in the deep
$\sim$50-100 arcmin$^2$ submm SCUBA field lying within the HDF-North
GOODS area (Pope et al.\ 2006).  We will use those $z\sim4$ sources to
set a lower limit on the contribution of highly obscured galaxies to
the SFR density at $z\sim4$.  Starting with the $\sim$5-8 ULIRGs
within this field that are plausibly at $z\sim4$ (e.g., Daddi et
al.\ 2009; Mancini et al.\ 2009) and then dividing by the comoving
volume at $z\sim4$, one derives a SFR density of $\sim$0.01-0.02
$M_{\odot} \textrm{yr}^{-1} \textrm{Mp}^{-3}$ (depending upon what
fraction of the candidates are actually at $z\sim4$).  Again, we can
correct our $z\sim4$ SFR density estimates for this population.  Not
surprisingly, this population only contributes $\sim10$\% to the SFR
density (see Figure~\ref{fig:wrelfrac}).

Of course, we might expect there to be highly obscured galaxies below
the $10^{12}$ $L_{\odot}$ luminosity limits that also contribute
substantially to the SFR density relevant to these probes, so one
might argue that 10\% is just a lower limit.  However, as we have
already argued, we should be able to account for the star formation of
these lower luminosity systems from LBG selections.  The results of
Reddy et al.\ (2006) suggest that the $UV$-corrected SFRs of galaxies
can be accurately estimated at $z\sim2$ to $10^{12}$ $L_{\odot}$, and
we would expect these corrections to be even easier to perform at
$z\sim4$ than at $z\sim2$ -- given the trend of high redshift galaxies
to show less and less dust extinction at a fixed SFR at increasingly
high redshifts (e.g., Reddy et al.\ 2006; Buat et al.\ 2007;
Burgarella et al.\ 2007).

Given that $>10^{12}$ $L_{\odot}$ ULIRGs only contribute $\sim$20\% of
the SFR density at $z$$\,\sim\,$2-3 and perhaps $\sim$10\% at
$z$$\,\sim\,$4, one might expect the contribution of such a population
at $z$$\,\sim\,$5-6 also to be quite small.  We can obtain a crude
estimate of this fraction, if we assume that IR luminous galaxies have
a bolometric luminosity function similar to that derived at $z\sim$2
(e.g., by Caputi et al.\ 2007 where $L_{bol}^{*}=10^{11.8\pm0.1}$) and
then evolve in a very similar way to the $UV$ LF at $z\geq3$.  We
therefore assume that $L_{bol}^{*}$ scales with redshift in the same
way as $L_{UV}^{*}$, i.e., as $10^{-0.4(0.36(z-3))}\sim(1/1.39)^{z-3}$
to match the observed $M_{UV}^{*} (z) = M_{UV}^{*} (z=4) + 0.36 (z-4)$
evolution in the $UV$ LF from $z\sim4$ to $z\sim7$ (e.g., Bouwens et
al.\ 2007; Bouwens et al.\ 2008).\footnote{Note that here we are
  implicitly assuming that $L_{bol}^{*}$ evolves from $z\sim4$ to
  $z\sim3$ despite the lack of evolution in $L_{UV}^{*}$ over this
  redshift range (e.g., Reddy et al.\ 2008).  Such a brightening in
  $L^*$ would be a natural continuation of the evolution in the $UV$
  LF from $z\sim7$ to $z\sim4$ (e.g., Bouwens et al.\ 2008).  The
  reason we may not observe this brightening in the $UV$ LF (i.e.,
  $L_{UV}^{*}$) beyond $z\sim4$ is because of the increasing
  importance of dust extinction in the most bolometrically luminous
  galaxies (e.g., see \S5.5 where we argue that dust extinction may be
  partially responsible for the cut-off in the $UV$ LF at
  $z\sim2-4$).} We then compare the fractional SFR density in $L_{bol}
> 10^{12} L_{\odot}$ galaxies with that from the star-forming
population as a whole.  We find that the fractional SFR density in
ULIRGs is 6\% at $z\sim5$ and 2\% at $z\sim6$.  For these assumptions,
the fractional SFR density in $L_{bol} > 10^{12} L_{\odot}$ galaxies
is 23\% and 10\% at $z\sim3$ and $z\sim4$, respectively, which would
seem to be a good match to the observations.  We emphasize that this
very low ULIRG estimate at $z\gtrsim$5 is consistent with the lack of
very red $UV$-continuum slopes $\beta$ at $z$$\,\sim\,$5-6
(Figure~\ref{fig:complete}: which is in contrast to the tail towards
redder [i.e., $\sim-1$ to 0] $\beta$'s at $z\sim4$).

\begin{deluxetable}{lccc}
\tablewidth{0pt} \tabletypesize{\footnotesize} \tablecaption{Inferred
  Star Formation Rate Densities, including the contributions from
  highly dust obscured galaxies
  (\S6.2-6.3).\tablenotemark{a,b}\label{tab:sfrdensulirg}} \tablehead{
  \colhead{Dropout} & \colhead{} &
  \multicolumn{2}{c}{$\textrm{log}_{10}$ SFR density ($M_{\odot}$
    Mpc$^{-3}$ yr$^{-1}$)} \\ \colhead{Sample} & \colhead{$<z>$} &
  \colhead{$L>0.3 L_{z=3}^{*}$} & \colhead{$L> 0.04 L_{z=3}^{*}$}}
\startdata & & \multicolumn{2}{c}{Dust-Corrected} \\ 
$U$ & 2.5\tablenotemark{c} & $-$0.93$\pm$0.03 & $-$0.78$\pm$0.03 \\ 
$B$ & 3.8 & $-$1.05$\pm$0.05 & $-$0.90$\pm$0.05 \\ 
$V$ & 5.0 & $-$1.72$\pm$0.06 & $-$1.57$\pm$0.06 \\ 
$i$ & 5.9 & $-$2.10$\pm$0.08 & $-$1.70$\pm$0.08 \\ 
$z$ & 7.4 & $-$2.37$\pm$0.21 & --- \\ 
& & \multicolumn{2}{c}{Dust-Corrected + ULIRG\tablenotemark{d}} \\ 
$U$ & 2.5\tablenotemark{c} & $-$0.77$\pm$0.03 & $-$0.66$\pm$0.03 \\ 
$B$ & 3.8 & $-$0.98$\pm$0.05 & $-$0.85$\pm$0.05 \\ 
$V$ & 5.0 & $-$1.66$\pm$0.06 & $-$1.53$\pm$0.06 \\
$i$ & 5.9 & $-$2.07$\pm$0.08 & $-$1.69$\pm$0.08 \\
$z$ & 7.4 & $-$2.37$\pm$0.21 & --- \\ \enddata
\tablenotetext{a}{Based upon LF parameters in Table 2 of Reddy \&
  Steidel (2009), Table 7 of Bouwens et al.\ (2007:see \S5.6), and
  Table 4 of Bouwens et al.\ (2008).}
\tablenotetext{b}{The SFR
  density here tabulated in terms of the Salpeter IMF.  Expressing
  these results in terms of a Kroupa (2001) IMF, one should divide the
  results given here by a factor of $\sim$1.7.}
\tablenotetext{c}{We
  adopt the Reddy \& Steidel (2009) UV LF at $z=3$ for computing the
  SFR density at $z\sim2.5$.}
\tablenotetext{d}{At $z\sim2.5$, we
  include the SFR density contribution from $>10^{12}$ $L_{\odot}$
  ultraluminous IR galaxies (ULIRG: \S6.2) by using the $z\sim2$ IR LF
  of Caputi et al.\ (2007) as Reddy \& Steidel (2009) do.  At
  $z\sim4$, we estimate their contribution based upon a small sample
  of $z\sim4$ ultra-luminous IR galaxy candidates (see Daddi et
  al.\ 2009) within the HDF-North GOODS SCUBA supermap (Pope et
  al.\ 2006).  Based upon the $z\sim4$ result, we assume that ULIRGs
  provide only a $\sim$5\% contribution to the SFR density at
  $z$$\,\sim\,$5-6.  We believe this assumption is a reasonable one
  given the trends towards much bluer $UV$-continuum slopes at high
  redshifts seen in both LBG and BBG samples (\S4.2) and thus
  apparently smaller fraction of galaxies with substantial dust
  obscuration.}
\end{deluxetable}

\begin{figure}
\epsscale{1.13}
\plotone{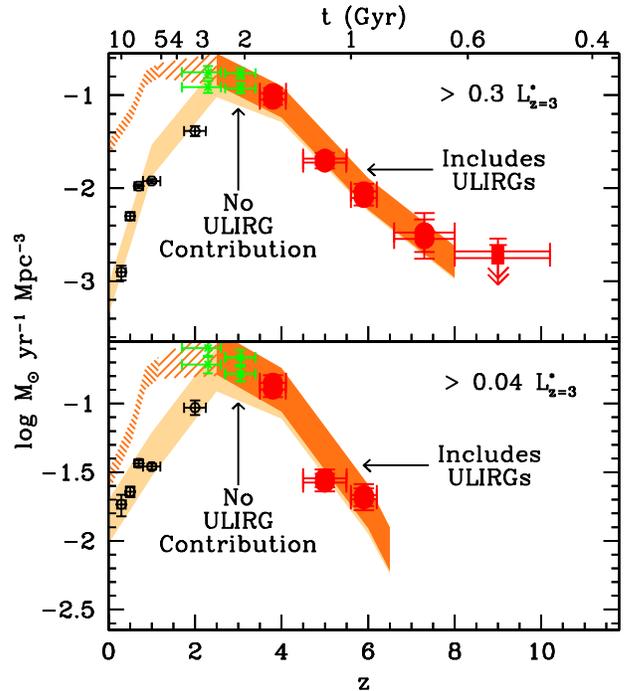}
\caption{Star formation history inferred with (\textit{upper set of
    points, darker orange contours}) and without (\textit{lower set of
    points, lighter orange contours}) a contribution from highly
  obscured ultra-luminous infrared bright ($>10^{12}L_{\odot}$)
  galaxies.  Otherwise similar to Figure~\ref{fig:sfrdens}.  The top
  panel is based upon the total $UV$ luminosity density integrated to
  0.3 $L_{z=3}^{*}$ and the bottom panel is based upon the total $UV$
  luminosity density integrated to $0.04L_{z=3}^{*}$.  At $z\sim2.5$,
  we follow Reddy \& Steidel (2009) in using the $z\sim2$ IR LF of
  Caputi et al.\ (2007) to include the contribution from these ultra
  luminous IR galaxies, while at $z\sim4$, we estimate their
  contribution based upon $\sim$5-8 good $z\sim4$ ULIRG candidates
  (see Daddi et al.\ 2009) within the HDF-North GOODS SCUBA supermap
  (Pope et al.\ 2006).  See also Table~\ref{tab:sfrdensulirg} and
  \S6.2.  The darker orange contours at $z\sim0-1.2$ show the SFR
  density derived using the deepest mid-IR/far-IR observations over
  the GOODS and Far Infared Extragalactic Legacy (FIDEL) fields
  (Magnelli et al.\ 2009).  While IR bright galaxies appear to add
  significantly to the SFR density at low redshift in the upper panel,
  this is because the lower set of points/orange contours only include
  galaxies at the bright end of the $UV$ LF (corresponding to the $0.3
  L_{z=3}^{*}$ limit) and the $UV$ LF there cuts off at much fainter
  magnitudes than at higher redshifts (compare this situation to the
  lower panel where the contribution of the mid-IR/far-IR galaxies is
  much less on a percentage basis).  From this figure, it is clear
  that the contribution of the luminous IR sources to the total star
  formation rate density at $z>2$ is only modest (see also
  Figure~\ref{fig:wrelfrac}).\label{fig:sfrdensulirg}}
\end{figure}

\begin{figure}
\epsscale{1.13}
\plotone{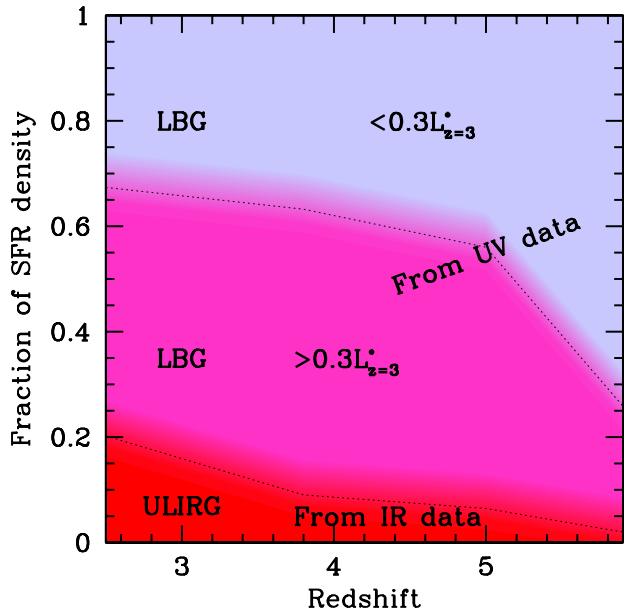}
\caption{Fraction of the total SFR density at high redshift in lower
  luminosity LBGs ($<0.3L_{z=3}^{*}$), higher luminosity LBGs
  ($>0.3L_{z=3}^{*}$), and ultra-luminous ($>10^{12} L_{\odot}$)
  infrared galaxies (ULIRGs).  While there will obviously be some
  overlap between LBG and ULIRG selections, we consider the ULIRG
  population here separately since mid-IR/far-IR data are required to
  fully account for the SFR in these galaxies.  The contribution of
  star-forming LBGs is included all the way down to zero luminosity.
  These fractions are derived based upon the discussion in
  \S5.4,\S6.1, and \S6.2, and estimates in Tables~\ref{tab:sfrdens}
  and \ref{tab:sfrdensulirg} and Figures~\ref{fig:sfrdens} and
  \ref{fig:sfrdensulirg}.  It is clear that virtually all the SFR
  density at $z\gtrsim$3-4 can be accounted for from LBG
  selections.\label{fig:wrelfrac}}
\end{figure}

Recent work on GRBs provide strong support for the general conclusion
that highly obscured ultra-luminous IR galaxies contribute at most a
small fraction of the SFR density at $z\gtrsim4$.  Since GRBs are
perhaps a good tracer of massive star formation in the high redshift
universe (e.g., Y{\"u}ksel et al.\ 2008; Li 2008; Kistler et
al.\ 2009) and may not be biased to lower metallicity environments
(Prochaska et al.\ 2007), we would expect GRBs to occur in galaxies
roughly in proportion to their share of the star formation activity in
the high redshift universe.  Follow-up work done on GRB host galaxies
show that these systems are quite faint at both rest-frame $UV$ (Chen
et al.\ 2009) and optical wavelengths (Chary et al.\ 2007a).  We can
therefore conclude that most of the star formation at high redshift
occurs in galaxies that are similarly faint.

\subsection{Fractional Light Output in the UV and mid-IR/far-IR}

Before using the extinction estimates for LBG samples
(Table~\ref{tab:effdust}: \S5.4) and the aforementioned estimates of
the ULIRG contribution (\S6.2) to estimate the SFR density, we first
use them to assess the relative flux output from galaxies in the UV
and that output in the mid-IR/far-IR, after dust reprocessing.  This
entire topic is important in thinking about how best to nail down the
star formation rate density and energy output in the high-redshift
universe and for our thinking about the contribution of the
high-redshift universe to the extragalactic background light.

In calculating the relative flux output in the UV and mid-IR/far-IR,
we assume that essentially all the light from the ULIRG population is
emitted in the mid-IR/far-IR and adopting the extinction estimates
from Table~\ref{tab:effdust} for the LBG population.  We include a
plot of this fraction versus redshift in Figure~\ref{fig:irfrac}.  Not
surprisingly the majority of the light output from star-forming
galaxies at $z\sim2.5-4$ occurs in the mid-IR/far-IR (after dust
reprocessing) as has been emphasized repeatedly (e.g., Hughes et
al.\ 1998; P{\'e}rez-Gonz{\'a}lez et al. 2005).  However, as we will
note in \S6.4, it appears that we can account for most of this energy
output using the LBG population.  Again, the situation appears to be
quite different at $z\gtrsim5$, with most of the light output
occurring in the UV.

\subsection{The Star Formation Rate Density at $z$$\,\sim\,$2-6 Including ULIRGs}

Including the contribution from highly obscured ultra-luminous IR
galaxies (\S6.2) in with that derived from the $UV$ LFs
(Table~\ref{tab:sfrdens} and Figure~\ref{fig:sfrdens}), our best
estimates for the star formation rate densities are given in
Table~\ref{tab:sfrdensulirg} and Figure~\ref{fig:sfrdensulirg}.  The
ULIRG population increases the SFR density estimated at $z\sim2.5$ and
$z\sim4$ (to faint limits) by $\sim20$\% and $10$\%, respectively.  We
show the fractional contribution of ULIRGs to the SFR density at
$z$$\,\sim\,$2-6 in Figure~\ref{fig:wrelfrac}.

The present dust-corrected SFR densities are less than has been
considered in many previous studies.  This is largely because many
previous studies corrected the star formation rate density by factors
of $\sim5-8\times$ (e.g., Giavalisco et al.\ 2004b) -- which is the
dust correction relevant for $L^*$ galaxies.  This has important
implications for studies (e.g., Hopkins \& Beacom 2006) featuring
comparisons of the stellar mass density with the star formation rate
density.

The claim has been that the star formation rate density -- after
integration -- exceeds the stellar mass density by factors of
$\sim3-4$ (e.g., by Hopkins \& Beacom 2006; Wilkins et al.\ 2008).
Adopting a small dust correction (as is appropriate including the
effect of the lower luminosity galaxies) substantially reduces the
inferred star formation rate density at high redshift (by factors of
$\sim$1.1, $\sim$1.1, $\sim$2.1, and $\sim$2.5 at $z\sim2.5$,
$z\sim4$, $z\sim5$, and $z\sim6$, respectively relative to that used
by Wilkins et al.\ 2008).  This should help somewhat to resolve the
apparent discrepancy in comparisons with the stellar mass density
inferred at high redshift (see also discussion in Reddy \& Steidel
2009).  Another consideration that may help in this regard is the
stellar mass in lower luminosity galaxies.  Reddy \& Steidel (2009)
argue that a consideration of the SFR density and stellar mass
densities to the same limiting luminosities largely resolves the
apparent discrepancy between the inferred stellar mass density and
that obtained by integrating SFR history (approximately doubling the
inferred stellar mass densities at $z$$\,\sim\,$2-3).

What stands out in the above discussion is the role of lower
luminosity galaxies and their apparently sizeable contribution to
various volume-averaged quantities like the SFR density and stellar
mass density.  Clearly, any reasonable examination of these quantities
(or comparisons between the stellar mass density and the integral of
the SFR density) must accurately account for this population or risk
ignoring those galaxies that are most central to making sense of the
cosmic evolution.

\section{Summary}

We use the deep optical and near-IR imaging data over the HUDF and
other deep, wide-area fields to quantify the distribution of
$UV$-continuum slopes $\beta$ (i.e.,
$f_{\lambda}\propto\lambda^{\beta}$) for star-forming galaxies over a
wide range in redshift ($z$$\,\sim\,$2-6) and luminosity
($0.1L_{z=3}^{*}$ to $2 L_{z=3}^{*}$).  $z$$\,\sim\,$2-6 galaxies are
selected through a $U$, $B$, $V$, and $i$ dropout technique while
$UV$-continuum slopes (1600-2300\AA$\,\,$rest-frame) are derived from
the optical and near-IR broadband colors.  We then corrected the
distribution of $UV$-continuum slopes for observational selection and
photometric errors and tabulate these slopes as a function of both
redshift and luminosity (Table~\ref{tab:uvslope} and
Figure~\ref{fig:colmag}).  We then discuss possible interpretations of
the trends we find (\S4).

Our conclusions are as follows:
\begin{itemize}
\item{The $UV$-continuum slope distribution of $UV$ bright $L_{z=3}^*$
  galaxies over the range $z$$\,\sim\,$2-6 has a mean $\beta$ ranging
  from $-1.2$ to $-2.4$, with a dispersion of $\sim$0.4
  (Table~\ref{tab:uvslope} and Figure~\ref{fig:colmag}).  As found in
  previous studies, we find that the mean $UV$-continuum slope $\beta$
  is bluer (by $\sim0.5$) at $z\sim6$ than it is at $z\sim3-4$.
  \textit{We also find that the mean $UV$-continuum slope $\beta$ is
    bluer (by $\sim$0.5) at lower luminosities than it is at higher
    luminosities (see also Meurer et al.\ 1999).}  In doing so, we
  establish the following correlation between $\beta$ and $M_{UV,AB}$
  (\S3.9): $\beta=(-0.20\pm0.04)(M_{UV,AB}+21)-(1.40\pm0.07\pm0.15)$
  at $z\sim2.5$ and
  $\beta=(-0.15\pm0.01)(M_{UV,AB}+21)-(1.48\pm0.02\pm0.15)$ at
  $z\sim4$.  The present quantification of these distributions is
  essential for accurate LF determinations at high redshift (see
  \S4.3).}
\item{The dropout color selections identify star-forming galaxies at
  $z$$\,\sim\,$2-6 with $UV$-continuum slopes $\beta$ as red as
  $\beta\sim0.5$ (see Figures~\ref{fig:sel} and \ref{fig:seleffect}).
  By contrast, the distribution of $UV$-continuum slopes $\beta$ that
  we observe has a mean value of $\sim-1.2$ to $-2.4$ with a $1\sigma$
  dispersion of $0.4$ -- bluer than the color selection limit (i.e.,
  $\beta\sim0.0-0.5$) by $\Delta\beta \gtrsim1.5$.  If the
  distribution of $UV$-continuum slopes has a simple form (i.e., has
  only a single mode), this suggests that our high-redshift dropout
  selections are largely complete (see also \S4.1 in Bouwens et
  al.\ 2007), and \textit{there is not a large population of
    dust-reddened galaxies at $z\gtrsim4$.}  This deficit of red
  galaxies is particularly prominent relative to the selection
  function at $z\gtrsim5$ (see Figure~\ref{fig:complete}: \S4.2).
  Independent evidence for this come from Balmer break selections at
  $z\sim3-4.5$ (which are biased towards older and more massive
  galaxies) where the distribution of $UV$-continuum slopes $\beta$ is
  also found to be quite blue (Brammer \& van Dokkum 2007: see
  \S4.2).}
\item{Stellar population models were used to investigate the effects
  that changes in dust, age, metallicity, and IMF would have on the
  $UV$-continuum slope $\beta$.  We found that a factor of $\sim$2
  (0.3 dex) decreases in the dust, age, and metallicity would make the
  $UV$-continuum slope $\beta$ bluer by $\sim$0.35, $\sim$0.1, and
  $\sim$0.05, respectively.  Changing the slope of the IMF of galaxies
  by $\sim0.5$ from Salpeter has a minimal effect, changing $\beta$ by
  $\lesssim$0.1.  The $UV$-continuum slope $\beta$ appears to be most
  sensitive to changes in the dust extinction.  \textit{Dust is likely
    the most significant cause for changes in $\beta$ as a function of
    redshift and luminosity (\S4.4).}}
\item{Assuming that the observed correlation between dust extinction
  and $UV$-continuum slope $\beta$ at $z\sim0-2$ (e.g., Meurer et
  al.\ 1995; Meurer et al.\ 1999; Burgarella et al.\ 2005; Laird et
  al.\ 2005; Reddy et al.\ 2006; Dale et al.\ 2007) holds out to
  $z\sim6$, we estimate the probable dust extinction in
  $z$$\,\sim\,$2-6 LBG selections (see Table~\ref{tab:effdust}).
  \textit{For $>0.3L_{z=3}^{*}$ galaxies at $z\sim2.5$, our estimate
    is 6.0$\pm$2.5} -- very similar to the values derived by Reddy et
  al.\ (2006) and Erb et al.\ (2006b) using a variety of different
  multiwavelength data and SFR calibrators, but $\sim$3$\times$ higher
  than Carilli et al.\ (2008) estimated for $>$0.2$L_{z=3}^{*}$
  $z\sim3$ $U$-dropouts by stacking radio observations over the COSMOS
  field.  \textit{At $z$$\,\sim\,$5-6 and to much lower luminosities
    ($>0.04L_{z=3}^{*}$), the inferred dust extinction is much less,
    i.e., $\lesssim$2-3$\times$.}  }
\item{Since lower luminosity galaxies dominate the luminosity density
  in the $UV$ at $z\gtrsim2$ (Bouwens et al.\ 2006, 2007; Yan \&
  Windhorst 2004; Beckwith et al.\ 2006; Reddy \& Steidel 2009: see
  Figure~\ref{fig:relfrac}), establishing the dust extinction of lower
  luminosity galaxies is crucial for estimates of the total star
  formation rate density.  As a result of the very low dust
  extinctions inferred for lower luminosity galaxies, \textit{we
    expect the average dust extinction for the star-forming population
    at $z$$\,\sim\,$2-6 integrated to very low luminosities to be
    quite small ($\lesssim2\times$: see Table~\ref{tab:effdust}).}
  This is similar to what Reddy \& Steidel (2009) concluded from the
  observations and various physical arguments (see Table 5 and Figure
  10 from that work).}
\item{We establish a conversion between $UV$ luminosities and
  bolometric luminosity based on the aforementioned correlation
  between dust extinction and $\beta$ and the observed correlation
  between $\beta$ and $M_{UV,AB}$: $L_{bol} = 10^{11.67 - 0.58
    (M_{UV,AB}+21)}$ $L_{\odot}$ (though we emphasize that this
  relationship may not work for galaxies where $L_{bol}>10^{12}$
  $L_{\odot}$).  We remark that it is striking that the UV LF cuts off
  at luminosities ($\sim$1-2 $L_{UV}^{*}$) that approximately
  correspond to that of ULIRGs ($\sim10^{12}$ $L_{\odot}$: see
  Figure~\ref{fig:colmagill}).  \textit{This suggests that
    $L_{UV}^{*}$ at $z\sim3$ is set by dust extinction -- or more
    precisely the luminosity at which dust obscuration becomes so
    significant as to offset the increase in energy output from
    stars.}  This may explain why the value of $L_{UV} ^*$ (e.g.,
  Steidel et al.\ 1999; Reddy et al.\ 2008) does not evolve much from
  $z\sim4$ to $z\sim2$ (see \S5.5).}
\item{We correct our SFR density estimates upperward to account for
  the contribution from dust obscured ULIRGs.  At $z\sim2.5$, we
  follow the analysis of Reddy \& Steidel 2009 in using the $z\sim2$
  IR LF of Caputi et al.\ (2007) to account for the contribution from
  these galaxies and make a similar correction at $z\sim4$ based upon
  $\sim5-8$ plausible submm candidates at $z\sim4$ in the HDF-North
  GOODS SCUBA field discussed by Daddi et al.\ (2009: see \S6.2).
  \textit{We find that this obscured population only increases the SFR
    density (and dust extinction) estimates by $\sim$20\% at
    $z\sim2.5$ and by just $\sim10$\% at $z\sim4$
    (Figure~\ref{fig:wrelfrac}).  Given the evolution towards bluer
    $UV$-continuum slopes $\beta$ at high redshifts in both LBG and
    BBG selections (\S4.2), we argue that the contribution of
    dust-obscured ULIRGs to the SFR density at $z\gtrsim5$ is even
    less than at $z\sim4$ (i.e., $\lesssim10$\%).  The model we
    construct in \S6.2 (assuming that $L_{bol}^{*}$ evolves similarly
    to $L_{UV}^{*}$) suggests that the fractional contribution of
    ULIRGs to the SFR density is only 6\% and 2\% at $z\sim5$ and
    $z\sim6$, respectively.}}
\end{itemize}
The distribution of $UV$-continuum slopes at $z$$\,\sim\,$2-6 provides
us with a key window of information for studying the formation and
evolution of star-forming galaxies from very early times.  The very
blue colors of galaxies at $z$$\,\sim\,$5-6 and at lower luminosities
suggest that they are indeed much less evolved than their lower
redshift or higher luminosity counterparts.  These results suggest
that the dust corrections at high redshifts are not particularly large
and that LBG (dropout) selections should give a substantially complete
census of the SFR density at $z\gtrsim5$ and probably even at
$z\gtrsim4$.  Moreover, highly dusty galaxies seem unlikely to provide
a significant contribution to the SFR density at these early times.

In the future, we expect to be able to constrain the distribution of
$UV$-continuum slopes $\beta$ at $z\sim1-2$ and $z\sim5-8$ with much
more precision and accuracy than is possible at present.  These
advances should come using data from the HST WFC3/UV and WFC3/IR
instrument to be installed on HST.  This instrument will allow us to
collect deep UV data needed to efficiently select star-forming
galaxies at $z\sim1-3$ and the near-IR data needed to estimate the
$UV$-continuum slopes $\beta$ for $z\sim4-8$ galaxies.\\

\acknowledgements

We would like to thank Louis Bergeron, Susan Kassin, Dan Magee,
Massimo Stiavelli, and Rodger Thompson for their assistance in the
reduction of NICMOS data which has been essential for quantifying the
$UV$-continuum slopes for faint star-forming galaxies at
$z$$\,\sim\,$5-6.  We thank Veronique Buat, Denis Burgarella, David
Elbaz, Daniel Schaerer and Naveen Reddy for stimulating conversations
and Roderik Overzier and Naveen Reddy for helpful feedback on our
submitted paper.  Alice Shapley kindly sent us an electronic copy of
her stacked spectrum of $z\sim3$ LBGs so we could compare it against
the model spectra we use to estimate the $UV$-continuum slopes
$\beta$.  We acknowledge Roelof de Jong and other scientists at STScI
for their efforts at characterizing the non-linearity in the NICMOS
detector.  We are appreciative to our referee for detailed and
insightful feedback which greatly improved this manuscript.  Finally,
we believe thanks are especially due to all those at NASA, STScI and
throughout the community who have worked so diligently to make Hubble
the remarkable observatory that it is today.  The servicing missions,
like the recent SM4, have rejuvenated HST and made it an
extraordinarily productive scientific facility time and time again,
and we greatly appreciate the support of policymakers, and all those
in the flight and servicing programs who contributed to the repeated
successes. We acknowledge support from NASA grants HST-GO09803.05-A
and NAG5-7697.

\begin{figure}
\epsscale{0.6}
\plotone{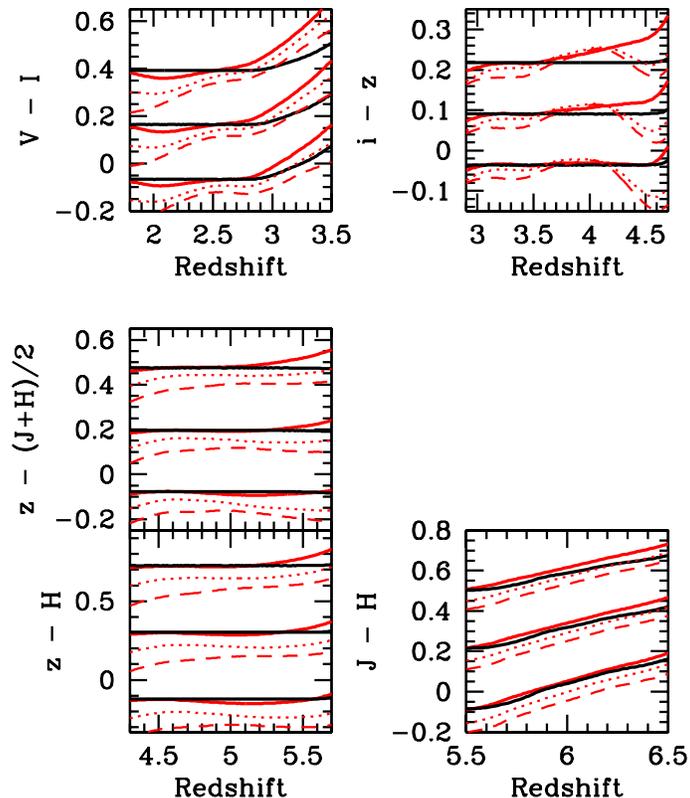}
\caption{Expected $UV$ colors for star-forming galaxies in our
  $z\sim2.5$ $U$, $z\sim4$ $B$, $z\sim5$ $V$, and $z\sim6$ $i$ dropout
  selections.  The black lines (from bluest to reddest) correspond to
  power-law SEDs with $\beta$'s of $-$2.2, $-$1.5, and $-$0.8,
  respectively.  The red lines (from bluest to reddest) correspond to
  SEDs computed assuming an $e^{-t/\tau}$ star formation history,
  $t=10$ Myr, $\tau=70$ Myr, $[Z/Z_{\odot}]=-0.7$, the Bruzual \&
  Charlot (2003) libraries, a Salpeter IMF, and dust extinctions
  $E(B-V)$ of 0.0, 0.15, and 0.3, respectively (Calzetti et
  al.\ 2000).  These latter models have effective $UV$-continuum
  slopes $\beta$ of $-$2.2, $-$1.5, and $-$0.8 over the range
  1600\AA$\,\,$to 2300\AA$\,\,$ -- the same as the black lines.
  However, the red curves should be much more realistic -- being based
  upon stellar population models (Bruzual \& Charlot 2003).  We
  reference the measured colors from each of our dropout samples to
  this fiducial set of SEDs (red lines) to account for intrinsic
  differences in the rest-frame wavelengths of the broadband imaging
  data available for each of our samples (Table~\ref{tab:pivotbands}
  and \S3.4).  Also shown (dashed and dotted lines, respectively) are
  the predicted colors for SEDs computed assuming a young stellar
  population (a 30 Myr constant star formation rate model) and a
  somewhat older stellar population (a model for which star formation
  proceeded at a constant rate for 100 Myr and then ceased for 10 Myr
  prior to observation).  Dust extinction for these alternate star
  formation histories is varied so that the UV-continuum slope $\beta$
  is equal to the same values ($-2.2$, $-1.5$, and $-0.8$) as for our
  fiducial SEDs.  In general, the $UV$-continuum slopes $\beta$ we
  derive from the $UV$ colors shown here show little dependence upon
  the redshift of sources in our dropout samples.  Small changes in
  the mean redshift of our selection (i.e., $\Delta z \sim 0.1$)
  result in shifts of $\sim0.02$, $\sim0.02$, $\sim0.01$, and
  $\sim0.05$ in the $\beta$'s derived for our $U$, $B$, $V$, and $i$
  dropout selections (see Appendix A).\label{fig:colz}}
\end{figure}

\appendix

\section{Deriving the $UV$-continuum slopes from the observed colors}

Here we describe our method for converting the rest-frame $UV$ colors
we measure for galaxies in our samples to their equivalent
$UV$-continuum slopes.  We base these conversions on model SEDs
(Bruzual \& Charlot 2003) calibrated to have $UV$-continuum slopes
$\beta$ of $-$2.2, $-$1.5, and $-$0.8 over the wavelength range
$\sim$1600\AA$\,\,$to $\sim$2300\AA$\,\,$(to match the wavelength
probed by the deep multiwavelength data available for our sample).
The model SEDs assume a $e^{-t/\tau}$ star formation history, $t=70$
Myr, $\tau=10$ Myr, $[Z/Z_{\odot}]=-0.7$, and a Salpeter IMF with
$E(B-V)$ dust extinction of 0.0, 0.15, and 0.3 applied using the
Calzetti et al.\ (2000) attenuation law.  We have calculated the color
of these model SEDs as a function of redshift by integrating these
SEDs across the filter sensitivity curves.  The results are shown in
Figure~\ref{fig:colz} as the solid red lines.  We also have calculated
the colors expected when the model SED has a perfect power law slope
$f_{\lambda}\propto\lambda^{-2}$ and cuts off at 912\AA$\,\,$(see
black lines on Figure~\ref{fig:colz}).

Then, knowing the effective $\beta$ of the model SEDs and the observed
colors of these SEDs at the mean redshift of our different dropout
samples, we can derive a relation between the measured colors and the
$UV$-continuum slope.  The derived relations are
\begin{eqnarray}
\beta = 3.04 (V-I) - 1.99~~&\textrm{(for U-dropouts)} \\
\beta = 5.30 (i-z) - 2.04~~&\textrm{(for B-dropouts)} \\
\beta = 2.45 (z-(J+H)/2) - 1.98~~&\textrm{(for V-dropouts with $J_{110}$-band data)} \\
\beta = 1.61 (z-H) - 1.96~~&\textrm{(for V-dropouts)} \\
\beta = 2.47 (J-H) - 2.27~~&\textrm{(for i-dropouts)}
\label{eq:ibeta}
\end{eqnarray}
The mean redshifts assumed here for our $U$, $B$, $V$, and $i$
dropouts in deriving these relations are 2.5, 3.8, 5.0, and 5.9
(Bouwens et al.\ 2004; Bouwens et al.\ 2007).  If the mean redshifts
of our dropout selections differ from the model redshifts by $\Delta
z\sim0.1$, the derived $\beta$'s would change by $\sim0.02$,
$\sim0.02$, $\sim0.01$, and $\sim0.05$ with respect to those quoted
here.

While the conversion formulas above were derived assuming a very
plausible star formation history for $z$$\,\sim\,$2-6 galaxies, the actual
star formation histories for individual galaxies in our samples likely
show considerable variation.  Some galaxies are likely in the midst of
starbursts when we observe them, while other galaxies have likely
undergone bursts some time in the past and so would have slightly
older stellar populations.  This results in some degree of variation in
the shape of the SEDs in our samples -- which is above and beyond what
we can parametrize using the model SEDs above, with a simple
one-parameter variation in the overall dust extinction.

To show the extent to which the observed colors depend on the assumed
star formation history, in Figure~\ref{fig:colz} we also show these
colors for a relatively young stellar population (with 30 Myr of
constant star formation) and a somewhat older stellar population (a
model for which star formation proceeded at a constant rate for 100
Myr and then ceased for 10 Myr prior to observation).  Dust
attenuation for these alternate star formation histories is varied so
that the UV-continuum slope $\beta$ is alternatively $-2.2$, $-1.5$,
and $-0.8$.  The colors of these model SEDs at the mean redshift of
our dropout samples are similar to that for our fiducial model (by
construction).  The only exception is the $i-z$ color for our
$B$-dropout selection which probes a slightly different wavelength
range (i.e., Table~\ref{tab:pivotbands}) than the default range
1650\AA$\,\,$to 2300\AA.

We can attempt to use these model SEDs to estimate the approximate
uncertainty in converting from the observed colors to $UV$-continuum
slope $\beta$.  Assuming the high-redshift population is composed of
an equal mix of the two alternate model SEDs and our fiducial model
SED and averaging over redshift, we estimate an error $\Delta \beta$
of 0.02, 0.10, 0.02, and 0.02 in the derived $\beta$'s at $z\sim2.5$,
$z\sim4$, $z\sim5$, and $z\sim6$.

\end{document}